\documentclass[a4paper, aps, prd, nofootinbib, notitlepage, 10pt]{revtex4-1}% twocolumn
\usepackage[OT1,T1]{fontenc}
\usepackage[latin1]{inputenc}
\usepackage{amsmath}
\usepackage{amsfonts}
\usepackage{tikz-cd}
\usetikzlibrary{positioning}
\usepackage{bbm}
\usepackage{accents}
\newcommand{\utilde}[1]{\undertilde{#1}}
\newcommand{\di}{\mathrm{d}} %Differential
\newcommand{\ou}[3]{{#1}{}^{#2}{}_{#3}} %Indexstellung
\newcommand{\uo}[3]{{#1}{}_{#2}{}^{#3}} %Indexstellung
\newcommand{\I}{\mathrm{i}} %imaginaere Einheit
\newcommand{\E}{\mathrm{e}} %Euler Zahl
\newcommand{\ellp}{{\ell_{\mathrm{P}}}} %Planck Laenge
\newcommand{\CC}{\mathrm{cc.}} % komplex konjugiertes
\newcommand{\lbreck}{[}

\newcommand{\vom}{\overset{v}{\omega}{}}
\newcommand{\vpi}{\overset{v}{\pi}{}}
\usepackage{dsfont}
\usepackage{mathrsfs}
\usepackage{color}
\newcommand{\qq}[1]{``#1''} %Anfuehrungszeichen
\usepackage{hyperref}
\newenvironment{subalign}{\subequations\align}{\endalign\endsubequations}
\newcommand{\Ref}[1]{(\ref{#1})}
\newcommand{\eref}[1]{(\ref{#1})}
\usepackage[scaled=0.97]{berasans}
\newcommand{\M}{\text{\sffamily\bfseries{M}}}
\newcommand{\C}{{\mathbb C}}
\newcommand{\N}{{\mathbb N}}
\newcommand{\R}{{\mathbb R}}
\newcommand{\Z}{{\mathbb Z}}
\newcommand{\T}{{\mathbb T}}

\newcommand{\cJ}{{\mathcal J}}

\newcommand{\cD}{{\mathcal D}}

\newcommand{\SU}{\mathrm{SU}}

\newcommand{\SL}{\mathrm{SL}}

\newcommand{\U}{\mathrm{U}}

\renewcommand{\sl}{{\mathfrak{sl}}}

\newcommand{\be}{\begin{equation}}
\newcommand{\ee}{\end{equation}}
\newcommand{\bea}{\begin{eqnarray}}
\newcommand{\eea}{\end{eqnarray}}
\newcommand{\nn}{\nonumber}

\newcommand{\w}{\wedge}

\newcommand{\f}{\frac}

\def\p{\partial}

\newcommand{\id}{\mathbbm{1}}

\newcommand{\re}{\mathrm{Re}}
\newcommand{\im}{\mathrm{Im}}

\newcommand{\la}{\langle}
\newcommand{\ra}{\rangle}
\newcommand{\bra}[1]{\la {#1}|}
\newcommand{\ket}[1]{|{#1}\ra}

\let\a=\alpha \renewcommand{\b}{\beta} \let\g=\gamma  
\renewcommand{\d}{\delta}  \let\eps=\epsilon
\let\th=\theta   
\let\m=\mu              \let\r=\rho \let\om=\omega
 \let\vphi=\varphi \let\Si=\Sigma 

\newcommand{\tpi}{\utilde{\pi}} \newcommand{\tom}{\utilde{\om}} \newcommand{\uPi}{\utilde{\Pi}}
\newcommand{\upi}{\utilde{\pi}} \newcommand{\uom}{\utilde{\om}} 
\newcommand{\J}{{\cal J}} \newcommand{\uJ}{\utilde{\cal J}}
 
\newcommand{\po}{\pi\om} \newcommand{\upo}{\tpi\tom}

\begin{document}
\title{The twistorial structure of loop-gravity transition amplitudes}
\author{Simone Speziale}
\author{Wolfgang M. Wieland}
%\email{simone.speziale@cpt.univ-mrs.fr}
%\email{wolfgang.wieland@cpt.univ-mrs.fr}
\affiliation{Centre de Physique Théorique,
   Campus de Luminy, Case 907,
   13288 Marseille, France, EU}
\thanks{Unité Mixte de Recherche (UMR 7332) du CNRS et des Univ. 
Aix-Marseille et Sud Toulon Var. Unité affiliée à la
FRUMAM.}

\date{\today}
\begin{abstract}\noindent
The spin foam formalism provides transition amplitudes for loop quantum gravity. Important aspects of the dynamics are understood, but many open questions are pressing on. In this paper we address some of them using a twistorial description, which brings new light on both classical and quantum aspects of the theory. 
At the classical level, we clarify the covariant properties of the discrete geometries involved, and the role of the simplicity constraints in leading to SU(2) Ashtekar-Barbero variables. We identify areas and Lorentzian dihedral angles in twistor space, and show that they form a canonical pair. The primary simplicity constraints are solved by \emph{simple} twistors, parametrized by SU(2) spinors and the dihedral angles. We construct an SU(2) holonomy and prove it to correspond to the (lattice version of the) Ashtekar-Barbero connection. We argue that the role of secondary constraints is to provide a non trivial embedding of the cotangent bundle of SU(2) in the space of simple twistors.
At the quantum level, a Schr\"odinger representation leads to a spinorial version of simple projected spin networks, where the argument of the wave functions is a spinor instead of a group element. We rewrite the Liouville measure on the cotangent bundle of SL(2,C) as an integral in twistor space. Using these tools, we show that the Engle-Pereira-Rovelli-Livine transition amplitudes can be derived from a path integral in twistor space. We construct a curvature tensor, show that it carries torsion off-shell, and that its Riemann part is of Petrov type D. Finally, we make contact between the semiclassical asymptotic behaviour of the model and our construction, clarifying the relation of the Regge geometries with the original phase space.

\end{abstract}

\maketitle

%--------------------------------------------------------------------------------------
\section{Introduction}\label{Intro}
%--------------------------------------------------------------------------------------
\noindent
An intriguing relation between loop quantum gravity and twistors has recently emerged in the literature \cite{twigeo2,WielandTwistors,IoHolo,IoTwistorNet}. It relies on a new parametrization of the classical phase space of holonomies and fluxes. In this paper we push this representation further and study dynamical properties of the theory, addressing a number of open questions. Among these, we show that transition amplitudes for loop quantum gravity can be written as path integrals in twistor space, and that torsion is present off-shell.

In the first part of the paper, we focus on classical aspects of the twistorial representation. We review the results appeared in \cite{twigeo2,WielandTwistors,IoHolo,IoTwistorNet}, and put them together in a coherent picture. Using the twistorial parametrization, we identify a pair of canonically conjugated variables, that corresponds geometrically to an area and a Lorentzian dihedral angle.
We study the algebra of primary simplicity constraints in twistor space, and show that the constraint surface is parametrized by SU(2) holonomies and fluxes, plus the dihedral angle. The \emph{simple} twistors solution of the constraints are determined by SU(2) spinors and the dihedral angle.
We then argue that solving the secondary constraints leads to a non-trivial embedding of $\SU(2)$ variables in the covariant space, a structure that reproduces at the discrete level what is achieved by the Ashtekar-Barbero variables. In fact, we also prove that the reduced SU(2) connection is the (lattice version of the) parallel transport with respect to the Ashtekar-Barbero connection.
Finally, we give the map from twistors to covariant twisted geometries, and show that the role of the simplicity constraints is to match the left and right metric structures, precisely as in the continuum theory.

In the second part of the paper, we use these structures to derive a series of results for the quantum theory.
In Section \ref{SecQuant}, we show that the classical phase space and its algebra of constraints can be quantized, leading to a Hilbert space of quantum twistor networks. This is achieved choosing a Schr\"odinger quantization of twistor space, as in \cite{WielandTwistors}. The resulting states are wave functions on spinors, instead of cylindrical functions on the group. A basis is given by the homogeneous functions appearing in the unitary irreducible representations of the Lorentz group. They carry a representation of the spinorial Heisenberg algebra, that includes the holonomy-flux algebra, and introduce a new framework for covariant loop quantum gravity. 
Treating the diagonal simplicity constraints as first class, and the second class off-diagonal constraints via a master constraint technique, we derive what we call simple quantum twistor networks. These are related to the simple projected spin networks of \cite{EteraLifting}, which appear \cite{AlexandrovNewVertex,IoCovariance} as boundary states of the EPRL spin foam model \cite{EPRL} and its generalizations to arbitrary cellular decompositions \cite{KKL,CarloGenSF}  (see also \cite{EPR,EPRlong,LS,LS2,FK,BahrOperatorSF,BarrettLorAsymp,HanZhangLor,Mikovic:2011zx}, and \cite{PerezLR} for a recent introduction). The change of representation between the spinorial wave functions and the cylindrical functions can be given explicity, and appears naturally in the construction of the spin foam dynamics.
The existence of an SU(2)-invariant scalar product and the equivalence with the Hilbert space of loop quantum gravity are natural results in our formalism.

In Section \ref{SecMeasure}, we use the twistorial parametrization to rewrite the Liouville measure of the cotangent bundle $T^*\SL(2,\C)$. This requires a Faddeev-Popov procedure for the area-matching constraint, which we provide explicitly, and prove gauge-invariance of the path integral. The result has applications to the spin foam formalism in general.

In Section \ref{secV} we derive the EPRL transition amplitudes as a path integral in twistor space, using the previous results: the quantized phase space, the Liouville measure, plus a discretization of the $BF$ action which is bilinear in the spinors. The result is an independent derivation of the model, where the wedge amplitude is like the infinitesimal step of a Feynman path integral, with the intermediate position eigenstates given by the constrained states, and a ``straight'' evolution given by the $BF$ action. The calculation is based on the properties of the homogeneous functions in the boundary states, and involves the evaluation of a certain complex integral. The final amplitude perfectly coincides with the (generalized) EPRL model, up to additional face factors that can be independently specified, for instance as argued in \cite{BojoPerez,Carlodimj}.

The EPRL model has the important property of reproducing the Regge action in the large spin limit \cite{BarrettLorAsymp} (see also \cite{ConradyFreidel2, ConradyFreidel3, BarrettEPRasymp, HanZhangEucl, HanZhangLor}). In Section \ref{secVI}, we present the relation between the semiclassical Regge behaviour and the covariant phase space. In doing so, we explain some key aspects of the large spin asymptotics, which have so far gone unnoticed. First, part of the saddle point equations determine a certain subset of the primary simplicity constraint surface. Second, the secondary constraints are solved by the remaining saddle point equations. This key step rests significantly on the restriction to triangulations and the assumption of flatness of the 4-simplices, a case in which a Levi-Civita connection is known from Regge calculus. 
We show that the dihedral angles entering the Regge action are precisely those of our phase space.
We discuss the ``semi-coherence'' of the model, in the sense that not the whole phase space structure plays a dynamical role. This shows up, among other things, in the fact that the areas are purely quantum numbers.
We also introduce a curvature tensor, and show that off-shell it has non-vanishing torsional components, consistently with the continuum formalism.
In the conclusions we summarize our results and discuss some of the new lines of research that this program proposes.

There are two appendices. The first gives a list of conventions on spinor calculus and useful formulae for the irreducible unitary representations of the Lorentz group,
while the second appendix contains the explicit evaluation of an integral entering the derivation of the amplitude.
Further concerning conventions, the papers \cite{twigeo2,IoHolo,IoTwistorNet} use an index-free notation, whereas \cite{WielandTwistors} uses the standard spinorial notation. Here, primi juvenes, we work with explicit indices. 
In the Appendix, we provide a translation to the index-free notation.
Accordingly, $A,B,C,\dots$ are spinor indices in the left-handed representations, their complex conjugate (i.e. their right handed counterparts) carry bars, i.e. we write $\bar A,\bar B,\bar C,\dots$. Also, $I,J,K,\dots$ label internal Minkowski vectors, $a,b,c,\dots$ are abstract indices in tangent space, and $i,j,k,\dots$ run from one to three. The metric signature is $(-,+,+,+)$, resulting in Minkowski vectors corresponding to anti-Hermitian matrices in the irreducible $(\boldsymbol{\frac{1}{2}},\boldsymbol{\frac{1}{2}})$ representation of $\SL(2,\mathbb{C})$. Brackets $(\cdots)$ and $[\cdots]$ surrounding the indices denote their normalised symmetrisation and anti-symmetrisation respectively, and $\epsilon^{0123}=1$ fixes the normalisation of the internal Levi-Civita tensor.

%--------------------------------------------------------------------------------------
\section{From twistors to twisted geometries \label{secII}}
%--------------------------------------------------------------------------------------

%--------------------------------------------------------------------------------------
\subsection{Twistors and $\boldsymbol{T^*\SL(2,\C)}$}\label{twiststar}
%--------------------------------------------------------------------------------------
\noindent
Following our previous works \cite{WielandTwistors, twigeo2,IoHolo,IoTwistorNet}, we begin with an abstract, oriented graph decorated by a pair of twistors $(Z,\utilde{Z})\in\T^2\simeq\mathbb{C}^8$ on each link, and associate $Z$ and $\utilde{Z}$ to the source and target points respectively. This will be a general rule for us, \qq{tilded} quantities always belong to the final point.
Each twistor is described as a pair of spinors, $Z=(\omega^A,\bar\pi_{\bar A})\in\mathbb{C}^2\oplus\bar{\mathbb{C}}^2{}^\ast=:\mathbb{T}$, where $\omega^A$ is left-handed, 
and the right-handed part $\bar\pi_{\bar A}$ lies in the complex-conjugate dual vector space. We equip the space $\T^2$ with an $\SL(2,\mathbb{C})$-invariant symplectic structure, whose non-vanishing Poisson brackets are
\be
\big\{\pi_A,\omega^B\big\}=\delta^B_A=-\big\{\utilde{\pi}_A ,\utilde{\omega}^B\big\}, \quad\text{and}\quad
\big\{\bar\pi_{\bar A} ,\bar\omega^{\bar B} \big\}=\delta^{\bar B}_{\bar A} =-\big\{\utilde{\bar\pi}_{\bar A},\utilde{\bar\omega}^{\bar B}\big\}.
\label{spinbrack}
\ee
We have chosen opposite signs between tilded and untilded Poisson brackets, but symmetric brackets would work as well.

As it is well known from the literature \cite{PenroseRindler1, PenroseRindler2, Penrose72}, $\T$ and $\T^2$ carry a representation of the Lorentz group, preserving the symplectic structure.
The generators of the action are the left-handed bispinors
\be\label{fluxpar}
\Pi^{AB}  = -\f12\om^{(A}\pi^{B)},
\qquad \utilde{\Pi}^{AB}  = \f12\utilde{\om}^{(A}\utilde{\pi}^{B)},
\ee
and their right-handed complex conjugates. That is, the Hamiltonian vector fields of $(\Pi, \bar\Pi)$ and $(\uPi, \bar\uPi)$ generate the canonical $\SL(2,\mathbb{C})$ action on $Z$ and $\utilde{Z}$.
The proof is a straightforward application of the brackets \Ref{spinbrack}, but also needs some additional $\SL(2,\C)$ structures. The first is the invariant antisymmetric $\epsilon$-tensor (its components fixed by requiring $\epsilon_{01}=\eps^{01}=1$) mapping contravariant spinors to their algebraic duals and vice versa,
\begin{equation}
\omega_A=\epsilon_{BA}\omega^B,
\qquad \omega^A=\epsilon^{AB}\omega_B, 
\qquad
\epsilon^{AC}\epsilon_{BC}=\uo{\epsilon}{B}{A}=\delta^A_B.
\end{equation}
The second are the anti-Hermitian $\mathfrak{sl}(2,\mathbb{C})$ generators 
$\ou{\tau}{A}{Bi}$, related to the Pauli matrices $\sigma_i$ through $2\I\tau_i:=\sigma_i$. These matrices induce a map between $\mathfrak{sl}(2,\mathbb{C})$ and $\C^3$:
\begin{equation}
{\Pi}\in\mathfrak{sl}(2,\mathbb{C}) \ : \ \ou{\Pi}{A}{B}=\Pi^i\ou{\tau}{A}{Bi}, \qquad \Pi^i\in \C^3.\label{selfcomp}
\end{equation}
Then,
\be
\big\{\Pi_i ,\Pi_j \big\}  =  -\uo{\epsilon}{ij}{k}\Pi_k,  \quad  
\big\{\uPi_i ,\uPi_j \big\}  =  -\uo{\epsilon}{ij}{k}\uPi_k,  
\quad \big\{\Pi_i ,\utilde{\Pi}_j \big\} = 0, \label{pibracks}
\ee
and the same for their conjugated. 
These are the Poisson brackets for (two copies of) the Lorentz algebra, in the chiral splitting. The chiral generators 
\begin{equation}\label{rotboost}
\Pi_i=\frac{1}{2}\big(L_i+\I K_i\big),\qquad \utilde{\Pi}_i=\frac{1}{2}\big(\utilde{L}_i+\I\utilde{K}_i\big)
\end{equation}
are complex, with real and imaginary parts, $L$ and $K$, generating rotations and boosts respectively.
We will later identify the $\mathfrak{sl}(2,\mathbb{C})$ elements $\Pi$ and $\utilde{\Pi}$ with the gravitational fluxes, i.e. the Plebanski 2-form smeared over 2-dimensional submanifolds in the hypersurface of initial data. Be this hypersurface space-like, we require $\Pi^{AB}\Pi_{AB}\neq 0$, implying linear independence of the spinors:
\be\label{nonull}
\pi\omega := \epsilon_{AB}\pi^A\omega^B=\pi_A\omega^A\neq 0, \qquad \upo \neq 0.
\ee
With this restriction, the pair $(\pi^A,\omega^A)$ forms a complete basis of $\mathbb{C}^2$, just as well as $(\utilde{\pi}^A,\utilde{\omega}^A)$. We introduce the linear map translating one to the other,
\begin{equation}
\ou{h}{A}{B}\omega^B=\utilde{\omega}^A,\quad\ou{h}{A}{B}\pi^B=\utilde{\pi}^A,\label{holdef}
\end{equation}
which we will later identify with the holonomy along the link.
For this map to be unimodular, i.e. $h\in \SL(2,\C)$, it must preserve the bilinear generated by $\epsilon_{AB}$,
hence
\begin{equation}\label{defC}
C 
:= \pi\omega-\utilde{\pi}\utilde{\omega}=0.
\end{equation}
In the following, we will refer to \Ref{defC} as the (complex) \emph{area-matching constraint}.
Furthermore, thanks to the restriction \Ref{nonull}, we can uniquely parametrise the holonomy in terms of the basis spinors as
\be
\ou{h}{A}{B}  =  \frac{\utilde{\omega}^A\pi_B-\utilde{\pi}^A\omega_B}{\sqrt{\utilde{\pi}\utilde{\omega}} \, \sqrt{\pi\omega}}.\label{holpar}
\ee
The functions $\Pi$, $\utilde{\Pi}$ and $h$ are related by
\be\label{offshellh}
\utilde{\Pi}=-\frac{\pi\omega}{\utilde{\pi}\utilde{\omega}}h\Pi h^{-1},
\ee
and span 14 out of 16 dimensions of $\T^2$. They obey the Poisson brackets 
\begin{subequations}\begin{align}\label{offshell}
& \big\{\Pi_i,h \big\}  =  -h \tau_i, \qquad \big\{\utilde{\Pi}_i,h \big\}  =  \tau_ih,  \\
& \big\{\ou{h}{A}{B},\ou{h}{C}{D}\big\}=-\frac{2C}{(\pi\omega)(\utilde{\pi}\utilde{\omega})}\big(\epsilon^{AC}\Pi_{BD}+\epsilon_{BD}\utilde{\Pi}^{AC}\big).
\end{align}\end{subequations}
On the constraint hypersurface $C=0$, two key properties hold: Firstly, the adjoint representation relates the fluxes, 
$\utilde{\Pi}=-h\Pi h^{-1}$. Secondly, the components of the holonomy commute. Hence, we recover the Poisson algebra of $T^*\SL(2,\C)$ with $\Pi$ and $\uPi$ as the (chiral) left- and right-invariant Hamiltonian vector fields on the group manifold:
\begin{equation}\label{holfluxal}
\big\{\Pi_i,\Pi_j\big\}=-\uo{\epsilon}{ij}{k}\Pi_k,
\qquad \big\{\Pi_i,h\big\}=-h \tau_i, \qquad \big\{\uPi_i,h\big\}=\tau_i h,
\qquad \big\{h^A{}_B,h^C{}_D \big\}\big|_{C=0}=0.
\end{equation}

In fact, we can easily see that this procedure amounts exactly to a symplectic reduction $T^*\SL(2,\C)\simeq\SL(2,\C)\times\mathfrak{sl}(2,\C)\simeq
\T^2/\!\!/C$. On the $C=0$ constraint hypersurface, the Hamiltonian vector field $\mathfrak{X}_C=\{C,\cdot\}$ generates the orbits
\begin{equation}
\exp{(z\mathfrak{X}_C+\bar{z}\mathfrak{X}_{\bar{C}})}:(\omega,\pi,\utilde{\omega},\utilde{\pi})\mapsto(\E^z\omega,\E^{-z}\pi,\E^z\utilde{\omega},\E^{-z}\utilde{\pi}),\quad z\in\mathbb{C}.\label{scaltrans}
\end{equation}
The functions $\Pi, \uPi$ and $h$ are invariant under such gauge transformations, and thus span the space obtained by symplectic reduction.

Let us add two more remarks to complete the analysis. First, the map between twistors and holonomy-flux variables 
 is 2--to--1, since exchanging spinors as
\begin{equation}
(\omega,\pi,\utilde{\omega},\utilde{\pi})\mapsto(\pi,\omega,\utilde{\pi},\utilde{\omega})\label{distrans}
\end{equation}
leaves both holonomy \eref{holpar} and flux \eref{fluxpar} unchanged. Hence, to arrive at the reduced phase space, we also have to divide out this residual $\Z_2$ symmetry \Ref{distrans}. 
Second, because of the restriction $\po\neq 0$, the spinorial parametrization cannot cover the submanifold of $(h,\Pi):\Pi^{AB}\Pi_{AB}=-\mathrm{Tr}\,\Pi^2=0$, and what we truly find is $T^\ast\SL(2,\mathbb{C})$ removed from all its null configurations. %$\mathrm{Tr}\,\Pi^2=0$. 
The complete isomorphism could be defined through a suitable treatment of the degenerate configurations, see e.g. the analogue situation in the $\SU(2)$ case \cite{twigeo}. However, below we will identify $\Pi$ with the Plebanski field smeared over 2-dimensional surfaces in a $t=\mathrm{const.}$ slice of initial data. Be this hypersurface space-like, the restriction is automatically fulfilled, and has no physical consequence for the following.

%---------------------------------------------------------------------------------
\subsection{Twistors from the LQG action}\label{ex}
%---------------------------------------------------------------------------------
\noindent 
The interest in the phase space of $\SL(2,\C)$ holonomies and fluxes comes from loop quantum gravity.
We work in the first-order tetrad formalism, and start from the often-called Holst action for general relativity. In terms of chiral variables, it is
\begin{equation}
S_{\mathrm{Holst}}[A,e]=\frac{\hbar}{\ellp^2}\frac{\beta+\I}{\I\beta}\int_M\ou{\Sigma}{A}{B}(e)\wedge \ou{F}{B}{A}(A)+\CC,\label{holstactn}
\end{equation}
where $\ellp=\sqrt{8\pi\hbar G_{\mathrm{N}}/c^3}$ is the Planck length,  $\beta>0$ is the Barbero--Immirzi parameter, and \qq{$\CC$} denotes complex conjugation of everything preceding. See e.g. \cite{CorichiHolstBoundary10,AshtekarBoundaryTerms, BianchiWielandSurf,SmolinGRasSFstate} for the case with boundary terms.
The action \Ref{holstactn} is a non-analytic functional of the left-handed $\mathfrak{sl}(2,\mathbb{C})$ connection $A$, and the four soldering forms $e$ transforming in the irreducible ${\bf (\tfrac{1}{2},\tfrac{1}{2})}$ representation of $\SL(2,\mathbb{C})$. Curvature $\ou{F}{A}{B}$ and Plebanski 2-form $\ou{\Sigma}{A}{B}(e)$ are uniquely determined by the equations
\begin{align}
\ou{F}{A}{B}(A)=\di \ou{A}{A}{B}+\ou{A}{A}{C}\wedge\ou{A}{C}{B}, \qquad
\ou{\Sigma}{A}{B}(e)=e^{A\bar C}\wedge e_{B\bar C} = 
\f{\I}2 \tau^A{}_{Bi} \left(\frac{\I}{2}\epsilon^i{}_{lm} e^l\wedge e^m + e^0\wedge e^i\right), \label{sigdef}
\end{align}
where in the last expression, we have explicitly put the projector onto the left-handed variables.

In order to read off the symplectic structure, a 3+1 split $M=\Sigma\times\mathbb{R}$ is needed.
We take the pullback of the $\mathfrak{sl}(2,\mathbb{C})$ connection onto the spatial hypersurface $\Sigma_t=\Sigma\times\{t\}$, call this (by a little abuse of notation) $\ou{A}{i}{a}$ and find its conjugate momentum to be
\begin{equation}
\uo{\Pi}{i}{a}=-\frac{\hbar}{\ellp^2}\frac{\beta+\I}{4\I\beta}\tilde{\epsilon}^{abc}\Sigma_{ibc},\label{pisigmarel}
\end{equation}
where $\tilde{\epsilon}^{abc}$ is the spatial Levi-Civita density\footnote{Itself constructed from the 4-dimensional Levi-Civita density $\tilde{\epsilon}^{dabc}$ via 
$\di t{\tilde{\epsilon}}^{abc}={\tilde{\epsilon}}^{dabc}\partial_dt$, where $t$ is the time coordinate.}
on $\Sigma_t$. 
The only non-vanishing Poisson brackets are
\begin{equation}
\big\{\uo{\Pi}{i}{a}(p),\ou{A}{j}{b}(q)\big\}=\delta^j_i\delta^a_b\tilde{\delta}(p,q)=\big\{\uo{\bar\Pi}{i}{a}(p),\ou{\bar A}{j}{b}(q)\big\},\label{poissklamm}
\end{equation}
where $\tilde{\delta}(p,q)$ is the three-dimensional Dirac distribution (a scalar density) on $\Sigma_t$, and $\delta^a_b$, $\delta^j_i$ are spatial Kronecker symbols.
The Cauchy hypersurface $\Sigma_t=\Sigma\times\{t\}$ of initial data carries a time normal $n_a$, allowing us to define a Hermitian metric for $\mathbb{C}^2$. We work in the time gauge, in which this normal gives (or rather: is represented by) the identity matrix, that is the one corresponding to the canonical $\SU(2)$ subgroup of the Lorentz group:
\begin{equation}
\delta^{A\bar A}:=-\I\sqrt{2}n^{A\bar A}=-\I\sqrt{2}\ou{e}{A\bar A}{a}n^a,\quad\text{time-gauge:}\quad
\delta^{0\bar 0}=1=\delta^{1\bar 1},\quad\;\delta^{0\bar 1}=0=\delta^{1\bar 0}.
\label{tgauge}
\end{equation}
It is this metric with respect to which the Pauli matrices \Ref{selfcomp}
are Hermitian, and it is this normal with respect to which the real and imaginary parts of $\Pi$ correspond to rotations and boosts \eref{rotboost} respectively.

The chiral splitting has led us to a complex phase space. To guarantee that the metric is real, reality conditions must be imposed. Using our gauge condition \Ref{tgauge} equation \Ref{sigdef} constrains in fact all components of $\tilde{\epsilon}^{abc}\Sigma_{ibc}$ to be real. 
This imposes the reality conditions 
\begin{equation}
\frac{1}{\beta+\I}\Pi_i+\CC=0 
\quad \Leftrightarrow \quad K_i+\beta L_i=0
\quad \Leftrightarrow \quad
\Pi_i= -\E^{\I\vartheta} \bar\Pi_i, \label{realcond}
\end{equation}
on the momentum $\Pi_i$. Here we have introduced the angle
\be
\E^{\I\vartheta}=\frac{\beta+\I}{\beta-\I},\qquad  \beta = \cot\tfrac{\vartheta}{2}.
\ee
The intermediate form $K+\beta L=0$ shows explicitly, as highlighted in \cite{Wieland1}, that the reality conditions amount to the canonical version of the primary simplicity constraints, in their linear version introduced in \cite{EPRL}. The final form further shows that they match the two chiral metric structures induced by $\SL(2,\C)$, as discussed in \cite{Capo2,MikeLR,Iobimetric}.

A complete canonical analysis of the action can be found e.g. in \cite{AshtekarBook,Barros,BuffenoirPleb,AlexandrovSO4cov,AlexandrovReality,Wieland1}. 
Preserving the primary constraints \eref{realcond} under Hamiltonian time evolution leads to secondary constraints, given by the vanishing of the spatial projection $\tilde{\epsilon}^{abc}D_b\ou{e}{i}{c}$ of the torsion 2-form. Once the primary constraints are solved, the secondary ones imply that the spatial part of the Lorentz connection is Levi-Civita.\footnote{Completing the canonical analysis \cite{AshtekarBook,Wieland1} shows that also the remaining components of the 4-dimensional torsion 2-form vanish, some of them implying evolution equations for the triad $e^i$ on the spatial hypersurface, 
the others fixing the boost component of the Lagrange multiplier $\Lambda^i=A^i(\partial_t)$ to the value $\im(\Lambda^i)=N^a\ou{K}{i}{a}+e^{ia}\partial_aN$, where $\ou{K}{i}{a}=\im(\ou{A}{i}{a})$ is the extrinsic curvature and $N$, $N^a$ and $\partial_t$ denote lapse, shift and the time flow vector-field.}
The system of primary and secondary simplicity constraints is second class, and 
canonical coordinates on the reduced phase space are provided by Ashtekar-Barbero variables\footnote{These are defined in the time-gauge, but fully covariant formulations exist \cite{Barros,AlexandrovChoice,Cianfrani,GeillerNewLook}. An alternative parametrization is suggested by Alexandrov \cite{AlexandrovChoice}.}
\cite{Ashtekar:1987gu,Barbero,Immirzi96real}. 

The crucial step towards loop quantisation \cite{ThiemannBook, AshtekarReport} is a certain ``covariant'' smearing of the continuous gravitational phase space. 
One introduces a graph $\Gamma$ in the spatial manifold, consisting of oriented links $\gamma,\gamma^\prime,\dots$, to each of which we assign a dual, i.e. an oriented surface $t, t^\prime, \dots$ transversally intersecting the corresponding links. We may think of the graph $\Gamma$ as being dual to a cellular decomposition of the spatial manifold, each node of $\Gamma$ dual to a 3-cell.
The elementary phase space variables are then smeared over these lower dimensional objects, obtaining a collection of holonomies and fluxes:
\begin{equation}
\SL(2,\mathbb{C})\ni h[t]= h_\gamma=\mathrm{Pexp}\Big( -\int_\gamma A\Big),\quad
\mathfrak{sl}(2,\mathbb{C})\ni\Pi[t]=\int_{p\in t} h_{q\rightarrow \gamma(0)}\Pi_p h_{\gamma(0)\rightarrow q}.
\label{smrdvar}
\end{equation}
Here $\mathrm{Pexp}$ is the path ordered exponential, and $h_{q\rightarrow \gamma(0)}$ denotes the holonomy parallely transporting from the integration variable $q\in t$ along $\gamma$ towards the initial point $\gamma(0)$ of the link dual to the surface $t$. %, in \cite{AshtekarQG3,ThiemannSymplectic,FreidelGeiller} this construction is made much more explicit.
Since each surface $t$ carries an orientation there is also the oppositely oriented element $t^{-1}$ which come along with
\begin{equation}
h[t^{-1}]=h[t]^{-1},\quad\text{and}\quad \Pi[t^{-1}]=-h[t]\Pi[t]h[t]^{-1}\equiv\utilde{\Pi}[t].\label{ornt}
\end{equation}
Under this smearing, the Poisson structure \Ref{poissklamm} reduces precisely to \Ref{holfluxal} on each link, while variables at different links commute. See \cite{AshtekarQG3, ThiemannBook} and also \cite{FreidelGeiller} for a more recent discussion.

The smeared phase space is thus the Cartesian product of $T^*\SL(2,\C)$ associated to each link, which we derived from a network of twistors supplemented by the complex area-matching condition \Ref{defC}.
Local Lorentz invariance is imposed as in lattice gauge theories by a closure condition on the  $\SL(2,\C)$ generators at each node, namely ${\cal G}_n = \sum_{t\in n}\Pi[t]=0$ and its complex conjugate.
Then, we have to realize the primary and secondary simplicity constraints, and find the analogue of the Ashtekar-Barbero variables for the space of solutions. This is a challenging problem, and a general solution is still unknown.
As pointed out long ago \cite{BarrettCrane,BarrettCraneLor,BaezBarrett}, a successful discretization can be found restricting attention to a 4-simplex, and further assuming that the interior is flat. In this case we are in the framework of Regge calculus:
the geometry is uniquely described by edge lengths, and the Levi-Civita connection is known.
In the general case, the situation is different.
The primary constraints remain tractable, as they do not involve the connection, and can be defined on the boundary of any 4-cell, without assumptions on its interior.
The secondary constraints, on the other hand, lack so far a general treatment. 
The main difficulty is that outside the framework of Regge geometries we do not know how to define the Levi-Civita connection. And in general, even on shell of the primary constraints, the data on the boundary graph do not describe Regge geometries, but a generalization going under the name of  twisted geometries \cite{twigeo,twigeo2,IoSigmaDiscrete}, which we will review below. These (i) are defined for a general cellular decomposition, (ii) allow for discontinuities in the metric (possibly related to torsion), and (iii) do not impose conditions on the interior curvature of the 4-cells. The Levi-Civita connection is so far not understood in this framework.
We will see in the rest of this Section how the twistor description of the phase space brings new light to many of these questions.

%-------------------------------------------------------
\subsection{Reality conditions and reduction to $\boldsymbol{\SU(2)}$ spinors}
%-------------------------------------------------------
\noindent
We will now rewrite the reality conditions in terms of the spinorial representation, and solve them explicitly.
The result will reduce twistors down to SU(2) spinors, with the emergence of the $\SU(2)$ holonomy of the $\beta$-dependent Ashtekar-Barbero connection.
We discretize \Ref{realcond} on both source and target variables of each link,
\be\label{Simpl}
\forall t: \qquad\Pi[t] = \E^{i\vartheta} {\Pi}^\dagger[t], \qquad\uPi[t] = \E^{i\vartheta} {\uPi}^\dagger[t],
\ee
where the Hermitian conjugate is taken according to
\begin{equation}
\ou{(\Pi^\dagger)}{A}{B}=\delta^{A\bar A}\delta_{B\bar B}\ou{\bar{\Pi}}{\bar B}{\bar A}.
\end{equation}
In the spinorial parametrization, the first equation in \Ref{Simpl} reads
\be
\om_{(A}\pi_{B)} = -\E^{\I\vartheta} \delta_{A\bar A}\delta_{B\bar B}\bar{\om}^{(\bar{A}}\bar\pi^{\bar{B})}.\label{simpltwist}
\ee
It apparently gives two equivalent decompositions of $\Pi_{AB}$ in terms of spinors and their complex conjugate. But the decomposition of a symmetric bispinor is unique up to exchange and complex rescaling of the constituents, therefore $\pi$ and $\om$ must be linearly related. Furthermore, part of the complex rescaling is fixed by the phase appearing explicitly in \Ref{simpltwist}, leaving only the freedom to real rescalings. Hence, we can parametrize the solutions as 
\be
\pi_A = r \E^{\I\frac{\vartheta}{2}}\delta_{A\bar A}\bar\om^{\bar A}, \qquad \om_A = -\frac{1}{r}\E^{\I\frac{\vartheta}{2}} \delta_{A\bar A}\bar\pi^{\bar{A}},\qquad r\in\R-\{0\}.\label{realsolvd}
\ee
The matching of left and right geometries as implied by \Ref{Simpl} immediately translates into the left and right spinors being proportional. The same conclusion holds in a general gauge, with a generic normal replacing the identity matrix, as in \Ref{tgauge}. Remarkably, the simplicity equations then take up the same form as Penrose's incidence relation. It would be intriguing to explore the existence of a deeper connection between these two notions.
That simplicity implies proportionality of the spinors is a key result, and was also derived in \cite{IoHolo}. It means that a \emph{simple} twistor, i.e. a twistor satisfying the simplicity constraints, is determined by a single spinor, plus a real number, whose meaning will become clear below. 

By contractions with $\omega$ and $\pi$, equation \Ref{simpltwist} can be conveniently separed in two parts,
\begin{align}
F_1=\frac{\I}{\beta+\I}\omega^A\pi_A+\CC=0,\qquad
F_2=\frac{\I}{\sqrt{2}}\delta^{A \bar A}\pi_A\bar{\omega}_{\bar A}=n^{A\bar A}\pi_A\bar{\omega}_{\bar A}=0.
\end{align}
Here, $F_1$ is real and Lorentz-invariant, while $F_2$ is complex but only $\SU(2)$ invariant. 
Following the literature, we will refer to $F_1$ as the \emph{diagonal} simplicity constraint, and $F_2$ as  \emph{off-diagonal}.
The constraints $F_i$, $i=1,2$, the corresponding $\utilde{F}_i$ for the tilded spinors, and the area matching $C$, form a system of constraints on the link space $\T^2\cong \C^8$.
The algebra can be easily checked to give
\begin{equation}
\{F_1, F_2 \} = -\frac{2\I\beta}{\beta^2+1}F_2, \quad \{F_2, \bar F_2\} = \I \, {\im}(\pi\omega),\quad \{C,F_1\}=0,\quad\{C,F_2\}=-F_2=-\{\bar{C},F_2\},\label{FCalg}
\end{equation}
and the same for tilded quantities.

The system should be supplemented with secondary constraints coming from a suitable Hamiltonian, and we will come back to this point below, because it plays an important role in the identification of the extrinsic curvature. Neglecting any secondary constraints for the moment, we conclude that the diagonal simplicity constraints $F_1$ and $\utilde{F}_1$  are of first class, as well as $C$, whereas $F_2$ and $\utilde{F}_2$ are second class. That some constraints are second class even in the absence of secondary constraints is a well-known consequence of the non-commutativity \Ref{holfluxal} of the fluxes.
The first class constraints generate orbits inside the constraint hypersurface.
The orbits of $C$ are given in \Ref{scaltrans}, whereas those generated by the diagonal simplicity constraints are found from
\be
\{F_1, \om^A\} = \f{\I}{\beta+ \I} \om^A, \qquad \{F_1, \pi_A\} = - \f{\I}{\beta+ \I} \pi_A.
\ee 
We also remark that the system is \emph{reducible}, since only three of the four constraints $F_1, \utilde{F}_1$ $\re( C )$ and $\im( C)$ are linearly independent. 
We thus have three independent first class constraints, and two, complex, second class constraints. The reduced phase space has $16-3\times 2-2-2=6$ real dimensions, and we will now prove it to be $T^*\SU(2)$. To that end, it is convenient to treat separately the area matching and the simplicity constraints, the order being irrelevant.
There are two convenient choices of independent constraints, depending on the order in which one solves them. If solving the simplicity first, we can choose
\be
\label{indepC}
C_{\mathrm{red}}=\frac{C}{\beta+\I}+\CC, \quad\underbrace{ F_1,\quad\utilde{F}_1, \quad F_2, \quad \utilde{F}_2}_{F}.
\ee
If instead we solve $C$ first, we can take
\be
\label{indepC2}
\re( C ), \quad \im( C), \quad \underbrace{D:=F_1+\utilde{F}_1, \quad F_2, \quad \utilde{F}_2}_{F_{\rm red}}.
\ee
The situation is summarised in Figure \ref{FigRed}.

\begin{figure}[ht]\begin{tikzpicture}[descr/.style={fill=white,inner sep=2.5pt}]
  \matrix(m)[matrix of math nodes, row sep=4em,  column sep=8em]
  {\T^2 & \C^2\times \C^2 \\
 T^*\SL(2,\C) & T^*\SU(2) \\ };
  \path[->,font=\small]
  (m-1-1)edge node[descr] {$F$} (m-1-2)
         edge node[descr] {$C$} (m-2-1)
  (m-2-1)edge node[descr] {$F_{\rm red}$} (m-2-2)  
  (m-1-2)edge node[descr] {$C_{\rm red}$} (m-2-2);
\end{tikzpicture}
\caption{Primary constraint structures between twistor and holonomy-flux spaces. $F$ and $C$ schematically denote the simplicity and area matching constraints, and arrows include division by gauge orbits, when relevant.}\label{FigRed}   
\end{figure}

Let us proceed solving the simplicity constraints first. For the untilded quantities, \Ref{realsolvd} solves all four $F=0$ constraints, however the expression is not $F_1$-gauge-invariant. For each half-link, gauge-invariant quantities live on the reduced space $\T/\!\!/F\simeq\C^2$, and are 
 parametrized by a single spinor, say $z^A\in\mathbb{C}^2$. 
Since the simplicity constraints introduce a Hermitian metric, we have a norm
$\|\omega\|^2=\delta_{A\bar A}\omega^A\bar\omega^{\bar A}$, and use it to define

\be
\mathcal{J} = \f{\|\omega\|^2}{\sqrt{1+\beta^2}} \, r, \label{Jdef}
\ee
which satisfies $\{F_1,\mathcal{J}\}=0$. In terms of $\cJ$, equation \Ref{realsolvd} gives
\begin{equation}
\pi_A=(\beta+\I)\frac{\mathcal{J}}{\|\omega\|^2}\delta_{A \bar A}\bar\omega^{\bar A},
\qquad \po = (\beta+\I)\J.\label{piparam}
\end{equation}
Then, the reduced spinor, $F_1$-gauge-invariant, can be taken to be
\begin{equation}
z^A = \sqrt{2\mathcal{J}}\frac{\omega^A}{\|\omega\|^{\I\beta+1}}, \qquad \|z\| = \sqrt{2\mathcal{J}}.
\label{spinordef1}
\end{equation}
Since we are assuming $\pi_A\omega^A\neq 0$, this implies $\cJ\neq 0$. We can further always assume $\mathcal{J}>0$: In the case $\mathcal{J}<0$, the sign is flipped by simultaneously exchanging $\pi$ with $\omega$ and $\utilde\pi$ with $\utilde\omega$, and we have already seen this operation to be a symmetry \Ref{distrans} of our spinorial parametrisation. 
Hence, selecting the sign of $\cJ$ removes the $\Z_2$ symmetry of the reduction.

The same results apply to the tilded quantities.
The reduced space $\T^2/\!\!/{F}\simeq\C^2\times\C^2$ is parametrised by the following spinors,
\begin{equation}
z^A = \sqrt{2\mathcal{J}}\frac{\omega^A}{\|\omega\|^{\I\beta+1}}, 
\qquad
\utilde{z}^A  = \sqrt{2\utilde{\mathcal{J}}}\frac{\utilde\omega^A}{\|\utilde\omega\|^{\I\beta+1}}.
\label{spinordef}
\end{equation}
Notice that they transform linearly under rotations, but not under boosts: they are SU(2) spinors. The Lorentzian structures are partially eliminated by the gauge-choice needed to define the linear simplicity constraints.

To get the Dirac brackets for the reduced SU(2) spinors, we introduce the embedding  $\iota$ of the $F=0$ constraint hypersurface into the original twistorial phase space, and compute the pullback of the symplectic potential. This gives
\begin{align}\label{reducedpot}\nn
\iota^\ast \Theta &=\iota^\ast \big(\pi_A\di\omega^A-\utilde{\pi}_A\di\utilde{\omega}^A+\CC\big)\\\nn
&=\iota^\ast \bigg[ \beta (\J + \uJ) \, \di \, \ln \left(\f{\|\om\|}{\|\tom\|}\right) + \beta (\J-\uJ) \, \di\,\ln (\|\om\| \|\utilde{\omega}\|)
+\bigg(\I\frac{\mathcal{J}}{\|\omega\|^2}\delta_{A\bar A}\bar{\omega}^{\bar A}\di\omega^A
-\I \frac{\utilde{\mathcal{J}}}{\|\utilde{\omega}\|^2}\delta_{A\bar A}\bar{\utilde{\omega}}^{\bar A}\di\utilde{\omega}^A+\CC\bigg)\bigg]\\
 &= \frac{\I}{2}\delta_{A \bar A}\big(\bar{z}^{\bar A}\di z^A-\bar{\utilde{z}}^{\bar A}\di \utilde{z}^A-\CC\big).
\end{align}
The induced Dirac brackets are the canonical brackets of four harmonic oscillators, 
\begin{equation}
\big\{\bar{z}^{\bar A},z^A\big\}_{\mathrm{D}}=-\I\delta^{\bar AA}=-\big\{\bar{\utilde{z}}^{\bar A},\utilde{z}^A\big\}_{\mathrm{D}}.\label{osci}
\end{equation}

This reduction is illustrated in the top horizontal line of Figure \ref{FigRed}. The next step is to implement the area-matching condition. As anticipated, part of $C=0$ is automatically satisfied on the surface of $F_1=\utilde{F}_1 =0$. Using \Ref{spinordef}, the independent  part $C_{\rm red}$ can be seen to give the real-valued $\SU(2)$ version of the area-matching condition introduced in \cite{twigeo2}, that is
\begin{equation}\label{Cred}
C_{\mathrm{red}} =\|z\|^2-\|\utilde{z}\|^2=0.
\end{equation}
The gauge orbits generated by $C_{\mathrm{red}}$ are $\U(1)$ phase transformations $z\mapsto\E^{\I \varphi} z$, for some angle $\varphi$.
As proven in \cite{twigeo2}, canonical variables on the reduced phase space $(\C^2\times\C^2)/\!\!/C_{\mathrm{red}}$ are SU(2) holonomies and fluxes, satisfying their canonical Poisson algebra. We are thus left with the phase space $T^\ast\SU(2)$, with elements $(U,\Sigma)\in\SU(2)\times\mathfrak{su}(2)$ parametrized as\footnote{With respect to the literature \cite{twigeo2,IoZako}, we have added the dimensional coefficents of the physical flux induced from the action. Also, the holonomy appearing here does not flip the spinors along the link, consistently with the definition of $h$. The alternative choice is to swap $\uom$ and $\upi$ in \Ref{holdef}. This allows to eliminate the opposite sign of the initial Poisson brackets, and induces an extra $\eps$ tensor in the the twisted geometries parametrization given below.}  \cite{twigeo2}
\be\label{SU2hf}
\ou{U}{A}{B}(z,\utilde{z}) = \frac{\utilde{z}^A\delta_{B\bar B}\bar{z}^{\bar B}+\delta^{A\bar A}\bar{\utilde{z}}_{\bar A}z_B}{\|z\|\|\utilde{z}\|},
\qquad \Sigma_{AB}(z,\utilde{z}) = \frac{\beta \ellp^2}{\hbar}\frac{\I}{2} z_{(A}\delta_{B)\bar B}\bar{z}^{\bar B}.  
\ee
This proves that the symplectic reduction of $\T^2$ by the area-matching and simplicity constraints gives $T^*\SU(2)$.

Let us conclude this Section with two important remarks.
The first is the identification of an abelian pair of canonically conjugated variables on $T^*\SL(2,\C)$. We introduce the quantity
\be\label{Xidef}
\Xi:=2 \ln \left(\frac{\|\omega\|}{\|\utilde{\omega}\|}\right).
\ee
An explicit calculation shows that at $C=0$
\be\label{repoXi}
\{\re (\po), \Xi\} = 1.
\ee
Also, from the second line of the symplectic potential \eref{reducedpot}, we immediately see that 
\be\label{JXi}
\{\cJ, \Xi\}=\f{1}{\beta}
\ee
on the surface of $F_i=\utilde{F}_i=0$.
The conjugated pair corresponds to the (oriented) area and (boost) dihedral angle associated with the dual face $t$.
In fact, from \eref{pisigmarel}, the squared area equals
\be
A^2[t] := \delta^{ij}\Sigma_i[t]\Sigma_j[t]=\frac{\ellp^4\beta^2}{\hbar^2} \mathcal{J}^2.
\ee
Notice also that the quantity $\re(\po)$ appearing in \Ref{repoXi} reduces to the area when the simplicity constraints are satisfied.
As for the dihedral angle, it is defined by the scalar product between the time-like normals of the two 3-cells sharing the face, that is $n$ and $\utilde{n}$. These are both related to the identity matrix by the time gauge \eref{tgauge}. The non-trivial information is then carried by the SO(1,3) holonomy $\Lambda(h_\g)$ between the two, needed to evaluate the scalar product in the same frame. A short calculation then gives
\begin{equation}
\begin{split}
\utilde{n}_I\ou{\Lambda (h_\gamma)}{I}{J}n^J&=\utilde{n}_{A\bar A}\ou{h_\gamma}{A}{B}\ou{\bar{h}_\gamma}{\bar A}{\bar B}n^{B\bar B}=-\frac{1}{2}\frac{1}{|\pi\omega|^2}\delta_{A\bar A}\delta^{B\bar B}\big(\uom^A\pi_B-\upi^A\omega_B\big)\big(\utilde{\bar\omega}^{\bar A}\utilde{\bar\pi}_{\bar B}-\utilde{\bar\pi}^{\bar A}\utilde{\bar\omega}_{\bar B}\big)=\\
&=-\frac{1}{2}\bigg(\frac{\|\uom\|^2}{\|\omega\|^2}+\frac{\|\omega\|^2}{\|\uom\|^2}\bigg)=-\mathrm{ch}(\Xi),
\end{split}
\end{equation}
valid on the constraint surface \eref{indepC}. 
The dihedral angle between 3-cells describes the extrinsic curvature in Regge calculus, therefore this abelian pair captures a scalar part of the ADM Poisson brackets, as we'll make clearer in the next Section.

The second remark concerns the orbits generated by $D$. Let us define the hypersurface of $T^*\SL(2,\C)$ solution of the simplicity constraints. From \Ref{indepC2}, we see that on the space reduced by $C=0$, that is $T^*\SL(2,\C)$, the independent simplicity constraints are $D=F_2=\utilde{F}_2=0$. 
These equations characterize a 7-dimensional constraint hypersurface within $\T^2$. 
From the previous construction, we know that six dimensions are spanned by the SU(2) holonomy-flux variables, or equivalently by the SU(2) spinors reduced by \Ref{Cred}. Since
\begin{equation}
\{D,z^A\}=0=\{D,\utilde{z}^A\},\quad\{D,\Xi\}=\frac{4}{1+\beta^2},\label{Dtrafos}
\end{equation}
the seventh dimension spreads along the orbits of $D$, each of which can be parametrized by the angle $\Xi$. Accordingly, we denote the constraint surface $T_\Xi$, and $T_\Xi\simeq T^*\SU(2)\times \R$.
This means that a pair of simple twistors, solutions of the area-matching and the simplicity constraints, are parametrized by the SU(2) spinors, plus the dihedral angle.

On $T_\Xi$, the Lorentz fluxes already coincide with the $\mathfrak{su}(2)$ Lie algebra elements introduced in \Ref{SU2hf}, providing a discrete counterpart of the continuum equation \eref{pisigmarel}. For the Lorentz holonomy we find, plugging \Ref{spinordef} and \Ref{Xidef} into \Ref{holpar}, 
\be\label{hred}
h_{\rm red}{}^{A}{}_{B}\equiv\ou{h}{A}{B}\big|_{F=0} = \frac{\E^{-\f12(\I\beta+ 1)\Xi}\utilde{z}^A\delta_{B\bar B}\bar{z}^{\bar B}+\E^{\f12(\I\beta+ 1)\Xi}\delta^{A\bar A}\bar{\utilde{z}}_{\bar A}z_B}{\|z\|\|\utilde{z}\|}.
\ee
This is still a completely general $\SL(2,\C)$ group element. If we now choose the specific $\Xi=0$ section through the orbits of $D$, it reduces to an SU(2) holonomy, and coincides with the $D$-invariant holonomy $U$.
The constraint hypersurface $T_\Xi$ plays an important role, because there we can distinguish the reduced Lorentz holonomy \Ref{hred} from the SU(2) holonomy \Ref{SU2hf}. The difference is captured by the orbits of the diagonal simplicity constraint.

%-------------------------------------------------------------------------------------
\subsection{Ashtekar-Barbero holonomy and extrinsic curvature}\label{secAB}
%-------------------------------------------------------------------------------------
\noindent
Consider the constraint hypersurface $T_\Xi$, and the two holonomies $U(z,\utilde{z})$ and $h_{\rm red}(z,\utilde{z},\Xi)$. While $h_{\rm red}$ describes the Lorentzian parallel transport, we now show that the $\SU(2)$ holonomy $U(z,\utilde{z})$ equals the holonomy of the real-valued Ashtekar-Barbero connection $A^{(\beta)}=\Gamma+\beta K$ (here $\Gamma$ and $K$ are the real and imaginary components of the selfdual $\SL(2,\C)$ connection $A=\Gamma+\I K$). 
Namely, that
\begin{equation}\label{UABprvn}
U(z,\utilde{z}) = U_\gamma:=\mathrm{Pexp}\Big(-\int_\gamma \Gamma +\beta K \Big).
\end{equation}
This identification is very important for the spin foam formalism, and the understanding of the relation between covariant and canonical structures. It is needed to match the boundary states appearing in spin foam models with the $SU(2)$ spin network states found from the canonical approach, see e.g. the discussions in \cite{AlexandrovNewVertex,IoCovariance,AlexandrovSigma,AlexandrovHilbert02}.

To prove \Ref{UABprvn}, let us first recall (see equation \Ref{smrdvar}) that $h$ is a left-handed group element corresponding to the parallel transport by the left-handed part of the Lorentz connection, $A=\Gamma+\I K$, where  $\Gamma$ represents the intrinsic covariant 3-derivative. This 3-derivative defines the SU(2) parallel transport
\be
G_\gamma:=\mathrm{Pexp}\Big(-\int_\g \Gamma^i\tau_i\Big)\in \SU(2).
\ee
The intrinsic and extrinsic contributions to the holonomies can be disentangled via an ``interaction picture'' for the path-ordered exponentials,\footnote{This can be explicitly proven by looking at the defining differential equation for the holonomy, which admits a unique solution for the inital conditions $U_{\gamma(0)}=\mathds{1}=h_{\gamma(0)}$.
It is the same type of equality that appears in the interaction picture used in time-dependent perturbation theory, with $\Gamma$ being the free Hamiltonian, and  $K$ the potential.}
\begin{align}\label{hintpic}
& h_\g = \mathrm{Pexp}\Big(-\int_\g \Gamma +\I K \Big) = 
G_\g \, \mathrm{Pexp}\Big(-\I\int_0^1\di t \, G_{\g(t)}^{-1}K_{\gamma(t)}(\dot\g) G_{\g(t)} \Big)
\equiv G_\g \, V_K, \\
& U_\gamma = \mathrm{Pexp}\Big(-\int_\g \Gamma +\beta K \Big) = G_\g\, \mathrm{Pexp}\Big(-\beta\int_0^1\di t \, G_{\g(t)}^{-1}K_{\gamma(t)}(\dot\g) G_{\g(t)} \Big)
\equiv G_\g \, V_K^\beta.
\end{align}
Both holonomies provide maps $\C^2\mapsto\C^2$ between tilded and untilded spinors, but while $h$ transports the covariant $\om^A$-spinors, $U$ transports the reduced spinors $z^A$. Let us introduce a short-hand ket notation,
\begin{equation}
|0\rangle\equiv \frac{z^A}{\|z\|},\quad |1\rangle\equiv\frac{\delta^{A\bar A}\bar{z}_{\bar A}}{\|z\|},\quad
|\utilde{0}\rangle\equiv\frac{\utilde{z}^A}{\|\utilde{z}\|},\quad |\utilde{1}\rangle\equiv\frac{\delta^{A\bar A}\bar{\utilde{z}}_{\bar A}}{\|\utilde{z}\|}.\label{braks}
\end{equation}
The holonomies can be thus characterized as the unique solutions to the equations
\be
\ket{\utilde{0}} = e^{(\I\beta+ 1)\Xi/2} h \ket{0} = U \ket{0}, 
\qquad \ket{\utilde{1}} = e^{(-\I\beta+ 1)\Xi/2} (h^\dagger)^{-1} \ket{1} = U \ket{1}.
\ee

Next, we recall that the source and target generators of the Lorentz algebra are related via the holonomy, see  \Ref{offshellh}. 
This relation, together with the simplicity constraints, implies that
\begin{equation}
\Pi=\E^{\I\vartheta}\Pi^\dagger=-\E^{\I\vartheta}(h^{-1}\utilde{\Pi} h)^\dagger=
-h^\dagger\utilde{\Pi}(h^{-1})^\dagger=
h^\dagger h\Pi(h^\dagger h)^{-1}.\label{pippo}
\end{equation} 
We see that the simplicity constraints automatically lead to a certain ``alignment'' between the holonomy and the generators, that immediately translates into an equation for the spinors:
\begin{equation}
\ou{(h^\dagger h)}{A}{B}\omega^B=\E^{-\Xi}\omega^A,
\qquad\ou{(h^\dagger h)}{A}{B}\pi^B=\E^{\Xi}\pi^A,\label{bstspin}
\end{equation}
with $\Xi$ given in \Ref{Xidef}.
Inserting \Ref{hintpic} in \Ref{bstspin}, we find
\begin{equation}
V_K^\dagger V_K |0\rangle=\E^{-\Xi}|0\rangle,
\qquad V_K^\dagger V_K |1\rangle=\E^{\Xi}|1\rangle.
\end{equation}
For small extrinsic curvature, we have that $V_K>0$ and $V_K^\dagger=V_K$ such that this eigenvalue equation has just one solution, given by\footnote{The solution is exact if the extrinsic curvature is covariantly constant along the link, i.e. $G_{\gamma(t)}^{-1}K_{\gamma(t)}(\dot\gamma)G_{\gamma(t)}$ is $t$-independent.}
\begin{equation}\label{Veigen}
V_K=\E^{-\Xi/2}|0\rangle\langle 0|+\E^{\Xi/2}|1\rangle\langle 1|.
\end{equation}
Within the same approximation, we also have 
\begin{equation}\label{Vbeigen}
V_K^\beta=\E^{\I\beta\Xi/2}|0\rangle\langle 0|+\E^{-\I\beta\Xi/2}|1\rangle\langle 1|.
\end{equation}
Finally, using the interaction picture in \Ref{pippo}, as well as properties \Ref{Veigen} and \Ref{Vbeigen}, we find 
\be\label{noname}
U \ket{0} = e^{(\I\beta+ 1)\Xi/2} h \ket{0} = G V_K^\beta \ket{0}, 
\qquad U \ket{1} = e^{(-\I\beta+ 1)\Xi/2} (h^\dagger)^{-1} \ket{1} =G V_K^\beta \ket{1},
\ee
and since $\ket{0}$ and $\ket{1}$ are a complete basis, this proves the desired result \Ref{UABprvn}.

We remark that what we have proved here is valid as an approximation for small curvature. That is, the precise statement is that the SU(2) holonomy $U$ provides the \emph{lattice version} of the Ashtekar-Barbero connection. As the phase space on a fixed graph only carries a notion of holonomy, and not of point-wise connection, our result is perfectly satisfying. If on the other hand one were interested in an exact continuous equivalence, this would require a projection on the simplicity constraint surface performed at every point of the graph \cite{AlexandrovNewVertex,AlexandrovHilbert02}. As pointed out above, it would be true also in the case of covariantly constant extrinsic curvature.

The equation \Ref{noname} provides a discrete counterpart to $A^{(\beta)}=\Gamma+\beta K = A +(\beta-\I)K$, with $\Xi$ playing the role of the extrinsic curvature. In this respect, notice also that from the linearized form of \Ref{Veigen}, and the continuum interpretation of $V_K$, we deduce
\begin{equation}
\Xi \approx \int_0^1\di s\,\ou{R^{(\mathrm{ad})}(G^{-1}_{\gamma(s)})}{i}{j}K^j_{\gamma(s)}(\dot\gamma)n_i[t],\label{bndryterm}
\end{equation}
where $\ou{R^{(\mathrm{ad})}(G)}{i}{j}\in SO(3)$ is the $\SU(2)$ element $G$ in the adjoint representation.
That is, the dihedral angle approximates the extrinsic curvature smeared over the dual link, projected down onto the direction $n^i[t]$ normal to the surface. 
As anticipated earlier, the canonical pairing \Ref{JXi} between $\Xi$ and the area $A[t]$ nicely describes the scalar part of the ADM phase space of general relativity,
where \cite{ThiemannBook} flux $\uo{\Sigma}{i}{a}$ and extrinsic curvature $\ou{K}{i}{a}$ are canonical conjugated.

We conclude that the $\SU(2)$ spinors $z$ and $\utilde{z}$ obtained from the symplectic reduction parametrise holonomies and fluxes of the SU(2) Ashtekar-Barbero variables. 
To prove this identification, it has been necessary to work on the covariant phase space, or at least on the constraint hypersurface $T_\Xi\cong T^*\SU(2)\times \R$, where we could disentangle extrinsic and intrinsic parts of the SU(2) holonomy. Therefore, to have a full geometric meaning, the SU(2) variables need to be embedded in $T_\Xi$.
This should not come as a surprise: from the continuum theory we know that one needs to embed the Ashtekar-Barbero connection into the space of Lorentzian connections in order to distinguish intrinsic from extrinsic contributions, and that the secondary constraints provide this embedding.
Similarly  in the discrete theory, we expect the secondary constraints to provide a non-trivial embedding of $T^*\SU(2)$ in  $T_\Xi$. More precisely, we expect the secondary constraints, and thus the embedding, to be defined only at the level of the complete graph, and not link by link, hence it is more correct to speak of an embedding of $T^*\SU(2)^L$ in
 $T_\Xi{}^L$.

Let us discuss this in more details. In the continuum theory, 
Ashtekar-Barbero variables, $(\Sigma, A^{(\beta)}=\Gamma+\beta K)$, are canonical coordinates on the reduced phase space, but are well-defined everywhere as functions on the original phase space. 
Then, solving the secondary constraints gives $\Gamma=\Gamma(\Si)$, and provides a specific embedding of the SU(2) variables into the original phase space. 
If one forgets about secondary constraints, and treats the linear primary constraints as a first-class system, one ends up with a quotient space of orbits $A^{(\beta)}=\mathrm{const.}$ (because of the brackets between the Lorentz connection and the reality conditions \Ref{realcond}), intersecting the constraint hypersurface transversally (because the Hamiltonian flow of second-class constraints always points away from the constraint hypersurface). 
Then, restoring the secondary constraints provides a non-trivial section, i.e. a gauge-fixing through these orbits, that is an embedding mapping any pair $(\Sigma,  A^{(\beta)})$ towards a point $(\Pi,A=\Gamma+\I K)$ in the original phase-space.
Such treatment of second-class constraints resonates with the gauge-unfixing ideas \cite{Anishetty,Mitra} recently applied to the framework of loop quantum gravity in \cite{Bodendorfer1,BodendorferSimpl}.  

At the discrete level, whatever the correct representation of the secondary constraints may be, it is reasonable to assume that they have the same effect on the constraint algebra, making $D$ second class.
Solving them, which typically can not be done link by link but requires knowing the graph, should provide a non-trivial section\footnote{The trivial section being $\Xi=0$.} through the orbits  \eref{Dtrafos} of $D$, 
that is a non-local function $\Xi_t(z_t,\utilde{z}_t)$ where each link dihedral angle is determined by spinors all over the graph.
This idea can be made explicit with the ubiquitous example of the flat 4-simplex. In this case, a metric geometry is defined by the ten edge lengths $\ell_e$. Then, all spinors are functions of these data (modulo gauges), and in particular, for each link, $\Xi_t=\Xi_t(\ell_e)$ give the dihedral angles, while $G_t(\ell_e)$ equal the holonomies of the Levi-Civita connection. Hence, on the graph phase space $T_\Xi{}^L$ there is a functional dependence $\Xi_t(z_t,\utilde{z}_t)$ between the ten dihedral angles and the twenty spinors, which provides the desired non-trivial section of the bundle $T_\Xi{}^{L}$. The dependence includes all the spinors and thus non-local on the graph, because each dihedral angle depends in general on all edge lengths. 
The constraint structure including the role of the secondary constraints is illustrated in Figure \ref{FigRed2}.

\begin{figure}[ht]\begin{tikzpicture}[descr/.style={fill=white,inner sep=2.5pt}]
  \matrix(m)[matrix of math nodes, row sep=4em,  column sep=8em]
  {\T^2 & \C^2\times\C^2\times\C & \C^2\times\C^2 \\
 T^*\SL(2,\C) & T^*\SU(2)\times\R & T^*\SU(2) \\ 
 & T^*\SU(2)& \\};

  \path[->,font=\small]
  (m-1-1)edge node[descr] {$F$} (m-1-2)  (m-1-2)edge node[descr] {$F_1,\utilde{F}_1$ orbits} (m-1-3)
  (m-1-1)edge node[descr] {$C$} (m-2-1)  (m-1-2)edge node[descr] {$C_{\rm red}$} (m-2-2)  (m-1-3)edge node[descr] {$C_{\rm red}$} (m-2-3)  
  (m-2-1)edge node[descr] {$F_{\rm red}$} (m-2-2)  (m-2-2)edge node[descr] {$D$ orbits} (m-2-3)
  (m-3-2)edge node[descr] {$\Gamma$-dependent secondary} (m-2-2);
\end{tikzpicture}
\caption{More detailed constraint structures and the role of the secondary constraints. In the presence of secondary constraints, the rightmost part of the diagram becomes irrelevant, as the orbits of $D$ are no longer pure gauges.
In the final step, we have reintroduced the graph structure, as a proper definition of the secondary constraints should not be local on the links.}\label{FigRed2}   
\end{figure}

Concerning the explicit form of the secondary constraints, we do not investigate them here, but hope to come back to it in future research. In particular, it has been argued in \cite{DittrichRyan,DittrichRyan2} that these constraints should be identified with the shape matching conditions of \cite{DittrichSpeziale}, the ones reducing twisted geometry to Regge geometry. A direct test of this claim would require commuting the primary constraints with a suitable Hamiltonian, but this has not been attempted yet. 
On the other hand, as mentioned at the end of section IIB, we expect that there should exist a notion of Levi-Civita connection (that is a solution of the secondary constraints) even in the absence of shape-matching conditions. If this is the case, then the situation would be different from the one advocated in \cite{DittrichRyan}.

%------------------------------------------------------------------------------------
\subsection{Twisted geometries}
%------------------------------------------------------------------------------------
\noindent
To complete the classical analysis, let us give the mapping between $\SL(2,\C)$ holonomy-fluxes and the variables of the twisted geometries parametrization \cite{twigeo,twigeo2,IoTwistorNet}. These variables consists of areas and angles associated to a cellular decomposition dual to the graph, and permit to interpret the classical phase space in terms of discrete geometries. In what follows, we always assume to be on-shell of the area matching, so $\po=\upo$.
We first notice that we can write the holonomy \Ref{holpar} as $h={g}(\uom,\upi) g(\om,\pi)^{-1}$, where
\be\label{totti}
g(\om,\pi) = \f1{\sqrt{\po}}\left(\begin{array}{cc} \om^0 & \pi^0 \\ \om^1 & \pi^1 \end{array}\right).
\ee
Following \cite{IoTwistorNet}, we use the Iwasawa decomposition for $\SL(2,\C)$ group elements,
\be\label{montella}
g=n(\zeta)T_\a e^{\Phi\tau_3}, \qquad (\zeta,\a,\Phi)\in\C^3,
\qquad n(\zeta)=\f1{\sqrt{1+|\zeta|^2}}\left(\begin{array}{cc} 1 & \zeta \\ -\bar\zeta & 1\end{array}\right),
\qquad T_\a=\left(\begin{array}{cc} 1 & \a \\ 0 & 1\end{array}\right).
\ee
Comparing \Ref{totti} and \Ref{montella}, we identify
\be\label{tgpar}
-{\bar\zeta}^{-1} = \f{\om^0}{\om^1}, \qquad \Phi = -2\arg(\om^0) +\I\ln\f{\|\om\|^2}{{\po}}, 
\qquad \a=\f{\bra{\om}\pi\ra}{\po}e^{2\I\arg(\om^0)},
\ee
where $\bra{\om}\pi\ra:=\d_{A\bar A} \pi^A{\bar\om}^{\bar A}$.
It is also convenient to define the angle
\be\label{defxi}
\xi:= 2\arg(\uom^0)-2\arg(\om^0) + \beta \Xi.
\ee
We then find
\begin{subequations}\label{covtg}\be\label{htwigeo}
h = n(\utilde{\zeta})T_{\utilde{\a}} e^{(-\xi + (\beta- \I) \Xi)\tau_3} T_\a^{-1} n^{-1}(\zeta).
\ee
Similarly, it is easy to verify that the fluxes \Ref{fluxpar}, in their matricial form \Ref{selfcomp}, read
\be
\Pi = -\f\I2 \po \, n(\zeta) T_\a \tau_3 T_\a^{-1} n^{-1}(\zeta),
\qquad 
\utilde{\Pi} = \f\I2 \po \, n(\utilde{\zeta}) T_{\utilde{\a}} \tau_3 T_{\utilde{\a}}^{-1} n^{-1}(\utilde{\zeta}),
\ee\end{subequations}
where the opposite sign is inherited directly from the opposite sign in \Ref{fluxpar}.

The parametrizations \Ref{covtg} of the covariant holonomy-flux variables gives a map 
\be
(\Pi, h) \mapsto (\zeta,\utilde{\zeta},\a,\utilde{\a},\xi,\Xi, \po),
\ee
in terms of the area of the face, $\po$, and a collection of angles, which we dub \emph{covariant} twisted geometries as in \cite{IoTwistorNet}. In these variables, the simplicity constraints \Ref{Simpl} are solved on the family of hypersurfaces 
\be\label{simpltg}
\a=\utilde{\a}=0, \qquad \po=(\beta+ \I)\cJ,
\ee
which corresponds to $T_\Xi$. 
On the trivial section $\Xi=0$, we recover   $T^*\SU(2)$ parametrized by twisted geometries (see \cite{twigeo,twigeo2}, adapted to the conventions of this paper),
\be\label{su2twigeo}
U = n(\utilde{\zeta}) e^{-\xi\tau_3} n^{-1}(\zeta), \qquad \Sigma = \f{\beta\ellp^2}{\hbar} \f\I2 \cJ n(\zeta)\tau_3n^{-1}(\zeta).
\ee
The map between the spinorial parametrization \Ref{SU2hf}  and the twisted geometry parametrization is
\be
\cJ = \f{\|z\|^2}2, \qquad \zeta = -\f{\bar z^1}{\bar z^0}, \qquad \xi = 2\arg(\utilde{z}^0)-2\arg({z}^0),
\ee
consistently with \Ref{spinordef} and \Ref{tgpar}.

The twisted geometry variables show explicitly the nature of the discrete geometries associated with the holonomy-flux algebra.
First of all, one should consider a cellular decomposition dual to the graph, and assign a 3d Cartesian frame within each 3-cell. Gauge invariance at the nodes, imposed by the closure condition $\sum_{t\in n} \Pi^i[t] =0$ and its complex conjugate, guarantees that we can apply Minkowski's theorem to infer the existence of a unique convex polyhedron around the node \cite{IoPoly}. The collection of polyhedra defines twisted geometries, a generalization of Regge geometries allowing discontinuous metrics \cite{twigeo,IoCarloGraph,IoPoly}. However, the left and right generators $\Pi$ and $\bar \Pi$ identify two different geometries. They coincide only when the simplicity conditions hold, \Ref{simpltg}.
This is precisely the role of the constraints also in the continuum theory: they match the left and right metric structures induced by the two Urbantke metrics \cite{Capo2,MikeLR,Iobimetric}. Twisted geometries show that the constraints have exactly the same role also at the discrete level.

At the level of reduced Dirac brackets, the area becomes conjugated to the angle $\xi$,
\be
\{\cJ, \xi \}=1,
\ee
which is now independent of the Immirzi parameter. 
Notice the remarkable analogy between \Ref{defxi}, relating the dihedral angle to the class angle of the SU(2) holonomy, and the canonical transformation in the continuum theory from the extrinsic curvature to the Ashtekar-Barbero connection, $A^{(\beta)}=A+(\beta-\I)K$. Equations \Ref{htwigeo} and \Ref{su2twigeo} make manifest the Ashtekar-Barbero interpretation of $U$ established in the previous Section.

%------------------------------------------------------------------------------------
\section{EPRL quantization of twisted geometries}\label{SecQuant}
%------------------------------------------------------------------------------------
\noindent
The classical phase space of twistors on a link can be quantized, leading to quantum twistor networks and quantized twisted geometries. In the following, we choose a procedure inspired by the EPRL model. We take a Schr\"odinger representation, and follow a Dirac procedure, quantizing first the unconstrained algebra, and then implementing the constraints. Our starting point is an auxiliary Hilbert space carrying a unitary representation of the canonical Poisson algebra \Ref{spinbrack}, 
\be\label{cancomm}
\big[\hat\pi_A,\hat\omega^B]=-\I \hbar \delta^B_A=-\big[\utilde{\hat\pi}_A ,\utilde{\hat\omega}^B\big].
\ee
Since the constraint structure is commutative, see Fig. \ref{FigRed}, let us first study the reduction by the simplicity constraints. That allows us to focus on a single half-link, and consider only the untilded quantities.
The Schr\"odinger representation is given by wave functions $f(\om)\in L^2(\mathbb{C}^2,d^4\om)$, where $d^4\om = (1/16)(\di\omega_A\wedge\di\omega^A\wedge\CC)$ is the canonical $\SL(2,\mathbb{C})$-invariant integration measure\footnote{We define the complex Lesbegue measure as $d^2x = (\I/2) \di \zeta\w \di\bar \zeta$. This normalization is responsible for the various powers of 2 appearing in later formulas.} on $\mathbb{C}^2$, and operators
\begin{equation}
\big(\hat\om^A f\big)(\omega^A)=\om^A f(\om^A), \qquad
\big(\hat\pi_A f\big)(\omega^A)=\frac{\hbar}{\I}\frac{\partial}{\partial \omega^A}f(\omega^A).\label{posreps}
\end{equation}
A ``momentum'' representation $\hat\pi^A=\pi^A$, $\hat\omega^A=\I\hbar\partial/\partial\pi_A$ is also possible. The two representations are related by the Fourier transform
\begin{equation}\label{FT}
{f}(\pi)=\frac{1}{\pi^2}\int_{\C^2}d^4{\omega}\,\E^{-\frac{\I}{\hbar}\pi\omega-\CC}f(\om),\qquad
f(\omega)=\frac{1}{\pi^2}\int_{\C^2}d^4{\pi}\,\E^{+\frac{\I}{\hbar}\pi\omega-\CC} {f}(\pi),
\end{equation}
whose properties are reviewed in Appendix  \ref{appdxA}. With the usual physicist's abuse of notation, we denote the Fourier transform in the same way as the original function.

Since the constraints involve the Euler dilatation operator, $\omega^A\partial/\partial \omega^A$, a convenient (generalized) basis is provided by its eigenfunctions. These are homogeneous functions $f^{(\rho,k)}(\om)$, 
parametrised by a pair $(\rho\in\R, 2k\in\mathbb{Z})$, such that
\begin{equation}
f^{(\rho,k)}(\lambda\om)=\lambda^{-k-1+\I\rho}\bar{\lambda}^{k-1+\I\rho} f^{(\rho,k)}(\om).
\end{equation}
In particular, it follows that
\begin{subequations}
\begin{align}
\omega^A\frac{\partial}{\partial\omega^A}f^{(\rho,k)}(\omega)&=(-k-1+\I\rho)f^{(\rho,k)}(\omega),\\
\bar\omega^{\bar A}\frac{\partial}{\partial\bar\omega^{\bar A}}f^{(\rho,k)}(\omega)&=(+k-1+\I\rho)f^{(\rho,k)}(\omega).
\end{align}\label{degr}
\end{subequations}
The auxiliary Hilbert space $L^2(\mathbb{C}^2,d^4\om)$ carries a unitary, reducible action of $\SL(2,\mathbb{C})$ with generators $\widehat{L}_i$ and $\widehat{K}_i$.
The homogeneous functions span irreducible (infinite dimensional) representations, with Casimirs
$$\widehat{L}^2-\widehat{K}^2 =  k^2-\r^2-1, \qquad \widehat{L}_i\widehat{K}^i = -k\r. 
$$
It is a generalized basis, since the homogeneous functions are distributions  in $L^2(\mathbb{C}^2,d^4\om)$, and not square integrable. This is similar to what happens for e.g. the action of the Euclidean group on $L^2(\mathbb{R}^3,d^3x)$, where the irreducible representations are labelled by the total energy $\propto\vec{p} \,{}^2$, but the basis-elements, being plane waves $\exp\I\vec{p}\cdot\vec{x}$, are not square-integrable.  
A finite, $\SL(2,\C)$-invariant Hermitian inner product is provided by a  2-dimensional surface integral on P$\C^2\subset\mathbb{C}^2$, with measure $d^2\om := \I/2 \omega_A\di\omega^A\wedge\bar{\omega}_{\bar A}\di\bar{\omega}^{\bar A}$. An orthonormal basis is then provided by elements $\{f^{(\rho,k)}_{j,m}\}$, labelled by spins $j=k,k+1;\dots$
and magnetic numbers $m=-j,\dots,j$ corresponding to the canonical $\SU(2)$ subgroup of $\SL(2,\mathbb{C})$,
\begin{equation}
\langle f^{(\rho,k)}_{j,m},f^{(\rho,k)}_{j',m'}\rangle=\frac{\I}{2}\int_{\mathrm{P}\mathbb{C}^2}\omega_A\di\omega^A\wedge\bar{\omega}_{\bar A}\di\bar{\omega}^{\bar A}
\overline{f^{(\rho,k)}_{j,m}(\omega^A)}f^{(\rho,k)}_{j',m'}(\omega^A)=\delta_{jj'} \d_{mm'} \label{fscalar}.
\end{equation}
The basis diagonalises $\widehat{L}^2$ and $\widehat{L}_3$. 
See Appendix \ref{appdxA} and \cite{Ruhl, GelfandLorentz, DucLorentz} for more details.
 Notice that thanks to the homogeneity of the integrand, \Ref{fscalar} is independent of the way $\mathrm{P}\mathbb{C}^2$ is embedded into $\mathbb{C}^2$. The existence of this Hermitian product plays a key role in the constraint reduction.

The representations with $(\rho,k)$ and $(-\rho,-k)$ are unitary equivalent. We can thus always restrict ourselves to $k>0$, which agrees with what we have done earlier: The half integers $k$ are the quanta of $\mathcal{J}$, choosing them positive fits our gauge condition $\mathcal{J}>0$ introduced in above. Reasons to consider both are given in \cite{CarloEdParity}.

%-------------------------------------------------------
\subsection{Simplicity constraints}
%-------------------------------------------------------
\noindent
Since $F_1$ is first class, it can be imposed strongly. However, the complex constraint $F_2$ is of second class, making a different procedure necessary. Motivated by deriving the EPRL model, we implement $F_1$ strongly and $F_2$ weakly. 
This can be done introducing a master constraint, a procedure \cite{ThiemannBook} quite common in LQG:
\begin{equation}
F_2=0 \ \Leftrightarrow \ \M=\bar{F}_2F_2,\quad\text{and}\quad\big\{F_1,\M\big\}=0.
\end{equation}
The new constraint algebra is abelian, and both $\M$ and $F_1$ can be imposed strongly.
Choosing a normal ordering as in \cite{WielandTwistors}, we find the following quantum constraints,
\begin{subequations}
\begin{align}
\widehat{F}_1 &= \frac{\hbar}{\beta^2+1}\Big[(\beta-\I)\omega^A\frac{\partial}{\partial\omega^A}-(\beta+\I)\bar\omega^{\bar A}\frac{\partial}{\partial\bar\omega^{\bar A}}-2\I\Big],\label{F1cons}\\
\widehat{F}_2&=\frac{\hbar}{\I}n^{A\bar A}\bar\omega_{\bar A}\frac{\partial}{\partial \omega^A},\\
\widehat{\M}&=\widehat{F}_2^\dagger\widehat{F}_2=\frac{\hbar^2}{4}\Big[\omega^A\frac{\partial}{\partial\omega^A}\frac{\partial}{\partial\bar{\omega}^{\bar A}}\bar{\omega}^{\bar A}-\big(\widehat{L}^2-\widehat{K}^2\big)+2\widehat{L}^2\Big].
\end{align}\label{realcondspin}
\end{subequations}
Both $\widehat{F}_1$ and $\widehat{\M}$ are diagonal on our canonical basis \Ref{fscalar}, and the action can be easily evaluated to give
\begin{align}
& \widehat{F}_1f^{(\rho,k)}_{j,m} = \frac{2}{\beta^2+1}\big(-\beta k+\rho\big)f^{(\rho,k)}_{j,m}\stackrel{!}{=}0  
 && \hspace{-2.5cm} \Leftrightarrow \hspace{1cm}  \rho=\beta k, \\
& \widehat{\M}f^{(\rho,k)}_{j,m} = \frac{1}{2}\big(j(j+1)-k(k+1)\big)f^{(\rho,k)}_{j,m}\stackrel{!}{=}0  
 && \hspace{-2.5cm} \Leftrightarrow \hspace{1cm} j=k.
\end{align}
That is, the non-Lorentz-invariant master constraint selects the lowest spin $j$ labels. The resulting wave-functions are
$ f^{(\beta j,j)}_{j,m}(\omega^A)$. From Appendix \ref{appdxA}, we find them to be
\begin{equation}\label{fsol}
f^{(\beta j,j)}_{j,m}(\omega^A)= \sqrt{\frac{2j+1}{\pi}}\sqrt{\binom{2j}{j+m}}
\|\omega\|^{2(\I\beta j-j-1)}(\bar\omega^{\bar 0})^{j+m}(\bar\omega^{\bar 1})^{j-m}.
\end{equation}
These functions are orthonormal with respect to the inner product on the Riemann sphere \Ref{fscalar}, and provide a map, often denoted $Y$-map in the literature, from the $j$-th irrep of $\SU(2)$ to the unitary irreducible $(\beta j,j)$-representation of $\SL(2,\C)$:
\be
Y:\mathcal{H}_j\ni |j,m\rangle\mapsto f_{j,m}^{(\beta j,j)}\in\mathcal{H}_{(\beta j,j)}.
\ee
In the following, we also use the notation $\ket{\beta;j,m}$ 
for the $\SL(2,\C)$ ket,\footnote{In the literature, the alternative notation $\ket{\beta j, j; j, m}\equiv \ket{\beta;j,m}$ is often found.} and write $\langle\omega\bar\omega|\beta;j,m\ra \equiv f^{(\beta j,j)}_{j,m}(\omega^A)$.

At this point, we would like to discuss two subtle aspects of the quantization. 
As the action generated by $F_1$ is non-compact, Dirac's quantization does not lead to a proper subspace of the auxiliary Hilbert space $L^2(\C^2,d^4\omega)$. Accordingly, the solution space spanned by \Ref{fsol} only makes sense in terms of \emph{distributions}, to be integrated over the previsouly defined P$\C^2$ inner product \Ref{fscalar}. The distribution by itself is not a function on the fully reduced phase space $\C^4/\!\!/F \simeq \C^2$, but depends also on the $F_1$ orbits, 
\be
\{F_1,f^{(\beta j,j)}_{j,m} \} \neq 0.
\ee
We can see this explicitly if we insert the parametrization \Ref{spinordef} in the definition of the wave function \eref{fsol}, which gives
\begin{equation}
f^{(\beta j,j)}_{j,m}(\omega^A)=\sqrt{\frac{2j+1}{\pi}}\sqrt{\binom{2j}{j+m}}
\f{(\bar z^{\bar 0})^{j+m}(\bar z^{\bar 1})^{j-m}}{\|\omega\|^2 \, \| z\|^{2j}},
\end{equation}
where the non-$F_1$-invariant quantity $\|\omega\|^2$ appears. On the other hand, the half-density \cite{ThiemannBook, AshtekarBook, Thiemannfermihiggs}
\begin{equation}\label{fred}
\sqrt{d^2\om} f^{(\beta j,j)}_{j,m}(\omega^A)= \sqrt{d^2z} \sqrt{\frac{2j+1}{\pi}}\sqrt{\binom{2j}{j+m}}
\f{(\bar z^{\bar 0})^{j+m}(\bar z^{\bar 1})^{j-m}}{\| z\|^{2j+2}},
\end{equation}
is $F_1$-invariant, thanks to the homogeneity of the measure $d^2\om$ on $\mathrm{P}\C^2$.
Hence, it is the half-densities that are properly defined on the reduced phase space.\footnote{A toy model may further illustrate this subtlelty. Consider a particle in $\mathbb{R}^3$. Our constraint be just the radial momentum constrained to vanish, that is $F=p_r=\vec{x}\cdot\vec{p}/|\vec{x}|=0$. The auxiliary Hilbert space is just the familiar $L^2(\R^3, d^3x)$. To make $\hat{p}_r$ self-adjoint its quantisation is given in terms of the covariant derivative operator $\hat{p}_r:=-\I\hbar(\partial_r+r^{-1})$. Solutions in the kernel of the constraint look like $\Psi(r,\vartheta,\varphi)=r^{-1}\psi(\vartheta,\varphi)$. As functions on the original phase-space, they fail to be gauge invariant, since $\{p_r,\Psi\}\neq 0$, while the half densities $\Psi\sqrt{d^3 x}=\psi\sqrt{\sin\vartheta \di r\di \vartheta \di \varphi}$  are.}

The second remark concerns the case of $j=0$. In fact, $j=0$ corresponds classically to the degenerate configurations $\cJ=0$, for which the twistorial description of the phase space breaks down. To complete the quantization, we need to provide independently the missing state. If we extrapolate \Ref{fsol} to $j=0$ we would find a non-trivial state, given by $\pi^{-1/2}\|\om\|^{-2}$. This choice could pose problems with cylindrical consistency, so we fix it by hand to be the trivial state,
 \be\label{f00}
 f^{(0,0)}(\om^A) \equiv 1.
 \ee
An argument in favour of this choice is that the primary simplicity constraints are all first class when $|L_i|=0$. Hence, one can identify \Ref{f00} as the unique invariant state satisfying \Ref{realcond} as strong operator equations.

A final comment before moving on. 
The reader may have also noticed a similarity between the quantum state \Ref{fsol} and the SU(2) coherent states. 
In fact, the state 
\be
\ket{j,z^A} = \sqrt{\f{2j+1}\pi}\sum_{m=-j}^j \sqrt{\binom{2j}{j+m}} \f{(z^0)^{j+m} (z^1)^{j-m}}{\|z\|^{2j}} \ket{j,m},\label{SU2cor}
\ee 
in the SU(2) $j$-th irreducible representation,
represents a coherent state peaked on the direction identified by $z^0/z^1$ on ${\rm P}\C^2\cong S^2$, normalized with respect to the measure $d^2z$  on $\mathrm{P}\C^2$. 
It is then easy to see that
\be\label{fCS}
\sqrt{d^2\om} f^{(\beta j,j)}_{j,m}(\omega^A) = \frac{\sqrt{d^2z}}{\|z\|^2} \la j,z^A \ket{j,m},
\ee
which implies that the $Y$-map endows SU(2) coherent states with the interpretation of functions on the solution space of the simplicity constraints. 
For later convenience, we also give the overlap 
\be\label{<fCS>}
\bra{\om\bar\om}Y|j,z^A\ra = \sum_{m=-j}^j f^{(\beta j,j)}_{j,m}(\omega^A) \bra{j,m}j,z^A\ra =
\f{2j+1}\pi \|\omega\|^{2(\I\beta j-1)} \left(\f{\bra{\om}z\ra}{\|\om\| \, \|z\|}\right)^{2j}.
\ee
where $\bra{\om}z\ra := \d_{A\bar A}z^A \bar \om^{\bar A}$.
Although we are labeling the SU(2) coherent states with spinors, as e.g. in \cite{BarrettLorAsymp}, it is only the Hopf section $z^0/z^1$ that carries a semiclassical meaning. The norm and the overall phase of the spinor have no physical counterparts from the point of view of SU(2).

%-------------------------------------------------------
\subsection{Area-matching constraints}
%-------------------------------------------------------
\noindent
For the tilded quantities, since we have an opposite sign in the Poisson brackets \Ref{cancomm}, it is convenient to take wave functions
$ \tilde{f}(\tpi)\in L^2(\mathbb{C}^2,d^4\tpi)$, and
\be
\big(\utilde{\hat{\pi}}_A   \tilde{f}  \big)(\tpi)=\tpi_A   \tilde{f}(\tpi), \qquad
\big(\utilde{\hat{\omega}}^A  \tilde{f}  \big)(\tpi)=\frac{\hbar}{\I}\frac{\partial}{\partial \tpi_A}   \tilde{f}(\tpi).\label{momreps}
\ee
Solutions to the constraints can be obtained as before, restricting to the homogeneous functions $f^{(\beta j,j)}_{j,m}(\upi)$.
Later aspects of the quantum theory make it convenient to work with a slightly different basis, given by a dual map and an extra phase $\eta_{\beta,j}$,
\be
\tilde{f}^{(\beta j,j)}_{\bar{\jmath},\bar{m}}(\upi) := 
\eta_{\beta,j} (-1)^{j+m} f^{(\beta j,j)}_{j,-m}(\upi)
= \eta_{\beta,j}\sqrt{\frac{2j+1}{\pi}}\sqrt{{2j}\choose{j+m}} \|\pi\|^{2(\I\beta j-j-1)}(\bar\tpi^{\bar 0})^{j-m}(-\bar\tpi^{\bar 1})^{j+m},
\ee
where
\be
\eta_{\beta,j}:=(\tfrac{\I}{2})^{\I\rho-j}(-\tfrac{\I}{2})^{\I\rho+j}\frac{\Gamma(j+1-\I\rho)}{\Gamma(j+1+\I\rho)}.
\ee 
See Appendix \ref{appdxA} for details on the $\SL(2,\C)$ dual map. 
We will also use a ket notation, as for the untilded wave-functions, and write 
$\bra{\beta;j,\bar{m}}\upi\bar{\upi}] = \tilde{f}^{(\beta j,j)}_{\bar{\jmath},\bar{m}}(\upi)$, where the square bracket keeps track of the fact that we are working with the dual basis.
The advantage of this choice of basis is to be related to the following Fourier transform,  
\begin{align}\label{fourtrans}
\bra{\beta; j,m}\tpi\bar\tpi] \equiv \tilde{f}^{(\beta j,j)}_{\bar{\jmath},\bar{m}}(\tpi_A)
&=\frac{1}{(2\pi)^2}\int_{\mathbb{C}^2}d^4\omega \, \E^{\frac{\I}{2}\tpi_A\tom^A-\CC}\overline{f^{(\beta j,j)}_{j,m}(\tom^A)}, 
\end{align}
which plays a role in the construction of the spin foam model.
Accordingly, we quantize the phase space $\T^2$ by means of a ``mixed'' Schr\"odinger representation, with wave functions 
$G(\om,\tpi) \in L^2(\mathbb{C}^2\times \C^2,d^4\om d^4\tpi)$, and operators \eref{posreps} and \eref{momreps}.

Since homogeneous functions diagonalize the quantum area-matching constraint, a solution is immediately provided by a restriction on the labels,
\be
\Big(\hat C f^{(\rho,k)} \otimes f^{(\utilde{\rho},\utilde{k})}\Big)(\om, \tpi) = 0 \quad \Rightarrow \quad (\utilde{\rho},\utilde{k}) = (\rho, k).
\ee
We recover here the reducibility of the system at the quantum level, since the constraint is redundant once we impose both diagonal simplicities. It then amounts to $\tilde{\jmath}=j$.
The space of solutions of both simplicity and area-matching quantum constraints is spanned by the functions
\begin{equation}\label{Gdef}
G^{(j)}_{\bar{\utilde{m}},m}(\utilde{\pi}^A,\omega^A) 
:=\tilde{f}^{(\beta j,j)}_{\bar\jmath ,\bar{\utilde{m}}}(\utilde{\pi}_A)f^{(\beta j,j)}_{j,m}(\omega^A)
\equiv \bra{\beta;j,\utilde{m}}\upi\bar\upi] \bra{\om\bar\om}\beta;j,m\ra.
\end{equation}
The argument contains position variables at the source, and momentum variables at the target. They satisfy all constraints,
\begin{subalign}
\big(\widehat{F}_1G^{(j)}_{\utilde{\bar{m}},m}\big)(\utilde{\pi}^A,\omega^A)&=0=\big(\widehat{\utilde{F}}_1G^{(j)}_{\utilde{\bar{m}},m}\big)(\utilde{\pi}^A,\omega^A),\\
\big(\widehat{\M}G^{(j)}_{\utilde{\bar{m}},m}\big)(\utilde{\pi}^A,\omega^A)&=0=\big(\widehat{\utilde{\M}}G^{(j)}_{\utilde{\bar{m}},m}\big)(\utilde{\pi}^A,\omega^A), \\
\big(\widehat{C}G^{(j)}_{\utilde{\bar{m}},m}\big)(\utilde{\pi}^A,\omega^A)&=\frac{\hbar}{\I}\Big(\omega^A\frac{\partial}{\partial\omega^A}
-\utilde{\pi}_A\frac{\partial}{\partial\utilde{\pi}_A}\Big)G^{(j)}_{\utilde{\bar{m}},m}(\utilde{\pi}^A,\omega^A)=0,
\end{subalign}
and depend only on the reduced phase space variables, namely
\begin{equation}\label{GCinv}
\{C,G^{(j)}_{\bar{\utilde{m}},m}\}=0=\{\bar{C},G^{(j)}_{\bar{\utilde{m}},m}\},
\qquad \left\{D, \sqrt{d^2\om}\sqrt{d^2\upi} \, G^{(j)}_{\bar{\utilde{m}},m}\right\}=0.
\end{equation}
The first brackets can be established thanks to the property
$G^{(j)}_{\bar{\utilde{m}},m}(\lambda\omega^A,\lambda^{-1}\utilde{\pi}^A)=G^{(j)}_{\utilde{\bar{m}},m}(\utilde{\pi}^A,\omega^A) \ \forall \lambda\in\mathbb{C}-\{0\}$,
whereas the second requires the use of half-density as described in the previous section.
As before, the states are restricted to $j\neq 0$, and for $j=0$ we independently fix 
$G^{(0,0)} \equiv 1$ for later cylindrical consistency of the spin foam model.

The boundary state functions carry an irreducible, unitary representation of the Lorentz group, with scalar product induced from \Ref{fscalar}, and the $Y$-map is explicitly implemented by the restriction on the irreps. The functions are very similar to projected spin networks \cite{EteraProj}, but not identical: the difference is that we are quantizing on the twistor vector space, and not on functions on the Lorentz group. The latter representation, and its basis of \emph{simple} projected spin networks will appear below when we study the EPRL spin foam model.

%------------------------------------------------------------------------------------
\section{Path integral measure in terms of twistors}\label{SecMeasure}
%------------------------------------------------------------------------------------
\noindent In this Section, we use the above framework to express the Liouville measure on the symplectic manifold $T^*\SL(2,\C)$ in a simple and straightforward way in twistor space. This measure plays an important role in the spin foam formalism, and will be used below in deriving the EPRL model.

%------------------------------------------------------------------------------------
\subsection{Definition of the integration measure}\noindent
%------------------------------------------------------------------------------------
On twistor space $\T^2$ there is a natural integration measure given by the symplectic volume,
\begin{equation}
d^{16}\mu= \Phi\wedge\bar\Phi\wedge\utilde{\Phi}\wedge\utilde{\bar\Phi},
\qquad \Phi:=\di\pi_A\wedge\di\pi^A\wedge\di\omega_B\wedge\di\omega^B.
\label{sympvol}
\end{equation}
We are interested in projecting this measure to the reduced phase-space $\T^2/\!\!/C$ of gauge orbits generated by the complex area-matching condition $C=0$. The constraint being first class, this can be done following the Faddeev-Popov method. However, since the gauge transformations generated by $C$ are just rescalings \Ref{scaltrans}, the Faddeev-Popov determinant is trivial, thus a $C$-gauge-invariant 12-dimensional integral can be written as
$\int d^{14}\m_{\rm gf} \,  \d_\C( C ) \, G.$
The gauge-fixed measure appearing on the right-hand side is obtained by taking the interior product of the Liouville measure \Ref{sympvol} with the generator $\mathfrak{X}_C=\{C,\cdot\}$ of the gauge orbits,
\begin{equation}
d^{14}\mu_{{\mathrm{gf}}}(Z,\utilde{Z}):=\I\Upsilon\wedge \bar\Upsilon, \qquad
\Upsilon:=\iota_{\mathfrak{X}_C}\big(\Phi\wedge\utilde{\Phi}\big).
\label{gfmeasure}
\end{equation}

We now prove that integrating any function on the reduced phase space $\T^2/\!\!/C$ against this measure gives a $C$-gauge invariant quantity. Let $G$ be defined on $\T^2/\!\!/C$, thus in particular constant along the orbits, $\{C,G\}=0=\{\bar C, G\}$.
Next, consider two gauge-fixing surfaces $\mathcal{D}_1$ and $\mathcal{D}_2$, that can continuously be deformed into one another, and intersect all gauge orbits exactly once.\footnote{This is always possible thanks to the abelianity of the transformation.} 
Applying Stokes' theorem to the region $\cal R$ bounded by $\mathcal{D}_{1}$ and $\mathcal{D}_{2}$, we find
\begin{equation}
\int_{\cD_1} d^{14}\m_{\rm gf} \, G - \int_{\cD_2} d^{14}\mu_{\rm gf} \, G=
\I\int_{\mathcal{R}}\di\wedge\big(\iota_{\mathfrak{X}_C}\Phi\wedge\iota_{\mathfrak{X}_{\bar C}} \bar\Phi \, G\big)=
\I\int_{\mathcal{R}}(\partial+\bar{\partial})\wedge\big(\iota_{\mathfrak{X}_C}\Phi\wedge\iota_{\mathfrak{X}_{\bar C}} \bar\Phi \,G\big),\label{stks}
\end{equation}
where
\begin{equation}
\partial=\di\omega^A\frac{\partial}{\partial\omega^A}+\di\pi_A\frac{\partial}{\partial\pi_A}+
\di\utilde{\omega}^A\frac{\partial}{\partial\utilde{\omega}^A}+\di\utilde{\pi}_A\frac{\partial}{\partial\utilde{\pi}_A}
\end{equation}
is the analytic part of the exterior derivative.
But since $\Phi$ is already an analytic form of highest degree and $\mathcal{L}_{\mathfrak{X}_C}\Phi=0=\mathcal{L}_{\mathfrak{X}_C}\bar\Upsilon$ together with $\iota_{\mathfrak{X}_C}\bar\Phi=0$, we immediately get that 
\begin{equation}
\begin{split}
\I\int_{\mathcal{R}}\partial\wedge\big(\iota_{\mathfrak{X}_C}&\Phi\wedge\iota_{\mathfrak{X}_{\bar C}} \bar\Phi G\big)=\I\int_{\mathcal{R}}\big(\partial\wedge\iota_{\mathfrak{X}_C}\big)\wedge\big(\Phi\wedge\iota_{\mathfrak{X}_{\bar C}} \bar\Phi G\big)=\I\int_{\mathcal{R}}\big(\partial\wedge\iota_{\mathfrak{X}_C}+\iota_{\mathfrak{X}_C}\wedge\partial\big)\wedge\big(\Phi\wedge\iota_{\mathfrak{X}_{\bar C}} \bar\Phi G\big)=\\\nn
&=\I\int_{\mathcal{R}}\mathcal{L}_{\mathfrak{X}_C}\big(\Phi\wedge\iota_{\mathfrak{X}_{\bar C}} \bar\Phi G\big)=\I\int_{\mathcal{R}}\Phi\wedge\iota_{\mathfrak{X}_{\bar C}} \bar\Phi \mathcal{L}_{\mathfrak{X}_C}G=\I\int_{\mathcal{R}}\Phi\wedge\iota_{\mathfrak{X}_{\bar C}} \bar\Phi\big\{C,G\big\}=0,
\end{split}
\end{equation}
and equally for the anti-analytic part. Therefore the integral is independent of the gauge-section chosen,
\begin{equation}
{\text{if\;}\{G,C\}=0=\{G,\bar{C}\}:\int_{\mathcal{D}_{1}}d^{14}\mu_{{\mathrm{gf}}} G=\int_{\mathcal{D}_{2}}d^{14}\mu_{{\mathrm{gf}}} G}.\label{eichin}
\end{equation}
This concludes the proof that \Ref{gfmeasure} is a measure on the reduced phase space $\T^2/\!\!/C$. Since we have already recalled the space is isomorphic to $T^*\SL(2,\C)\simeq\T^2/\!\!/C$, the measure can also be shown to be equivalent to the standard measure on $T^*\SL(2,\C)$ given by the product of the Lesbegue measure on the algebra, times the Haar measure on the group.
It is useful to prove this equivalence explicitly, which we do next. 
%-----------------------------------------------------------
\subsection{Equivalence with canonical measure on $\boldsymbol{T^*\SL(2,\C)}$}
%-----------------------------------------------------------
\noindent
To that end, we use the parametrizations \Ref{fluxpar} and \Ref{holpar}. Then, an explicit calculation shows that the analytic part of the Lesbegue measure on the $\mathfrak{sl}(2,\mathbb{C})$ reads
\be
d^3\Pi := -\f23 {\rm Tr}\big(\di\Pi\wedge\di\Pi\wedge\di\Pi\big)=
\frac{1}{4}\omega_A\di\omega^A\wedge\pi_B\di\pi^B\wedge\di(\pi_C\wedge\omega^C).
\ee
The spinorial decomposition of the analytic part of the Haar measure is more elaborate.
We first introduce a complex basis in $\mathfrak{sl}(2,\mathbb{C})$:
\be
\Pi^{AB}=-\frac{1}{2}\pi^{(A}\omega^{B)},\quad \mathrm{Q}_+^{AB}=\omega^A\omega^B,\quad\mathrm{Q}_-^{AB}=\pi^A\pi^B,
\ee
to work out the analytic part of the Maurer-Cartan form in terms of spinors:
\begin{equation}
\begin{split}
\iota^\ast(h^{-1}\di h)=&(\pi\omega)^{-2}\iota^\ast\Big[(\omega_A\di\omega^A-\utilde{\omega}_A\di\utilde{\omega}^A)\mathrm{Q}_-+
(\pi_A\di\pi^A-\utilde{\pi}_A\di\utilde{\pi}^A)\mathrm{Q}_+
+4(\pi_A\di\omega^A-\utilde{\pi}_A\di\utilde{\omega}^A)\Pi
\Big],
\end{split}
\end{equation}
where we used $C=0$, and $\iota$ is the embedding of the $C=0$ hypersurface into $\mathbb{T}^2$.
With this decomposition, the left-handed part of the Haar measure gives
\begin{equation}\begin{split}\label{Haarm}
d^3h:=& -\f23\mathrm{Tr}\big(h^{-1}\di h\wedge h^{-1}\di h\wedge h^{-1}\di h\big) =\\  =&
\f4{(\pi\omega)^3}\iota^\ast\big((\omega_A\di\omega^A-\utilde{\omega}_A\di\utilde{\omega}^A)\wedge(\pi_B\di\pi^B-\utilde{\pi}_B\di\utilde{\pi}^B)\wedge(\pi_C\di\omega^C-\utilde{\pi}_C\di\utilde{\omega}^C)\big).
\end{split}\end{equation}
Putting the two quantities together one recovers
\begin{equation}
\frac{1}{2^9}\int_{\T^2_{\mathrm{gf}}} d^{14}\mu_{\mathrm{gf}}(Z,\utilde{Z})\,\delta_{\C}\big(C(Z,\utilde{Z})\big) 
G\big(h({Z,\utilde{Z}}),\Pi(Z)\big)=\int_{T^\ast \SL(2,\mathbb{C})}\!\!\!\!\!\!\!\!\!\!\!\!d^3\Pi\wedge d^3h\wedge d^3\bar{\Pi}\wedge d^3\bar{h}\,G(h,\Pi).\label{gfxdint}
\end{equation}
Here $\Pi(Z)$ and $h(Z,\utilde{Z})$ are short-hand notations for the twistorial parametrisation (\ref{fluxpar}, \ref{holpar}).
Notice that \Ref{Haarm} provides a definition of the Haar measure in terms of spinors. Alternative definitions have been given in \cite{EteraTamboSpinor} for $\SU(2)$, and \cite{IoTwistorNet} for $\SL(2,\C)$. They involve an unconstrained integration, with Gaussian measures instead of $\d$ functions and gauge-fixing. Both approaches work as well, since one is integrating functions which do not have a dependence along the Gaussian slope.

\medskip

The result shows that the basic Liouville measure on $T^*\SL(2,\C)$ can be expressed in terms of twistors. Therefore, any gauge invariant added to it fits into this framework. In the following, we will just consider the basic $BF$ measure to derive the EPRL model, but the reader should keep in mind that any further non-trivial term, such as those induced by secondary constraints \cite{AlexandrovSimplClosure}, could also be described in this language.

%-------------------------------------------------------------------------------
\section{Constructing the EPRL spinfoam model}\label{secV}
%-------------------------------------------------------------------------------

\noindent 
In this Section we study a specific way to provide quantum dynamics to this system, which leads us to the EPRL spin foam model \cite{EPRL}.
The dynamics can be derived in three steps. 
The first step is a decomposition of the spacetime manifold into 4-cells, and the assignement of the Hilbert space of states \Ref{Gdef} to the boundary graph of each cell.
The second step is to give dynamics in the bulk with exponentials of the $BF$ action, suitably discretized in terms of twistors. This is the framework of the Plebanski action that describes gravity as $BF$ plus simplicity constraints \cite{Capo2,MikeLR,Iobimetric,PerezLR}. 
Finally, we integrate the boundary states weighted by the $BF$ action against the measure previously defined. The result reproduces the transition amplitudes of the EPRL model, and thus provides a new and independent derivation thereof, based on the twistorial representation of loop quantum gravity. 
We will refer specifically to triangulations, with 4-simplices as fundamental cells, but the results immediately generalize to arbitrary cellular decompositions, and what we recover is the generalized EPRL model of \cite{KKL, CarloGenSF}.

%-----------------------------------------------------------------
\subsection{Discretizing the action}
\begin{figure}[ht]
     \centering
     \includegraphics[width= 0.35\textwidth]{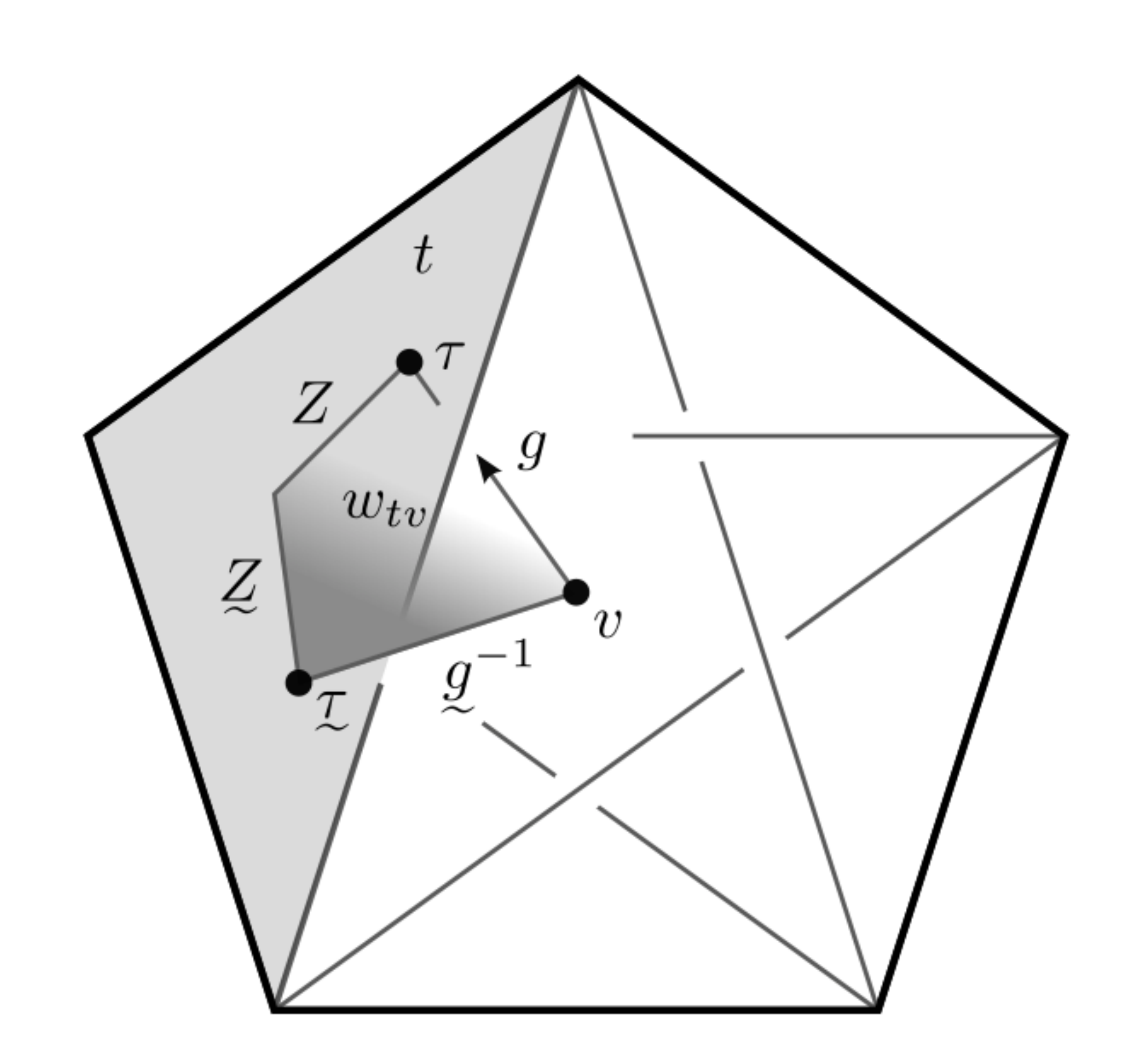}
     \caption{In a given 4-simplex two tetrahedra share a triangle $t$, that is in turn dual to a wedge $w_{tv}$. The wedge is bounded by a loop. Half of the loop  lies in the boundary of the 4-simplex, and  connects the two tetrahedra $\tau$ and $\protect\utilde{\tau}$, piercing through the triangle $t$. The other half enters the bulk and passes through the center $v$ of the 4-simplex.}
     \label{vertx}
\end{figure}

\begin{figure}[ht]
     \centering
\includegraphics[width= 0.70\textwidth]{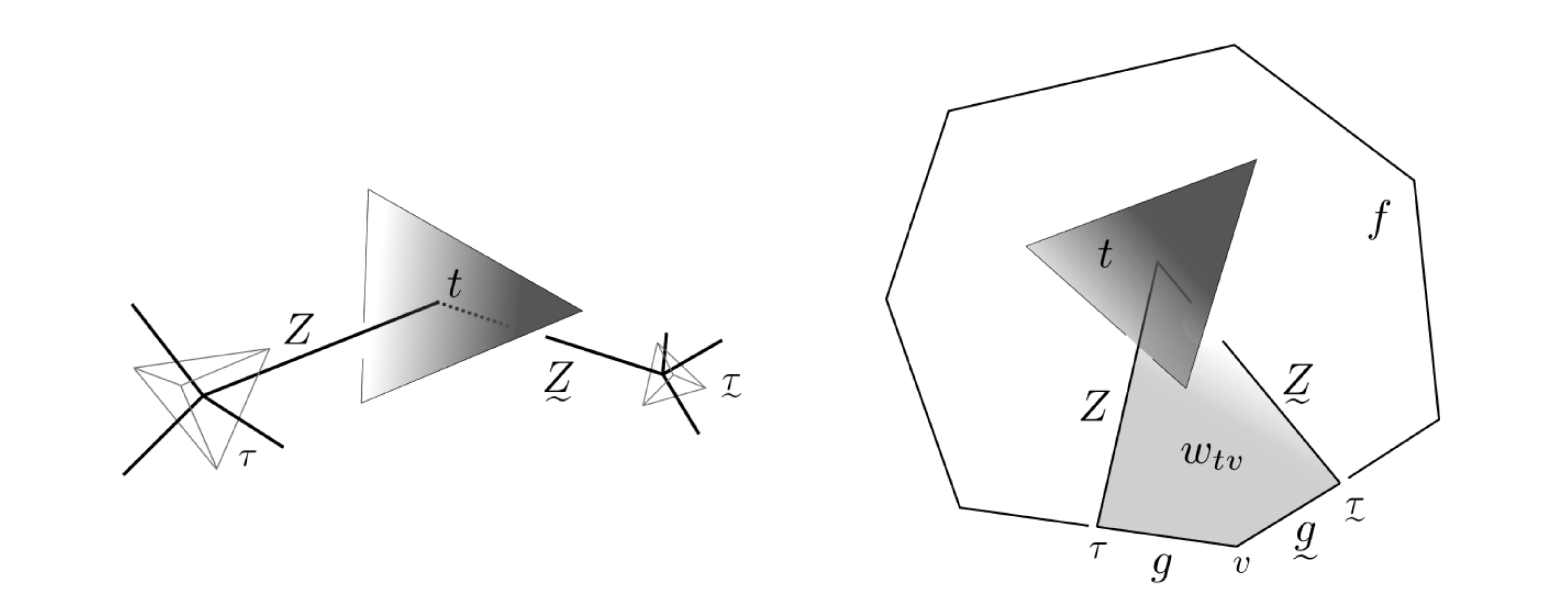}
     \caption{\emph{Left:} a triangle $t$ in the spatial hypersurface bounds two tetrahedra. Twistors $Z$ and $ \protect\utilde{Z}$ are attached to the underlying spinnetwork graph. \emph{Right:} The same triangle seen from a 4-dimensional perspective. The triangle $t$ is dual to a spinfoam face $f$ consisting of several wedges $w_{tv}$ one for each adjacent vertex $v$.}
     \label{spnfoamface}
\end{figure}

%-----------------------------------------------------------------
\noindent
The Holst action \Ref{holstactn} is equivalent to the following Plebanski action,
\begin{equation}
S_{\mathrm{Plebanski-Holst}}[\Sigma,A,\Psi]=\frac{\hbar}{\ellp^2}\frac{\beta+\I}{\I\beta}\int_M\ou{\Sigma}{A}{B}\wedge \ou{F}{B}{A}[A]
+\Psi\cdot {\cal S}(\Sigma)+\CC,\label{plebactn}
\end{equation} 
where $\Psi$ is a Lagrangian multiplier imposing the simplicity constraints ${\cal S}(\Sigma)$. See \cite{Capo2,MikeLR,Iobimetric,PerezLR} for details. In particular, the phase space structure is the same \cite{BuffenoirPleb,Wieland1}. This action is particularly advantageous to discretize, since it does not involve the tetrad, but bivectors which we know already how to treat as fluxes.
The first step to build the model is a cellular decomposition $\cal C$ of the spacetime manifold. For simplicity, we will refer to a simplicial decomposition, but our construction immediately generalize to arbitrary decompositions, in which case we recover the generalization of the EPRL models appeared in \cite{KKL,CarloGenSF}. For the case of a simplicial decomposition, $\cal C$ is made of 4-simplices, each of which consists of five tetrahedra and 10 triangles. 
Every triangle $t$ hinges several 4-simplices. Its dual face $f$ subdivides in wedges $w_{tv}$  \cite{MikeLeft}, each of which intersect one of those simplices. 
In every 4-simplex a triangle $t$ separates two tetrahedra in the corresponding 3-boundary. The wedge $w_{tv}$ is now bounded by a line, that starts at one of these tetrahedra, intersects $t$ transversally, reaches the other adjacent tetrahedron and passes through the center of the 4-simplex before finally closing to a loop. See Figures \ref{vertx} and \ref{spnfoamface} for illustrations.  The orientation of the loop $\partial w_{tv} $ is fixed by requiring the relative orientation $\epsilon(t,w_{tv})$ to be positive.

The wedge allows us to bridge between the boundary and the bulk of each 4-simplex. 
Half of the wedge boundary coincides with a link in the boundary graph of the 4-simplex, and carries the $\T^2$ phase space. We have smeared fluxes $\Sigma[t_w]$ and $\utilde{\Sigma}[t_w]$ associated with the two tetrahedra sharing the triangle $t_w$. The wedge label keeps track of the 4-simplex to which the triangle belongs. Reintroducing physical constants, we have, hiding the wedge labels,
\begin{equation}
\Sigma^{AB}=-\frac{\ellp^2}{\hbar}\frac{2\I\beta}{\beta+\I}\Pi^{AB}=\frac{\ellp^2}{2\hbar}\frac{\I\beta}{\beta+\I}\big(\omega^A\pi^B+\omega^B\pi^A\big),
\end{equation}
and the same for $\utilde{\Sigma}$. 
%In this way, spinors always belong to wedges, nonetheless we dot not introduce an extra label to keep notation simple. 
The other half of the phase space variables is the holonomy \Ref{holpar} along the boundary link.
Let us also introduce two new holonomies, $g$ and $\utilde{g}$ paralelly transporting from the center of the 4-simplex to the two boundary tetrahedra incident to the wedge. In this way, we can also form a wedge loop holonomy, 
given by the product of the phase space holonomy \Ref{holpar}, and the auxiliary bulk holonomies,
\begin{equation}
\ou{h}{B}{A}[\partial w]=-\ou{g}{B}{C}\uo{\utilde{g}}{D}{C}\frac{\utilde{\omega}^D\pi_A-\utilde{\pi}^D\omega_A}{\sqrt{\pi_E\omega^E}\sqrt{\utilde{\pi}_F\utilde{\omega}^F}}.
\end{equation}
This holonomy extends from the boundary to the bulk, and it is a mixed quantity depending on the phase space variables and the additional bulk holonomies.

Notice that we have introduced the phase space individually for each wedge, i.e. for each 4-simplex. Hence, the same link viewed from adjacent 4-simplices carries independent copies of the phase space.
Using these variables, we can discretize the BF part of the action as a sum of wedge contributions, 
\begin{equation}
\begin{split}
S_{BF}[\Sigma,A]&=\frac{\hbar}{\ellp^2}\frac{\beta+\I}{\I\beta}\int_M\ou{\Sigma}{A}{B}\wedge \ou{F}{B}{A}[A]+\CC=\\
&\approx\frac{\hbar}{\ellp^2}\frac{\beta+\I}{\I\beta}\sum_{w}\ou{\Sigma}{A}{B}[t_w]\ou{h}{B}{A}[\partial w]+\CC
\equiv \sum_w S_{\rm wedge}[\utilde{g}_w,g_w;\utilde{Z}_w,Z_w]
\end{split}\label{approxactn}
\end{equation}
Taking into account the presence of the two fluxes $\Sigma[t_w]$ and $\utilde{\Sigma}[t_w]=\Sigma[t^{-1}_w]$, each wedge contribution can be written as 
\begin{equation}
\begin{split}
S_{\mathrm{w}}[\utilde{g},g;\utilde{Z},Z]&=\frac{\hbar}{\ellp^2}\frac{\beta+\I}{2\I\beta}\Big(\ou{\Sigma}{A}{B}[t_w]\ou{h}{B}{A}[\partial w]+\ou{\Sigma}{A}{B}[t^{-1}_w]\ou{h}{B}{A}[\partial w^{-1}]\Big)+\CC=\\
&= \f{N_w}2 \ou{\big(g\utilde{g}^{-1}\big)}{A}{B}\big(\utilde{\omega}^B\pi_A+\utilde{\pi}^B\omega_A\big)+\CC
\end{split}\label{BFwedge}
\end{equation}
where
\be
N_w := \frac{1}{2}\bigg(\tfrac{\sqrt{\po}}{\sqrt{\upo}}+\tfrac{\sqrt{\upo}}{\sqrt{\po}}\bigg)
\ee
equals 1 on the $C=0$ constraint surface. Notice the factor $1/2$ in \Ref{BFwedge}. Its presence will lead to an extra phase in the spin foam amplitude, which we decided to reabsorb in the boundary states via the phase $\eta_{\beta,j}$.

%---------------------------------------------------------------------------
\subsection{The EPRL amplitude}
%---------------------------------------------------------------------------

\noindent
We construct the quantum amplitude taking the $BF$ action $S_{\rm wedge}$ to propagate the constrained boundary states \Ref{Gdef} along the bulk of the wedge. 
This leads to the following path integral, 
\begin{equation}\label{pathdef}
\big\langle G^{(j)}_{\bar{\utilde{m}},m}\big\rangle({\utilde{g},g}) :=
\frac{1}{2^{15}\pi^6}\int_{\T^2_{\mathrm{gf}}}d^{14}\mu_{\rm gf} \, \delta_{\mathbb{C}}( C ) \,
\bra{\beta; j,\utilde{m}}\utilde{\pi}\utilde{\bar\pi}]
\E^{\I S_{\mathrm{w}}(\utilde{g},g;\utilde{Z},Z)} \langle\omega\bar\om \ket{\beta;j,m},
\end{equation}
where the numerical factors have been chosen for later convenience, and 
we dropped the subscript $w$ in the variables since we are considering a single wedge at a time.
The integral is over a hypersurface $\T^2_{\rm gf}$ of $\T^2$ defining a gauge section of $C$, and it is invariant of the gauge chosen thanks to the invariance properties \Ref{GCinv} and \Ref{eichin}.
The integral expression is only defined for $j\neq 0$, since at the degenerate value the twistorial description of $T^*\SL(2,\C)$ does not work. Therefore, we need to independently fix the wedge amplitude for $j=0$. We do so by requiring cylindrical consistency of the final amplitude, which is satisfied by $\big\langle G^{(0)}_{00}({\utilde{g},g})\big\rangle \equiv 1$. In the following, we assume that this is the definition at $j=0$. 

The structure of \Ref{pathdef} resembles the infinitesimal step of a Feynman path integral, where the position eigenstates are represented by the \emph{constrained} boundary states, propagated by the $BF$ action.
The main result of this section is that the integrals can be explicitly performed leading to the Wigner matrices of simple projected spin networks, that is,
\be\label{key}
\big\langle G^{(j)}_{\bar{\utilde{m}},m}\big\rangle({\utilde{g},g}) = \mu(j) \, D^{(\beta j,j)}_{j\utilde{m}\, jm}(\utilde{g}g^{-1}),
\ee
where $\mu(j)$ is an $o(1/j)$ overall function. The results also provides the transformation between quantum twistor network states and $\SL(2,\C)$ cylindrical functions.

Before proving the result, let us briefly review how this quantity leads to the EPRL spin foam model.
First, we define amplitudes $A_f$, associated with the faces of the 2-complex. These are obtained taking the product of wedge amplitudes belonging to the face, and summing over the intermediate states. This gives an SU(2) trace  ${\rm Tr}_j$ on the magnetic quantum numbers $m$, which automatically implies that all spins in the face match, and a sum over the overall $j$,
\begin{subequations}\be\label{fampl}
A(g) = \sum_{j} \mu_f(j) \, {\rm Tr}_j \left(\prod_{w\in f} D^{(\beta j,j)}(\utilde{g}_wg^{-1}_w)\right).
\ee
Finally, we multiply the face amplitudes together and integrate over the remaining connection variables,
\label{EPRLdef}
\begin{equation}
Z_{\mathcal{C}} =  \int \prod_{v,\tau} dg_{v\tau} 
\prod_{f}A_f(g)
\end{equation}\end{subequations}
where 
$dg$ denotes the Haar measure on $\SL(2,\mathbb{C})$, and redundant integrations should be dropped to guarantee finiteness of the 4-simplex amplitude \cite{EnglePereiraFiniteness}.
This specific way to express the EPRL partition function through face amplitudes can be found for instance in \cite{CarloZako}, together with the other equivalent formulations.

\bigskip
In the rest of this section, we prove \Ref{key}. 
As a first step, let us write the $C$ constraint using a Lagrange multiplier $z$, and the integral representation of the complex $\d$-function
\begin{equation}
\delta_{\mathbb{C}}( C )=\frac{\I}{8\pi^2}\int_{\mathbb{C}}\di z\wedge\di\bar{z}\, \E^{\frac{\I}{2}z C-\CC},
\end{equation}
thus
\be\nn
\big\langle G^{(j)}_{\bar{\utilde{m}},m}\big\rangle ({\utilde{g},g})
=\frac{\I}{2^{18}\pi^8}\int_{\T^2_{\mathrm{gf}}} d^{14}\mu_{gf} \int_{\mathbb{C}}\di z\wedge\di\bar{z}\,
\exp\bigg[\frac{\I}{2}z C + \f{\I}{2} N_w \ou{\big(g\utilde{g}^{-1}\big)}{A}{B}\big(\utilde{\omega}^B\pi_A+\utilde{\pi}^B\omega_A\big)-\CC\bigg]G^{(j)}_{\utilde{\bar{m}},m}(\utilde{\pi},\omega).
\ee
Next, we introduce spinors at the center of the 4-simplex obtained parallel transporting with $g$ and $\utilde{g}$,
\begin{equation}
\begin{split}\label{spinv}
\vom^A=\ou{(g^{-1})}{A}{B}\omega^B,
& \qquad\vpi^A=\ou{(g^{-1})}{A}{B}\pi^B,\\
\utilde{\vom}^A=\ou{(\utilde{g}^{-1})}{A}{B}\utilde{\omega}^B, 
& \qquad\utilde{\vpi}^A=\ou{(\utilde{g}^{-1})}{A}{B}\utilde{\pi}^B.
\end{split}
\end{equation}
Thanks to the invariance under $\SL(2,\C)$ transformations of $C$ and the measure, we can change variables in the integral, from the wedge spinors to the $v$ spinors, and, dropping the supscript $v$, we find
\begin{equation}
\big\langle G^{(j)}_{\bar{\utilde{m}},m}\big\rangle({\utilde{g},g}) =
\frac{\I}{2^{18}\pi^8}\int_{\mathbb{C}}\di z\wedge\di\bar{z}\int_{\T^2_{\mathrm{gf}}}d^{14}\mu_{gf} \, \E^{\frac{\I}{2}\pi_A(\utilde{\omega}^A+z\omega^A)-\frac{\I}{2}\utilde{\pi}_A(\omega^A+z\utilde{\omega}^A)-\CC}
G^{(j)}_{\utilde{\bar{m}},m}\big((\utilde{g}\utilde{\pi})^A,(g\omega)^A\big).\label{pathdef3} 
\end{equation}
To perform the integrals on $\T^2_{\rm gf}$, we now specify which particular gauge section the 14-dimensional surface $\T^2_{\mathrm{gf}}$ corresponds to. It has to intersect all gauge orbits \Ref{scaltrans} exactly once, and the final result is independent of this choice thanks to the manifest gauge invariance of the integrand. We set:
\begin{equation}
\T^2_{\mathrm{gf}}\subset\mathbb{T}^2:\pi^A,\utilde{\omega}^A,\utilde{\pi}^A\in\mathbb{C}^2\;\text{arbitrary}\;,\text{but:}
\;\omega^A\in {\rm P}\C^2:\left\{\begin{split}
\omega^0&=\E^{\I\varphi}\cos\frac{\vartheta}{2}\\
\omega^1&=\sin\frac{\vartheta}{2}
\end{split}\quad\text{for}:\;\varphi\in(0,2\pi),\;\vartheta\in(0,\pi).
\right.
\end{equation}
In the coordinates chosen the integration measure decomposes according to
\begin{equation}
\f1{2^{15}} \int_{\T^2_{\mathrm{gf}}}d^{14}\mu_{\mathrm{gf}} \, G= \f\I2
\int_{\mathcal{S}}\omega_A\di\omega^A\wedge\bar{\omega}_{\bar{A}}\di\bar{\omega}^{\bar{A}}\int_{\mathbb{C}^2}d^4\pi\int_{\mathbb{C}^2}d^4\utilde{\omega}\int_{\mathbb{C}^2}d^4\utilde{\pi} \, G.
\end{equation} 
We see in \Ref{pathdef3} the appearence of a $\d$-function from the $d^4\pi$ integration,
\begin{equation}
\delta_{\mathbb{C}^2}(\uom^A+z\om^A)=\f{1}{(2\pi)^4}\int_{\mathbb{C}^2}d^4\pi \, \E^{\f{\I}{2}\pi_A(\uom^A+z\om^A)-\CC}.
\end{equation}
Writing this $\d$ explicitly, and inserting the expression  \Ref{Gdef} of $G$, \Ref{pathdef3} reads
\begin{align}
\nonumber&\big\langle G^{(j)}_{\bar{\utilde{m}},m}\big\rangle(\utilde{g},g)=-\frac{1}{\pi^4}\int_{\mathbb{C}}\di z\wedge\di{\bar z}\int_{\mathcal{S}}\omega_A\di\omega^A\wedge\bar{\omega}_{\bar{A}}\di\bar{\omega}^{\bar A}\int_{\mathbb{C}^2}d^4\utilde{\omega}\int_{\mathbb{C}^2}d^4\utilde{\pi}
\, \delta_{\mathbb{C}^2}(\utilde{\omega}^A+z\omega^A)\\
\nonumber&\qquad\qquad\qquad\E^{-\frac{\I}{2}\utilde{\pi}_A(\omega^A+z\utilde{\omega}^A)-\CC}
\tilde{f}^{(\beta j,j)}_{\bar{\jmath},\bar{\utilde{m}}}\big((\utilde{g}\utilde{\pi})_A\big)
f^{(\beta j,j)}_{j,m}\big((g\omega)^A\big)\\ \nn
&=-\frac{4}{\pi^2}\int_{\mathbb{C}}\di z\wedge\di{\bar z}\int_{\mathcal{S}}\omega_A\di\omega^A\wedge\bar{\omega}_{\bar{A}}\di\bar{\omega}^{\bar A}\int_{\mathbb{C}^2}d^4\utilde{\omega}
\delta_{\mathbb{C}^2}(\utilde{\omega}^A+z\omega^A)
 \overline{f^{(\beta j,j)}_{j,\utilde{m}}\big((\utilde{g}(\omega+z\utilde{\omega}))^A\big)}f^{(\beta j,j)}_{j,m}\big((g\omega)^A\big)\\
&=-\frac{4}{\pi^2}\int_{\mathbb{C}}\di z\wedge\di{\bar z}\int_{\rm P\C^2}\omega_A\di\omega^A\wedge\bar{\omega}_{\bar{A}}\di\bar{\omega}^{\bar A}
 \overline{f^{(\beta j,j)}_{j,\utilde{m}}\big((1-z^2)(\utilde{g}\omega)^A\big)}f^{(\beta j,j)}_{j,m}\big((g\omega)^A\big),
 \label{<G>}
\end{align}
  where in the second line we used the Fourier transform \Ref{FT}, and in the final line we used the $\d$ function to eliminate an integral.
Thanks to the homogeneity property of the canonical basis functions, we can rewrite this expression as 
\be
\big\langle G^{(j)}_{\bar{\utilde{m}},m}\big\rangle(\utilde{g},g) = 
\mu(j) \,
\f\I2 \int_{\rm P\C^2}\omega_A\di\omega^A\wedge\bar{\omega}_{\bar{A}}\di\bar{\omega}^{\bar A}
 \overline{f^{(\beta j,j)}_{j,\utilde{m}}\big((\utilde{g}\omega)^A\big)}f^{(\beta j,j)}_{j,m}\big((g\omega)^A\big),\label{pathdef2}
\ee
where
\be\label{defmu}
\mu(j) := \f{8\I}{\pi^2} \int_{\mathbb{C}}\di z\wedge\di{\bar z}(1-\bar{z}^2)^{-j-1-\I\beta j}(1-z^2)^{+j-1-\I\beta j}.
\ee
The integral on the complex projective plane gives precisely the Wigner matrices for the irreducible unitary representations of the Lorentz group, as a consequence of the scalar product \Ref{fscalar},
\be \label{Abarrett}
\f\I2\int_{\mathrm{P}\mathbb{C}^2}\omega_A\di\omega^A\wedge\bar{\omega}_{\bar A}\di\bar{\omega}^{\bar A}\overline{f^{(\beta j,j)}_{j,\utilde{m}}(\utilde{g}\omega^A)}f^{(\beta j,j)}_{j,m}(g\omega^A)=
D^{(\beta j,j)}_{\bar\jmath\bar{\utilde{m}}\,jm}(\utilde{g} g^{-1}).
\ee
The non-analytic complex integral is explicitly computed in the Appendix \ref{appdxB}, and gives 
\be\label{mu}
\m(j) = \frac{4}{\sqrt{1+\beta^2}}\frac{1}{\pi j}\E^{\I\Delta(\beta,j)},
\ee
where $\E^{\I\Delta(\beta,j)}$ is a phase and quickly converges to 1 as $j$ becomes large. 

This completes the proof of \Ref{key}, with $\mu(j)$ given in \Ref{mu}. 
Up to this factor, we recover the fundamental structure of the EPRL model.
Our derivation shows that its building block, the specific Wigner matrices appearing in \Ref{key}, have the interpretation of path integrals over twistor space.
Concerning the $\mu(j)$ factors, these come as a direct consequence of the integral over the $z$ Lagrange multiplier, and are thus absent in derivations of the model not based on twistor space. On the other hand, there is the freedom to assign extra holonomy-independent face amplitudes, and this is typically exploited to recover a factor of $2j+1$ needed to guarantee the convolution property of the transition amplitudes at fixed graphs \cite{BojoPerez,Carlodimj}.
If one so wishes, the same freedom can be exploited here to introduce extra face weights given by $(2j+1)/ \mu(j)$, thus completing the matching with the EPRL model.

%-----------------------------------------------------------------
\section{Semiclassical properties}\label{secVI}
%-----------------------------------------------------------------

\noindent
In this Section, we study some semiclassical properties of the model, using the twistorial formalism.
We introduce a notion of curvature and torsion tensors written in terms of the spinors. 
Next, we will relate our twistors to the spinors used in the asymptotic analysis of Barrett et al. \cite{BarrettLorAsymp}, thereby embedding the large spin behaviour in the original phase space. 

%--------------------------------------------------------------------
\subsection{Curvature tensor in terms of spinors} 
%--------------------------------------------------------------------
\noindent 
An important application of our construction is that it allows us to introduce a curvature \emph{tensor}, and study its decomposition in terms of irreducible components and its Petrov classification.
To that end, we work with the wedge spinors at the center of the 4-simplex defined in equation \Ref{spinv}, and reintroduce the $v$ supscript.
Once we have performed in \Ref{pathdef} the integration over $\vpi$, there is a Lagrange multiplier $z$ in a $\d$-function on $\mathbb{C}^2$,  imposing
\begin{equation}
\utilde{\vom}^A=-z\vom^B.
\end{equation}
The transformation mapping $(\vpi,\vom)$ to $(\utilde{\vpi},\utilde{\vom})$ must be a proper $\SL(2,\mathbb{C})$ element, but $\vpi_A\vom^A\neq 0$ implying that this is a complete basis in $\mathbb{C}^2$. We can thus decompose $\utilde{\vpi}$ into the $\vpi$ and $\vom$ spinors to get:
\begin{equation}
\utilde{\vpi}^A=-z^{-1}\vpi^A-{u}\, \vom^A,
\end{equation} 
where $u \in\mathbb{C}$ is the component of $\utilde{\vpi}$ with respect to $\vom$, hence depends on the bulk holonomies $g$ and $\utilde{g}$.
If we now look at the wedge holonomy, we get
\begin{equation}
\ou{h}{A}{B}[\partial w]=-\frac{1}{\po}\Big[z\vom^A\vpi_B-(z^{-1}\vpi^A+u\, \vom^A)\vom_B\Big]
\approx\delta^A_B+\ou{F}{A}{B}[w],
\end{equation}
at first order in the coordinate area of the wedge.
Lowering one index, 
\begin{equation}
F_{AB}[w]\approx-\frac{1}{\po}\Big[(z-z^{-1})\vom_{(A}\vpi_{B)}- u \, \vom_{(A}\vom_{B)}\Big]\label{curvat}.
\end{equation}

At this point, we can use the spinors to introduce a (complex) null tetrad in the internal Minkowski space, analogue to the Newman-Penrose tetrad. This is localised at the center of the 4-simplex, and can be defined as follows:
\bea\label{NP}
&& \ell^I \equiv\I\vpi^A\overset{v}{\bar{\pi}}{}^{\bar A},\quad k^I \equiv\I\vom^A\overset{v}{\bar{\omega}}{}^{\bar A},\quad
m^I \equiv\vom^A\overset{v}{\bar{\pi}}{}^{\bar A},\quad\bar{m}^I \equiv-\overset{v}{\bar{\omega}}{}^{\bar A}\vpi^A, \\\nn
&& \ell^2 = k^2=m^2=0, \qquad \ell^I k_I = m^I \bar m_I = - |\overset{v}{\pi}\overset{v}{\om}|^2.
\eea
Here we used abstract index notation, and 
$X^{A\bar A}\equiv X^I:X^{A\bar A}=\frac{\I}{\sqrt{2}}\ou{\sigma}{A\bar A}{\,I}X^I\in \C^2\otimes\bar\C^2$ (see Appendix). We took capital latin letters starting at $I$ for internal Lorentz indices, as customary in spin foam literature. These should not be confused with spinorial indices, starting at the beginning of the alphabet. 
We remark that $m$ and $\bar m$ identify the plane of the triangle in the reference frame of the 4-simplex. In fact,
\begin{equation}
\uo{\Lambda(g)}{I}{K}\uo{\Lambda(g)}{J}{L}\Sigma_{KL}[t]\propto\I m_{[I}\bar{m}_{J]},
\end{equation}
where $\Sigma_{IJ} \propto (\id-\b\star)\Pi_{IJ}$, 
and $\ou{\Lambda(g)}{I}{J}=\ou{g}{A}{B}\ou{\bar g}{\bar A}{\bar B}$ denotes the proper Lorentz transformation associated to the bulk holonomy $g\in \SL(2,\mathbb{C})$ previously introduced.
Consequently, the bivector $\ell_{[I} k_{J]}$ spans a time-like plane orthogonal to the triangle in the flat reference metric.
By construction, the wedge is parallel to this plane. 
Then, we can define the following curvature tensor in the internal Lorentz indices,
\begin{equation}
F_{CD\,IJ}[w] := a_w F_{CD}[w]\ell_{[I}k_{J]},\label{curv}
\end{equation}
where the overall scale factor $a_w$ measures the area of the wedge
and depends a priori on all the details of the geometry of the 4-simplex.
The expression \Ref{curv} defines a curvature tensor whose only non-vanishing components lie in the plane of the wedge.
This is the usual set-up of Regge calculus and loop quantum gravity, and it is consistent with the tetrahedra bounding the 4-simplex being flat.
In this way, we have achieved something new for the spin foam formalism, that is a description of the full curvature tensor. 

Having introduced a (chiral) curvature tensor,\footnote{The remaining right-handed components are obtained by complex conjugation.} we can decompose it into its $\SL(2,\mathbb{C})$ irreducible parts,  
$(\Psi,T,\Phi, \Phi_0) \in {\bf (2,0)\oplus(1,0)\oplus(0,0)\oplus(1,1)}$, using the $\eps$ tensor: 
\begin{equation}
\begin{split}\label{Fdec}
F_{CD\, IJ}\equiv F_{CD\, A\bar{A}B\bar{B}}&=
\Big[\Psi_{ABCD}+(T_{DB}\epsilon_{CA}+T_{CA}\epsilon_{DB})+\Phi_0 (\epsilon_{CA}\epsilon_{DB}+\epsilon_{DA}\epsilon_{CB})\Big]\bar\epsilon_{\bar A\bar B}+\epsilon_{AB}\Phi_{CD\bar A\bar B}.
\end{split}
\end{equation}
Notice that here the ${\bf (0,0)\oplus(1,1)}$ components are themselves chiral, in the sense that they contain the left-handed projector as in \Ref{sigdef}. For instance, $\Phi_0$ contains both $\d^{I[K}\d^{L]J}$ and $\eps^{IJKL}$ traces.
From the continuum theory, we know that if the connection is Levi-Civita the curvature coincides with the Riemann tensor. In this case, 
$\Psi$ is the (chiral) Weyl tensor, $(\Phi$, $\Phi_0)$ give the Ricci tensor with $\Phi_0$ as its trace, and the algebraic Bianchi identities guarantee that the ${\bf (1,0)}$ component vanishes. Conversely, any non-zero contorsion contributes to all components.

Applying this decomposition to \Ref{curv}, expressed in terms of spinors through \Ref{curvat} and \Ref{NP}, we obtain a spinorial description of the various irreducible components. Explicitly, 
\begin{subequations}\label{Fparam}
\bea
&& \Phi_0=-\frac{a_w}{24}(z-z^{-1})|\po|^2,
\\
&& \Phi_{CD\bar A\bar B}=\frac{a_w}{2}\Big[(z-z^{-1})\vom_{(C}\vpi_{D)}-u\,\vom_{(C}\vom_{D)}\Big]\overset{v}{\bar{\pi}}_{(\bar A}\overset{v}{\bar{\omega}}_{\bar B)},
\\
&& \Psi_{ABCD}=\frac{a_w}{2}\frac{\bar\pi\bar\om}{\po}\Big[(z-z^{-1})\vom_{(A}\vom_B\vpi_C\vpi_{D)}
-u\,\vom_{(A}\vom_B\vom_C\vpi_{D)}\Big], \label{Weyl}
\\ && T_{AB}=\frac{a_w}{8}u\,\bar{\pi} \bar\om \, \vom_A\vom_B.
\eea
\end{subequations}
We see that all components are non-vanishing, hence the off-shell wedge curvature carries contorsion.
This is rather welcomed: what we have defined is an off-shell quantity, and in loop quantum gravity the contorsion is an independent component of the connection, thus can take any value off-shell. 

To further interpret the equations, consider the case of vanishing torsion. Then the algebraic Bianchi identities impose
\begin{equation}
\Phi_0 \stackrel{!}{=}\bar{\Phi}_0,\quad T_{AB}\stackrel{!}{=}0,\quad\overline{\Phi_{AB\bar{C}\bar{D}}}\stackrel{!}{=}\Phi_{CD\bar A \bar B}.
\end{equation}
Looking at \Ref{Fparam}, we see that the conditions are fulfilled provided
\begin{equation}\label{notor}
z\stackrel{!}{\in}\mathbb{R},\quad u\stackrel{!}{=}0.
\end{equation}
In this way we are able to identify a component of the bulk holonomy, $u$, which describes contorsion. At the same time, the Lagrange multiplier $z$ picks up a geometric interpretation on the $C=0$ constraint surface, where $\re(z)$ is related to the Riemann tensor, and $\im(z)$ to contorsion.
Finally, let us discuss the algebraic classification of the resulting Riemann tensor. 
When \Ref{notor} are satisfied, we see from \Ref{Weyl} that the Weyl part has two principal null directions. It is thus of Petrov type D, precisely as posited in Regge calculus \cite{Regge}. 
Unlike in ordinary Regge calculus, we are able to describe additional contorsion components.

In the EPRL model, the on-shell wedge is flat, thus all components of this tensor vanish. One should then look at a curvature tensor associated to faces, which can be also constructed with our methods. They are in fact rather general, all is required is the existence of a Lorentzian phase space structure. 

%---------------------------------------------------------------
\subsection{Wedge flatness and large spin limit}\label{SecSemi}
%---------------------------------------------------------------
\noindent
A key property of the EPRL model is to reproduce exponentials of the Regge action in the large spin limit \cite{BarrettLorAsymp}.
The proof heavily rests upon the use of spinors, and an extra input of wedge (i.e. 4-simplex) flatness.
What we want to show in this Section is the relation between the spinors appearing in the semiclassical analysis and the twistor phase space description of our paper.
The starting point of the semiclassical analysis is the ``propagator'' \Ref{Abarrett}, associated to each wedge. This can be written either in terms of the original phase space spinors, or those transported to the center of the 4-simplex via \Ref{spinv},
\be \label{Abarrett1}
P = \int_{\mathrm{P}\mathbb{C}^2} d^2\vom \,\overline{f^{(\beta j,j)}_{j,\utilde{m}}(\utilde{g} \vom^A)}f^{(\beta j,j)}_{j,m}(g\vom^A)
=\int_{\mathrm{P}\mathbb{C}^2} d^2\om \,\overline{f^{(\beta j,j)}_{j,\utilde{m}}(\utilde{g}g^{-1} \om^A)}f^{(\beta j,j)}_{j,m}(\om^A).
\ee
By direct comparison, we identify $\vom$ with the spinor denoted $z$ in \cite{BarrettLorAsymp}.
The Hermitian scalar product appearing in \Ref{Abarrett} can be written as a \emph{double} integral, over both $\om$ and $\pi$, if one uses the dual map given by the complex structure as shown in Appendix \ref{appdxA}. Upon doing so, we can identify $(\vom,\vpi)$ with the spinor pair denoted $(z,w)$ in  \cite{BarrettLorAsymp}, and bring in the full phase space structure. However, $\pi$ plays no role in the semiclassical analysis in the literature, an aspect we will comment upon below. Instead, the expression \Ref{Abarrett1} is used, with the help of a different type of additional spinors.
Following \cite{LS}, the traces over the magnetic indices in \Ref{fampl} are replaced by resolutions of the identity written in terms of the spinorial coherent states, that is
\begin{equation}\label{resunit}
\sum_{j=-m}^m|jm\rangle\langle jm|=\int_{S^2}d^2\xi\,|j,\xi^A\rangle\langle j,\xi^A|,
\end{equation}
where $S^2$ is parametrised in
 stereographic coordinates by the Hopf section $\xi^0/\xi^1$,
 and $-2\I d^2\xi=\xi_A\di\xi^A\wedge\CC$ is the canonical measure. Here we took the SU(2) spinors of unit norm, which we can do without any loss of information in the following. 
The only semiclassical information contained in these states concerns directions in $\R^3$, identified by the Hopf section, or equivalently
\begin{equation}
n_i[t]\uo{\tau}{AB}{i}=\frac{1}{2\I}\xi_{(A}\delta_{B)\bar B}\bar\xi{}^{\bar B},
\end{equation}
and interpreted as unit vectors normal to the triangle in the frame of the source tetrahedron.

 With the resolution of the identity in terms of coherent states, \Ref{Abarrett1} contains the overlap $\la\om\bar\om\ket{j,\xi^A}$, computed in \Ref{<fCS>}.
Thanks to the factorization property of this overlap, \Ref{Abarrett1} can be rewritten as
the exponential of an action where the spin appears linearly,
\be\label{actionw}
P = \f{2j+1}\pi \int {d^2\vom} \, \f{\exp (s_w)}{\|g\vom\|^2 \|\utilde{g}\vom\|^2}, \qquad 
s_w [\vom, g, \utilde{g}; \xi,\utilde{\xi}] = \beta j \left(\ln\f{\|g\vom\|^2}{\|\utilde{g}\vom\|^2} + \Phi\right), 
\ee
where
\be\label{Phidef}
\Phi :=  \f2\beta \ln \left(\f{\bra{g\vom}\xi\ra}{\|g\vom\|} \f{\bra{\utilde{\xi}}\utilde{g}\vom\ra }{\|\utilde{g}\vom\|} \right).
\ee

Now comes the key input of wedge flatness. If the wedge is flat, the bulk holonomies add up to the phase space $\SL(2,\C)$ holonomy,  $h(\om,\uom,\pi,\upi)=\utilde{g}g^{-1}$, and thus
\begin{equation}
\omega^A=\ou{g}{A}{B}\vom^B,\qquad \utilde{\omega}^A=\ou{\utilde{g}}{A}{B}\vom^B.
\end{equation}
This immediately leads to
\begin{equation}\label{XiB}
\frac{\|g\vom\|^2}{\|\utilde{g}\vom\|^2}=\frac{\|\om\|^2}{\|\utilde{\om}\|^2}=\E^\Xi.
\end{equation}
We have recovered the quantity identified with the extrinsic curvature in the classical study of the phase space, Section \ref{secAB}. Hence, the action in \Ref{actionw} cointains the area-angle term $\beta j \Xi$.
This is a quite satisfying state of affairs, as it shows that the model contains the correct structure of the Regge action with areas $\beta j$.

Notice however that the equations of motion of this action have little in common with discrete general relativity. What is needed at this point are constraints relating the connection and area-angle variables among each other, so to recover the edge lengths as fundamental variables, and the specific functional dependence of areas and dihedral angles upon them.
Crucial help in this direction comes from the large spin limit. The action \Ref{actionw} is complex and with negative real part, and in the large spin limit the path integral \Ref{actionw} is dominated by configurations where the real part of the action vanishes. By inspection of \Ref{Phidef}, these are located at $\om$ proportional to $\xi$, that is
\begin{equation}\label{saddleeq}
\xi^A=\E^{\I\varphi}\frac{\omega^A}{\|\omega\|},
\qquad\utilde{\xi}^A=\E^{\I\utilde{\varphi}}\frac{\utilde{\omega}^A}{\|\utilde{\omega}\|}.
\ee
Using again the wedge flatness, these equations imply
\be
\utilde{\xi}^A=\E^{\Xi/2}\E^{-\I(\vphi-\utilde{\vphi})}\ou{(\utilde{g}g^{-1})}{A}{B}\xi^B. \label{saddleeq1}
\end{equation}
Remarkably, the saddle point equations relate the phase space spinors to the SU(2) boundary spinors in the same way as the primary simplicity constraints relate them to the reduced SU(2) spinors \Ref{spinordef1}. However, there is an important catch, in that the relative phase is not fixed, as it is instead in \Ref{spinordef1}. 
Therefore, we can identify the Hopf sections of the $\xi$'s with our $z$'s, but not the phases.
This is consistent with the fact that the phases of the SU(2) coherent states carry no semiclassical information.
Notice also from \Ref{saddleeq1} that the $\xi$'s are not parallel transported by the Ashtekar-Barbero holonomy $U$, as it is for the $z$'s.
The information about the extrinsic curvature is not in the boundary spinors, but in the bulk holonomies, through \Ref{XiB}.
The actual dependence of the action \Ref{actionw} on $\Xi$, a $D$-dependent quantity, may seem puzzling at first sight, since the boundary states implement \emph{strongly} the $D$ constraint.
It is the wedge $BF$ action that breaks the $D$-symmetry, and so reintroduces a dependence on $\Xi$ in the path integral.\footnote{Accordingly, the saddle point equations define an hypersurface in (the $\om\uom$-polarization of) the original phase space. This hypersurface depends on the relative phases $\vphi$ and $\utilde\vphi$, and it is \emph{not} invariant under $D$. }
This is an important point, because as we discussed at length, a non-trivial embedding of SU(2) in the larger space is necessary to properly talk about extrinsic curvature and Ashtekar-Barbero holonomy. So although the boundary states are insensitive to the extrinsic curvature, it enters the picture via the bulk dynamics.

We have shown that in the large spin limit we can embed the boundary data in the holonomy-flux phase space. The Hopf sections of the SU(2) coherent states are mapped to their equivalent of the reduced spinors \Ref{spinordef1}, and the $\Xi$ is mapped in the integration variables through \Ref{XiB}. On the other hand, the classical areas $\|z_t\|$ do not appear in the action \Ref{actionw} nor in the semiclassical analysis. What is interpreted as the areas are purely quantum numbers, the spins $j_t$.
 
To complete the saddle point analysis, it remains to impose the vanishing of the gradient of \Ref{actionw}.
A crucial ingredient at this point is to restrict attention to foams given by duals to  triangulations. Then, the wedge flatness implies the flatness of each 4-simplex in the triangulation. In this case, it was shown in \cite{BarrettLorAsymp} that the  vanishing of the gradient of \Ref{actionw} can only be satisfied if the boundary data $(j_t,\xi^A_t)$ correspond to a Regge 4-simplex, $(j_t(\ell_e),\xi^A_t(\ell_e))$. At the saddle point, the group elements $g$ are the Regge-Levi-Civita holonomies, and $\Xi=\th(\ell_e)$ is the dihedral angle between 4-normals. 
Notice in particular that as a consequence, the $h$ holonomies solve the secondary constraints at the saddle point.

This gives the right functional dependence in \Ref{actionw}, and the correct Regge dynamics is recovered if the term in $\Phi$ does not contribute. This was proved in \cite{BarrettLorAsymp}, where it was put to zero as a phase choice on the boundary of the 4-simplex. In fact, that this term does not contribute can be shown more generally, face by face on the foam.
From \Ref{fampl}, the action on a face is simply $s_f = \sum_w s_w$.
If $\utilde{g}g^{-1}$ is a pure boost, as at the saddle point, then we immediately have $\sum_w \Phi_w=0$, and so on each face,
\be
s_f = \beta j(\ell_e) \Xi_f(\ell_e), \qquad \Xi_f := \sum_{w\in f} \Xi_w.
\ee
The fact that the second term of $s_w$ vanishes exactly if the connection is Levi-Civita makes the expression look very much like a discrete form of the Holst action.

The discussion shows that the classical interpretation of the large spin asymptotics is consistent with our phase space structure. 
Furthermore, it provides an instance of how a physical connection, solution of both primary and secondary simplicity constraints, provides a non-trivial embedding of SU(2) in the twistor space.
The embedding depends on the spins, not on the norms of the SU(2) spinors, because in this model the areas are quantized.
In this sense, the model is ``semi-coherent''. The directions are represented by classical quantities in phase space, but the areas are quantum numbers. 
This semi-coherence shows up in the way the $\pi\upi$ half of the phase space immediately drops out of the analysis.
Their completely auxiliary role is evident also from the starting point \Ref{pathdef3}, where we showed that the EPRL wedge amplitude is a path integral on twistor space: If we interpret the integrand as the exponential of an action, the latter has no interesting dynamics, only trivial solutions. Technically, this is due to the lack of gluing of the phase space spinors, and it is not a problem for the semiclassical analysis of \cite{BarrettLorAsymp} because as explained, it is the spins and the boundary data spinors that carry the gluing and the classical interpretation of tetrahedra. 
The lack of gluing is in turn inherited from imposing the simplicity constraints as restrictions on the spin labels of the boundary states, so again it is a sign of the ``semi-coherence'' of the model. A face amplitude respecting all gluing conditions would not involve any boundary spin labels but be just an integral over twistors, one for each tetrahedron adjacent to the spin foam face, and would not use independent variables on each wedge. Our formalism provides the means to formulate the LQG dynamics directly in these more covariant terms. %both $\om$ and $\pi$ on equal footing. 
Amplitudes obtained in this way should not deviate much from the EPRL model, but we leave this open for future work.
The basic question here is whether it is possible to formulate a spin foam model as a path integral with an action manifestly discretizing general relativity, thus making the naive semiclassical limit immediate. 
This idea has been addressed in various ways in the literature \cite{DittrichRyan, DittrichRyan2, Bonzom:2008ru, Bonzom:2009hw, Bonzom:2009wm, BaratinOriti,AlexandrovSimplClosure}, and a similar approach has been developed for the Euclidean theory in \cite{EteraHoloEucl}.
We think our twistorial framework provides tools rich enough to make progress in this direction.

%---------------------------------------------------------------------------
\section{Conclusions and perspectives}
%---------------------------------------------------------------------------
\noindent
The twistorial description of loop quantum gravity is a powerful tool to investigate both classical and quantum aspects of the theory. 
As previously shown in \cite{WielandTwistors,IoHolo,IoTwistorNet}, twistors can be used to describe the theory's covariant phase space on a given graph, that is holonomies and fluxes of $\SL(2,\C)$. This is achieved assigning a pair of twistors with equal norms to each link of the graph. 
In this way, we embed the non-linear holonomy-flux algebra in a much simpler algebra of canonical Darboux form. The first advantage of doing so shows up in dealing with the simplicity constraints. In the usual path to the quantum theory, one solves the (primary and secondary) simplicity constraints at the continuum level, and then smears the resulting SU(2) variables. Here we have shown that swapping reduction and smearing is also possible. One smears the covariant $\SL(2,\C)$ variables, and the SU(2) variables are recovered solving the discretized simplicity constraints. As in the continuum, the process requires solving the primary and secondary constraints in successive steps. 
The primary constraint surface is a 7-dimensional hypersurface in $T^*\SL(2,\C)$, parametrized by SU(2) holonomies and fluxes, plus the dihedral angle $\Xi$ between the normals to the source and target 3-cells, in the time gauge. In a general gauge, the picture is unchanged, with an additional boost on each 3-cell transforming the normal out of its canonical time gauge.
From the twistorial perspective, the constraint surface is parametrized by what we call \emph{simple} twistors, parametrized by SU(2) spinors and the dihedral angle, through equations \Ref{piparam} and \Ref{Xidef}. The familiar notion of simple bivectors translates elegantly into simplicity of twistors.

We also found that the dihedral angle is a good coordinate along the orbits generated by the diagonal simplicity constraint. This has important consequences: Assuming that the secondary constraints turn the diagonal simplicity constraints into second class, their solution is then provided by a specific, physical, gauge-fixing section through the orbits. Whatever the gauge section is, we proved that the SU(2) holonomy corresponds to the Ashtekar-Barbero connection, with $\Xi$ measuring the extrinsic curvature projected along the normal to the face. The proof introduces a nice discrete counterpart to the continuum formula $A^{(\beta)}=A+(\beta-\I)K$. It is given by the relation between the SU(2) holonomy and the initial Lorentzian holonomy, equations \Ref{UABprvn} and \Ref{noname},
or equivalently by the relation between the SU(2) class angle and the dihedral angle, equation \Ref{defxi}.
The results show that a consistent symplectic reduction can be obtained after smearing, without any outsourcing from the continuum theory, and have important applications for the interpretation of the theory and the construction of spin foam models.

It remains to formulate an explicit discretization of the secondary constraints, and study the gauge sections they identify. This has been an important open question in the field for many years. The twistorial formalism offers a way to address it, and we hope to come back to this in future research. For the moment, we verified our treatment of the secondary constraints using the simple case of a flat 4-simplex, which is also the one relevant for the EPRL spin foam model. Unlike the case of primary constraints, the solution to the secondary constraints involves a non-local graph structure, and can not be found on each link separately.

Twistors lead to significant insights also in the quantum theory. We quantize the phase space and its Poisson algebra  with a Schr\"odinger picture, and obtain quantum twistor networks, instead of cylindrical functions on the group.
The new states are the homogeneous functions appearing as the canonical basis of the unitary representations of the Lorentz group. They still carry a representation of the holonomy-flux algebra as a sub algebra, and achieve a separation of the source and target structures of the node, which are entangled in the usual holonomy representation. 
Proceeding with Dirac, implementing the diagonal simplicity constraints strongly and the off-diagonal constraints weakly, selects the subspace of ``simple'' irreps $(\r=\g j, k=j)$. In this way we obtain a representation for the Hilbert space of loop quantum gravity, where the argument of the wave-function is a pair of spinors instead of a group element.
The representation is related to the simple projected spin networks \cite{EteraLifting,EteraProj} that appear as boundary states of the EPRL spin foam model \cite{EteraLifting,AlexandrovNewVertex,IoCovariance}.
In fact, we show that the translation between the spinorial wave functions and the cylindrical functions is provided precisely by the spin foam wedge amplitude.

A future goal of our research is to evaluate radiative corrections of quantum gravity transition amplitudes. These have so far provided very hard to compute, and the hope is to improve computational power through the complex analysis methods 
made available by the twistor language.
To that end, we established some preliminary results in this paper. The first concerns rewriting the Liouville measure of $T^*\SL(2,\C)$ in terms of twistors. This is given by equation \Ref{gfxdint}.
The second is a discretization of the $BF$ action as a bilinear in the spinors, see \Ref{BFwedge}. 
Using these tools we gave an independent derivation of the EPRL model, where each individual wedge amplitude is a path integral in twistor space resembling the infinitesimal step of a Feynman path integral, with the position eigenstates replaced by the quantum states solution to the constraints, propagated with the $BF$ action along the bulk of the wedge.
We then investigated some of the semiclassical properties of the EPRL model.
We defined a Newman-Penrose frame through the spinors, and used it to compute the curvature tensor. We computed its  irreducible components, and found two interesting results. The first is the presence, in the off-shell formalism, of torsional components. This is important for the consistency of the theory, because the models are meant to be a quantization of first-order gravity. After imposing a condition of vanishing torsion through the Bianchi identities, we identified the Weyl and Ricci components of the (now Riemannian) curvature tensor, and showed the latter to be of Petrov type D, as it is in Regge calculus.
In the EPRL model, the on-shell value of the wedge curvature is zero, and it is the face curvature tensor that carries the dynamical information. This can also be studied with our techniques.
The flatness of the wedge is also crucial in deriving the well-known asymptotic behaviour of the EPRL model \cite{BarrettLorAsymp,HanZhangLor}: In the large spin limit, the amplitude on a 4-simplex reproduces exponentials of the Regge action. An interesting question concerns the relation between the Regge behavior and the phase space.
We answered the question showing explicitly that: (i) the saddle point equations capture the spinorial version of the simplicity constraints, and (ii) the secondary constraints are solved by the Levi-Civita connection of Regge calculus.
This defines an embedding of the SU(2) variables in the covariant phase space, with a non-trivial section of the dihedral angles, $\Xi(j_t)$. The embedding is a function of the Regge data $j_t$, that is the areas of the triangles.
As a particular feature of the model, these areas are quantized, and not classical quantities. This reflects the semi-coherence of the boundary states, which are semiclassical in the directions, but sharp on the areas, and it also shows up in the lack of a real dynamical role for the momentum $\pi$. In this respect, it would be interesting to look at a version of the path integral where the areas are classical data. The construction of such path integral, and its precise relation to the EPRL model, we leave for future work.

%---------------------------------------------------------------------------
\section*{Acknowledgments}\noindent
Discussions with Sergey Alexandrov, Eugenio Bianchi, Ma\"{i}té Dupuis, Jurek Lewandowski, Alejandro Perez, Carlo Rovelli and Mingyi Zhang are gratefully acknowledged. We would also like to thank the Perimeter Institute, and the Physics Departments of the Universities of Warsaw and Erlangen for hospitality during part of the time this project was developed. This work is partially funded by the ANR Program Blanc LQG09.

\appendix

%---------------------------------------------------------------------------
\section{Spinors, the Lorentz group and its unitary representations}\label{appdxA}\noindent
In this Appendix we review and collect properties of the Lorentz group and its representations that were used in the paper. Further details can be found in \cite{Ruhl, GelfandLorentz, DucLorentz} .

\subsection{Index-free notation}
%---------------------------------------------------------------------------
\noindent
We used explicit spinorial indices in most of the formulas. It is also convenient to introduce a ket notation and dispose of the indices. We define
\be
\ket{\om}=\om^A, \qquad \bra{\om} = \ket{\om}^\dagger = \d_{A\bar A}\bar\om^{\bar A}, \qquad \|\om\|^2=\bra{\om}\om\ra,
\ee
where, again, the Hermitian conjugate is taken with respect to the time normal introduced in \eref{tgauge} and it is SU(2) invariant but not $\SL(2,\C)$ invariant. A further SU(2) quantity is the complex structure $J$ (or ``parity''), which allows us to introduce the dual spinor $(J\om)^A$, which we denote $|\om]$ for brevity:
\be
|\om] := -\eps \ket{\bar\om} = -\d^{A\bar B}\bar{\epsilon}_{\bar B\bar A} \bar\om^{\bar A}, \qquad [\om| = \om_A = \om^B \eps_{BA}, 
\qquad \lbreck\pi\ket{\omega} = \eps_{AB} \pi^A \om^B = \po.\label{dctnry}
\ee
The latter bi-linear is related to the SU(2)-invariant norm by the SU(2) complex structure,
\be\label{tirol}
\lbreck\pi \ket{\om} = \bra{J\pi}\om\ra.
\ee
In the index-free notation, the holonomy-flux variables are
\be
\Pi^i = [\om|\tau^i\ket{\pi}, \qquad h = \f{\ket{\uom}[\pi| - \ket{\upi}[\om|}{\sqrt{[\pi\ket{\om}} \, \sqrt{[\upi\ket{\uom}}},
\ee
and the simplicity constraints
\be
[\pi| = r\E^{\I\f\th2} \bra{\om}.
\ee

The papers  \cite{IoHolo,IoTwistorNet,IoZako} use the index-free notation, but with different conventions. Among these, the left-handed spinor $\omega^A$ is denoted as $\ket{t}$, as here, while $\pi_A$ is written as $\bra{u}$, so that the Lorentz bilinear reads $\bra{u}t\ra$. 

%---------------------------------------------------------------------------
\subsection{Spinors and the Lorentz group}
%--------------------------------------------------------------------------------------------

\noindent 
Consider the group of special orthochronous Lorentz transformations $L_+^\uparrow$ and its universal cover, that is $\SL(2,\C)$. The intertwining $\sigma$-matrices provide the relation between them two, that is the map
\begin{equation}
\Lambda:\SL(2,\C)\ni g\mapsto\Lambda(g)\in L_+^\uparrow:
\ou{g}{A}{B}\ou{\bar{g}}{\bar A}{\bar B}\ou{\sigma}{B\bar B}{\,I}=
\ou{\Lambda(g)}{J}{I}\ou{\sigma}{A\bar A}{\,J}.\label{unicover}
\end{equation}
These intertwiners consist of the identity and the three Pauli matrices $\ou{\sigma}{A}{Bi}$, we set
\begin{equation}
\ou{\sigma}{A\bar A}{0}=\delta^{A\bar A},\quad\ou{\sigma}{A\bar A}{i}=\ou{\sigma}{A}{Bi}\delta^{B \bar A}.\label{soldrng}
\end{equation}
But there is another important invariant structure, given by the $\epsilon$-tensors
\begin{equation}
\epsilon^{AB}=-\epsilon^{BA},\quad \epsilon_{AB}=-\epsilon_{BA},\quad
\epsilon^{AB}\epsilon_{CB}=\uo{\epsilon}{C}{A}=\delta^A_C,\quad\epsilon_{01}=\epsilon^{01}=1.
\end{equation}
Being $\SL(2,\C)$ invariant, we use them to move spinor indices:
\begin{equation}
\pi_A=\pi^B\epsilon_{BA},\quad\pi^A=\epsilon^{AB}\pi_B.
\end{equation}
For if $X^I$ be a point in Minkowski space the intertwiners \eref{soldrng} provide an antilinear bijection into the space of anti-Hermitian $2\times 2$ matrices:
\begin{equation}
X^{A\bar A}\equiv X^I:X^{A\bar A}=\frac{\I}{\sqrt{2}}\ou{\sigma}{A\bar A}{\,I}X^I\in \C^2\otimes\bar\C^2.\label{ismrphsm}
\end{equation}
This anti-isomorphism respects the rules for index moves for both spinors and Minowski vectors:
\begin{equation}
X_{A\bar A}Y^{A\bar A}=-X^0Y^0+X^1Y^1+X^2Y^2+X^3Y^3=X_IY^I.
\end{equation}
The differential map introduced in \eref{unicover}, that is the push-forward $\Lambda_\ast$
gives another isomorphism. It maps $\sl(2,\C)$ into $\mathfrak{so}(1,3)$ according to
\begin{equation}
\Lambda_\ast:\mathfrak{sl}(2,\C)\ni\ou{\Omega}{A}{B}=\frac{1}{2}\ou{\Sigma}{A}{B\,IJ}\Omega^{IJ}\mapsto\ou{\Omega}{I}{J}\in\mathfrak{so}(1,3),\label{diffmap}
\end{equation}
where we have introduced the $\Sigma$-matrices
\begin{equation}
\ou{\Sigma}{A}{B\,IJ}=-\frac{1}{2}\ou{\sigma}{A\bar C}{\,[I}
{\bar\sigma}_{\bar CB\,J]},\label{genrts}
\end{equation}
which are selfdual
\begin{equation}
\frac{1}{2}\ou{\epsilon}{MN}{IJ}\Sigma_{MN}=\star\Sigma_{IJ}=
\I\Sigma_{IJ}
\end{equation}
for $\epsilon^{0123}=1$. In \eref{genrts} the anti-symmetrisation is meant to be over Minkowski indices $I,J$ only. 
These generators are sometimes a bit cumbersome to work with, if we introduce the anti-Hermitian matrices $\ou{\tau}{A}{Bi}=\tfrac{1}{2\I}\ou{\sigma}{A}{Bi}$ we can write,
\begin{equation}
\frac{1}{2}\Omega^{IJ}\Sigma_{IJ}=\tau_i\big(\tfrac{1}{2}\epsilon{}_l{}^i{}_m\Omega^{lm}+\I\ou{\Omega}{i}{o}\big)=:\tau_i\Omega^i,
\end{equation}
and call $\Omega^i\in\mathbb{C}^3$ the selfdual components of $\ou{\Omega}{I}{J}\in\mathfrak{so}(1,3)$.

The group elements of $L_+^\uparrow$ are generated by the exponential of its Lie algebra:
\begin{equation}
\Lambda(\Omega)=\exp\Big(\frac{1}{2}\Sigma_{IJ}\Omega^{IJ}\Big)\in L_+^\uparrow,\quad
\text{where}\quad\ou{\big[\Sigma_{IJ}\big]}{M}{N}=2\delta^M_{[I}\eta_{J]N}
\end{equation}
are the generators of the algebra, and $\eta_{IJ}$ denotes the internal Minkowski metric. We introduce the generators of rotations and boosts respectively
\begin{equation}
L_i=\frac{\I}{2}\epsilon{}^l{}_i{}^m\Sigma_{lm},\quad K_i=\I\Sigma_{io},\label{boostrot}
\end{equation}
together with their complex combinations
\begin{equation}
\Pi_i:=\frac{1}{2}\big(L_i+\I K_i\big),\quad\bar\Pi_i:=\frac{1}{2}\big(L_i-\I K_i\big).
\end{equation}
The isomorphism \eref{diffmap} maps them towards the selfdual generators:
\begin{equation}
\Lambda_\ast\Pi_i=\I\tau_i,\quad \Lambda_\ast\bar\Pi_i=0
\end{equation}
The commutation relations split into two sectors of opposite chirality:
\begin{equation}
[\Pi_i,\Pi_j]=\I\uo{\epsilon}{ij}{m}\Pi_m,\quad [\bar\Pi_i,\bar\Pi_j]=\I\uo{\epsilon}{ij}{m}\bar\Pi_m,\quad
[\Pi_i,\bar\Pi_j]=0.
\end{equation}
The two Casimirs of the Lorentz algebra are:
\begin{equation}
\frac{1}{4}\epsilon^{IJMN}\Sigma_{IJ}\Sigma_{MN}=\frac{1}{2}\star\Sigma^{IJ}\Sigma_{IJ}=-2L_iK^i,\quad\text{and}\quad
\frac{1}{2}\Sigma_{IJ}\Sigma^{IJ}=K_iK^i-L_iL^i
\end{equation}

%----------------------------------------------------------------------------------
\subsection{Unitary representations}\noindent
%----------------------------------------------------------------------------------
We will give here a short overview over the unitary irreducible representations of the Lorentz group, key for understanding the EPRL model. Further reading may be found in \cite{Ruhl,  GelfandLorentz, DucLorentz}.
The representation
\begin{equation}
\SL(2,\mathbb{C})\ni g:\omega\in\mathbb{C}^2\mapsto g\omega\equiv\ou{g}{A}{B}\omega^B.\label{lactn}
\end{equation}
of $\SL(2,\mathbb{C})$ on $\mathbb{C}^2$ is already irreducible, but not unitary. The induced representations on functions $f:\mathbb{C}^2\rightarrow\mathbb{C}$, with the natural $L^2(\mathbb{C}^2,d^4\omega)$ inner product is unitary though reducible. This immediately follows from the homogeneity and unimodularity of the transformation. Irreducible unitary representations are then built just from homogenous functions on $\mathbb{C}^2$.

For the principle series, the weights of homogeneity are parametrised by a half integer $2k\in\mathbb{Z}$ and some $\rho\in\mathbb{R}$. That is we are dealing with functions
\begin{equation}
\forall\lambda\neq 0,\,\omega^A\in\mathbb{C}^2-\{0\}: f(\lambda\omega^A)=\lambda^{-k-1+\I\rho}\bar\lambda^{+k-1+\I\rho}f(\omega^A).\label{densweights}
\end{equation}
From this formula we can easily see for if the pair $(\rho,k)$ label an irreducible unitary representation, its complex conjugate is labelled by $(-\rho,-k)$.
A canonical basis in this infinite-dimensional space is given by the following functions,
\begin{equation}
f^{(\rho,k)}_{j,m}=\sqrt{\frac{2j+1}{\pi}}\|\omega\|^{2(\I\rho-1)}\ou{R^{(j)}(U^{-1}(\omega))}{j}{m},\label{baselmnts}
\end{equation}
where $j\geq k$ and $m=-j, \ldots, j$, and
\begin{equation}
\ou{R^{j}(U)}{m}{n}=\langle j,m|R^{(j)}(U)|j,n\rangle, \quad\text{for}\quad U(\omega)=\frac{1}{\|\omega\|}\begin{pmatrix}\omega^0 & -\bar{\omega}^{\bar{1}} \\ \omega^1 & \phantom{-}\bar{\omega}^{\bar{0}}\end{pmatrix}\in \SU(2),
\end{equation}
are the entries of the spin $j$ Wigner matrix for the $\SU(2)$ element $U(\omega)$ constructed from the spinor. 
The basis elements \eref{baselmnts} diagonalise a complete set of commuting  operators:
\begin{subalign}
&\big(\widehat{L}^2-\widehat{K}^2\big)f^{(\rho,k)}_{j,m}=(k^2-\rho^2-1)f^{(\rho,k)}_{j,m}, &&
\widehat{L}_i\widehat{K}^if^{(\rho,k)}_{j,m}=-k\rho f^{(\rho,k)}_{j,m}\\
&\widehat{L}^2f^{(\rho,k)}_{j,m}=j(j+1)f^{(\rho,k)}_{j,m},
&&\widehat{L}_3f^{(\rho,k)}_{j,m}=mf^{(\rho,k)}_{j,m}
\end{subalign}
where $\widehat{L}_i$ and $\widehat{K}_i$ are the quantisation of the generators earlier.

It is quite convenient to introduce a multi-index notation to group the pair $(j,m)$ into a single index $\mu$. We will also use the notation $\bar\mu$ to keep track of the complex conjugate representation, and use Einstein's summation convention for the $\mu$ indices. With our choices, the matrix representation of the group is the right action, defined according to:
\begin{equation}
\big(D(g)f^{(\rho,k)}_\mu\big)(\omega^A):=f^{(\rho,k)}_\mu\big(\ou{(g^{-1})}{A}{B}\omega^B\big)=
f^{(\rho,k)}_\mu(-\omega^B\uo{g}{B}{A})=\ou{D^{(\rho,k)}(g)}{\nu}{\mu}f^{(\rho,k)}_\nu(\omega^A).
\end{equation}
Since the representation is unitary, it admits an $SL(2,\mathbb{C})$-invariant Hermitian inner product. This is defined as a surface integral on $\mathrm{P}\mathbb{C}^2\subset \mathbb{C}^2$,\footnote{Because of the homogeneity, integrating over all of $\C^2$ would lead to divergences.}
\begin{equation}
\langle f^{(\rho,k)}_\mu|f^{(\rho,k)}_\nu\rangle=\frac{\I}{2}\int_{\mathrm{P}\mathbb{C}^2}\omega_A\di\omega^A\wedge\bar{\omega}_{\bar A}\di\bar{\omega}^{\bar A}
\overline{f^{(\rho,k)}_\mu(\omega^A)}f^{(\rho,k)}_\nu(\omega^A)=\delta_{\bar\mu\nu}\label{orthgnlty},
\end{equation}
its value being independent of the way $\mathrm{P}\mathbb{C}^2$ is embedded into $\mathbb{C}^2$  thanks to the homogeneity of the integrand. 

The $\SL(2,\mathbb{C})$ group locally represents the group of special orthochronous transformations. To recover the full Lorentz group we also need parity 
\begin{equation}
\big(Pf^{(\rho,k)}_\mu\big)(\omega^A)=f^{(\rho,k)}_\mu(\delta^{A\bar A}\bar\omega_{\bar A}),
\end{equation}
and time reversal 
\begin{equation}
\big(Tf^{(\rho,k)}_\mu\big)(\omega^A)=\epsilon_{\alpha\mu}\delta^{\alpha\bar\nu}\overline{f^{(\rho,k)}_\nu(\delta^{A\bar A}\bar\omega_{\bar A})},
\end{equation}
both of which have recently gained  \cite{CarloEdParity} some interest in LQG.
From \eref{densweights} we can realize parity and time reversal map the irreducible unitary representation of labeles $(\rho,k)$ to those of
$(\rho,-k)$ and $(-\rho,k)$ respectively.

In each representation space there are two invariants, the first one is the Hermitian inner product \eref{orthgnlty} introduced in above, the second one is the $\epsilon$-invariant 
\begin{equation}
\begin{split}
\big\lbreck f^{(\rho,k)}_\mu\big|f^{(\rho,k)}_\nu\big\rangle :=&
\f{k-\I\r}{4\pi} \int_{P\mathbb{C}^2\times P\mathbb{C}^2}\!\!\!\!\!\!\!\!\!\!\!\!\!\!\!\!\!\!\!\!\omega_A\di\omega^A\wedge\bar{\omega}_{\bar{A}}\di\bar{\omega}^{\bar{A}}\wedge\pi_A\di\pi^A\wedge\bar{\pi}_{\bar{A}}\di\bar{\pi}^{\bar{A}}
(\pi_A\omega^A)^{k-1-\I\rho}(\bar\pi_{\bar A}\bar\omega^{\bar A})^{-k-1-\I\rho}\\
&\qquad\cdot f^{(\rho,k)}_\mu(\pi^A)f^{(\rho,k)}_\nu(\omega^A)=\eps_{\mu\nu}.\label{epsbilin}
\end{split}
\end{equation}
Its matrix elements are
\begin{equation}
\epsilon_{(j,m)(j^\prime,m^\prime)}=(-1)^{k-m}\delta_{j,j^\prime}\delta_{m,-m^\prime}\frac{\Gamma(k+1-\I\rho)}{\Gamma(k+1+\I\rho)}\frac{\Gamma(j+1+\I\rho)}{\Gamma(j+1-\I\rho)},
\end{equation}
where Euler's $\Gamma$ function appears. Though infinite dimensional, each of the invariants comes with an inverse, and
\begin{equation}
\delta^{\mu\bar\alpha}\delta_{\nu\bar\alpha}=\delta^\mu_\nu=\epsilon^{\mu\alpha}\epsilon_{\nu\alpha},\quad
\delta^{\mu\bar\nu}=\overline{\delta^{\nu\bar\mu}},\quad\epsilon^{\mu\nu}=\delta^{\mu\bar\alpha}\delta^{\nu\bar\beta}\bar\epsilon_{\bar\alpha\bar\beta},\quad\text{and}\quad \epsilon^{\mu\nu}=(-1)^{2k}\epsilon^{\nu\mu}.\label{barrelations}
\end{equation}

Thanks to the completeness of the basis,  \Ref{epsbilin} and \Ref{orthgnlty} imply for each irreducible subspace $(\rho,k)$ a relation between the the ket and its dual,
\begin{equation}\label{[map]}
\lbreck f_\mu|=\frac{\pi}{\I\rho-k}\epsilon_{\mu\nu}\delta^{\nu\bar\a}\bra{f_\a}.
\end{equation}
Since both $\delta^{\nu\bar\nu}$ and $\epsilon_{\mu\nu}$ are invariant this map commutes with the group action, and implicitly shows the representation labelled by $(\rho,k)$ is unitarily equivalent to its complex conjugate, that is the $(-\rho,-k)$ representation. 

The map \Ref{[map]} allows us to relate the bilinear invariant \Ref{epsbilin} to the Hermitian inner product \Ref{orthgnlty}, which we used in section \ref{SecSemi} to compare our formulas to those of \cite{BarrettLorAsymp}. This is analogue to the finite dimensional case, see \Ref{tirol}, except that now both quantities are $\SL(2,\C)$ invariant.

The dual vector can be obtained also by Fourier transform, up to a phase, as was used in the main text in \Ref{fourtrans}. In fact, we have
%In \Ref{fourtrans} we have also introduced the Fourier transform of the basis vectors. The integral gives explicitly
\begin{equation}
{f}^{(\rho,k)}_{\bar{\mu}}(\pi_A):=\frac{1}{\pi^2}\int_{\mathbb{C}^2}d^4\omega\E^{\I\pi_A\omega^A-\CC}\overline{f^{(\rho,k)}_\mu(\omega^A)}=\E^{-\I\pi k}\frac{\Gamma(k+1-\I\rho)}{\Gamma(k+1+\I\rho)}\delta_{\mu\bar\mu}\epsilon^{\mu\nu}f^{(\rho,k)}_\nu(\pi^A),\label{fouriertrafo}
\end{equation} 
and defines an antilinear map from the $(\rho,k)$ representation onto itself, whereas complex conjugation maps the $(\rho,k)$ towards the $(-\rho,-k)$ representation, implicitly showing that $(\rho,k)$ and $(-\rho,-k)$ are unitarily equivalent.
We will give a detailed proof of this integral elsewhere, but let us mention the basic strategy behind: First, thanks to the $\SL(2,\C)$ invariance of the integral, one can realize the left hand side equals the right hand side up to a constant. This constant can only depend on the labels $\rho$ and $k$. Next, one shows, this constant has unit norm. Calculating the integral for the states of spin labels $k=j=m$, eventually gives the phase appearing in \eref{fouriertrafo}.

%-----------------------------------------------------------------------
\section{The integral determining $\mu(j)$}\label{appdxB}
%-----------------------------------------------------------------------

\begin{figure}[ht]
     \centering
     \includegraphics[width= 0.3\textwidth]{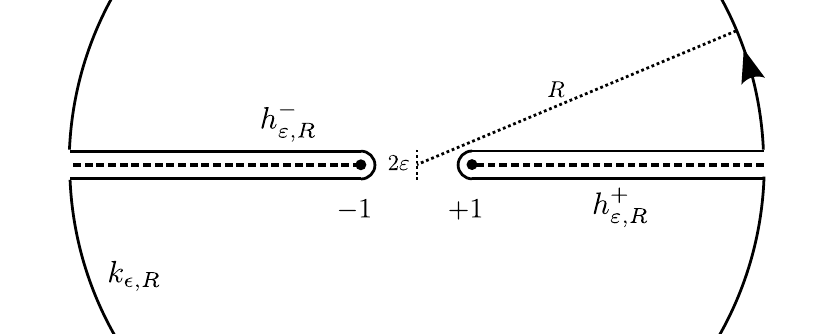}
     \caption{The integration domain is bounded by a Hankel contour around the branch cuts in the complex plane.}
     \label{contr}
\end{figure}

\noindent
In section \ref{secV} we encountered an integral over the complex plane that contributes to the loop quantum gravity face amplitude. In this Appendix, we present the details of its evaluation, as well as the explicit form of the phase $\Delta(\beta,j)$. The integral of interest is
\begin{equation}
I_{\beta,j}=\frac{\I}{2}\int_{\mathbb{C}}\di z\wedge\di \bar{z}(1-z^2)^a(1-\bar{z}^2)^b,\quad\text{for: }
\left\{\begin{split}
a=-\I\beta j+j-1,\\
b=-\I\beta j-j-1,
\end{split}\label{Iint}
\right.
\end{equation}
and $2j\in \N^+$.
%For $j=0$ this integral is ill-defined, for $2j\in\mathbb{N}_>$ we proceed as follows. Let us first recognise
Although the integrand is not holomorphic, the integral can be manipulated to a contour integral using Stokes' theorem. To do that, we write
\be\label{arturo}
(1-z^2)^a = \f12 \f{\p}{\p z} \Big[zF(-a,\tfrac{1}{2};\tfrac{3}{2};z^2)\Big],
\ee
where \cite{abramstegun}
\begin{equation}\label{hypergeo}
F(a,b;c;z)=\frac{\Gamma(c)}{\Gamma(b)\Gamma(c-b)}\int_0^1\di t\,t^{b-1}(1-t)^{c-b-1}(1-tz)^{-a},\quad \mathrm{Re}(c)>\mathrm{Re}(b)>0,
\end{equation}
is the hypergeometric function. \Ref{arturo} can be verified using
\begin{equation}
\int_0^x\di t(1-t^2)^a=\frac{x}{2}\int_0^1\di t\, t^{-\frac{1}{2}}(1-x^2t)^a.
\end{equation}
Hence,
\begin{equation}
I_{\beta,j}=\frac{\I}{4}\int_{\mathcal{D}}\di z\wedge\di \bar{z}\frac{\partial}{\partial z}\Big[zF(-a,\tfrac{1}{2};\tfrac{3}{2};z^2)\Big](1-\bar{z}^2)^b=
\frac{\I}{4}\int_{\mathcal{D}}\di\wedge\Big[\di{\bar{z}}\,zF(-a,\tfrac{1}{2};\tfrac{3}{2};z^2)(1-\bar{z}^2)^b\Big].\label{totalderiv}
\end{equation}
The hypergeometric function $F(\cdot,\cdot;\cdot;z^2)$ is analytic (hence differentiable) unless $z^2\in\{x\in\mathbb{R}|x\geq 1\}$, where there is a brach cut, so the integration domain $\mathcal{D}$ denotes the complex plane cut along the real $x$-axis from $-\infty$ to $x=-1$ and $x=1$ to $x=\infty$. Applying Stoke's theorem, we find
\begin{equation}
I_{\beta,j}=\frac{\I}{4}\lim_{\epsilon\searrow 0}\lim_{R\rightarrow\infty}\Big[\int_{h_{R,\varepsilon}^+}\!\!\!+\int_{h_{R,\varepsilon}^-}\!\!\!+\int_{k_{R,\varepsilon}}\Big]\di \bar{z}\,zF(-a,\tfrac{1}{2};\tfrac{3}{2};z^2)(1-\bar{z}^2)^b,
\end{equation}
where $h_{\infty,\varepsilon}^\pm$ are Hankel contours encircling the points $\pm 1$, while $k_{R,0}$ denotes a circle of radius $R$ around the origin. Orientation and shape of the contour are fixed in Fig.\ref{contr}. 
The contribution from the integral around the circle at infinity vanishes. 
To prove it, we use  \cite{abramstegun} 
\begin{equation}\nn
\begin{split}
F(a,b;c;z)=\frac{\Gamma(c)\Gamma(b-a)}{\Gamma(b)\Gamma(c-a)}&(-z)^{-a}F(a,1-c+a;1-b+a;z^{-1})+\\
&+\frac{\Gamma(c)\Gamma(a-b)}{\Gamma(a)\Gamma(c-b)}(-z)^{-b}F(b,1-c+b;1-a+b;z^{-1}).
\end{split}
\end{equation}
For $j>0$ and $a=-\I\beta j+j-1$ we find the limits
\begin{equation}\nn
\lim_{z\rightarrow 0}F(-a,-\tfrac{1}{2}-a;\tfrac{1}{2}-a,z)=0=
\lim_{z\rightarrow 0}F(\tfrac{1}{2},0;a+\tfrac{3}{2},z)
\end{equation}
Since ${\re}(a+b+1)=-1<0$ and ${\re}(2b+1)=-2j-1<0$ we get 
\begin{align}\nn
&\nonumber\lim_{R\rightarrow\infty}\Big|\int_{k_{R,0}}\di \bar{z}\,zF(-a,\tfrac{1}{2};\tfrac{3}{2};z^2)(1-\bar{z}^2)^b\Big|\leq 
\lim_{R\rightarrow\infty}\int_{-\pi}^\pi\di\varphi|R^{2b+2}|\\
&\nonumber\times\bigg(\frac{1}{|2a+1|}
|R^{2a}|\,\big|F(-a,-\tfrac{1}{2}-a;\tfrac{1}{2}-a,R^{-2}\E^{-2\I\varphi})\big|+\frac{\Gamma(\tfrac{3}{2})|\Gamma(-a-\tfrac{1}{2})|}{|\Gamma(-a)|}R^{-1}\big|F(\tfrac{1}{2},0;a+\tfrac{3}{2},R^{-2}\E^{-2\I\varphi})\big|
\bigg)=0
%\\&\qquad=0
\end{align}
Therefore, in the limit $R\rightarrow\infty$ only the Hankel contours $h^+_{\varepsilon,\infty}$ and $h^-_{\varepsilon,\infty}$ can contribute to the integral $I_{\beta,j}$. A moment of reflection reveals them to be equal:
\begin{equation}
\int_{h^-_{\varepsilon,R}}\di \bar{z}\,zF(-a,\tfrac{1}{2};\tfrac{3}{2};z^2)(1-\bar{z}^2)^b=
\int_{h^+_{\varepsilon,R}}\di \bar{z}\,zF(-a,\tfrac{1}{2};\tfrac{3}{2};z^2)(1-\bar{z}^2)^b.
\end{equation}
The problem when trying to evaluate these integrals is that 
$F(\cdot,\cdot;\cdot;z)$ is generally not single valued for $z$ real and $z>1$. However there is now a remarkable identity relating the hypergeometric function around the branch cut to those regions where it behaves perfectly regular, it reads \cite{abramstegun}:
\begin{equation}
\begin{split}
F(a,b;c;z)&=\frac{\Gamma(c)\Gamma(c-a-b)}{\Gamma(c-a)\Gamma(c-b)}F(a,b;a+b-c+1;1-z)+\\
&\qquad+(1-z)^{c-a-b}\frac{\Gamma(c)\Gamma(a+b-c)}{\Gamma(a)\Gamma(b)}F(c-a,c-b;c-a-b+1;1-z)
\end{split}\label{ident}
\end{equation}
We insert this identity into $I_{\beta,j}$, and eventually get:
\begin{equation}
\begin{split}
I_{\beta,j}=&\frac{\I}{2}\lim_{\varepsilon\searrow 0}\int_{h_{\epsilon,\infty}^+}\di \bar{z}\,z(1-\bar{z}^2)^b\bigg(\frac{\Gamma(\tfrac{3}{2})\Gamma(a+1)}{\Gamma(a+\tfrac{3}{2})}F(-a,\tfrac{1}{2};-a;1-z^2)+\\
&-\frac{1}{2a+2}(1-z^2)^{a+1}F(\tfrac{3}{2}+a,1;a+2;1-z^2)\bigg)
\end{split}\label{int2}
\end{equation}
In the limit $\epsilon\rightarrow 0$ the second part of this equation vanishes: For if $a-b\in \mathbb{Z}$, the function $(1-\bar{z})^b(1-z)^{a+1}F(\,,\,;\,;1-z)$ is single valued (though not analytic) around the branch cut. But we can split the Hankel contour in two parts, one lying in the upper half of the complex plane, the other one in the lower half. Since the integrand is single-valued around the cut, and both contributions appear with opposite signs they cancel in the limit of $\epsilon\rightarrow 0$.
Concerning the first part of \eref{int2}, 
\begin{equation}
F(-a,\tfrac{1}{2};-a;1-z^2)=F(\tfrac{1}{2},-a;-a;1-z^2)=\big(z^2\big)^{-\frac{1}{2}}.
\end{equation}
Hence
\begin{equation}
I_{\beta,j}=\frac{\I}{4}\frac{\Gamma(\tfrac{1}{2})\Gamma(a+1)}{\Gamma(a+\tfrac{3}{2})}\lim_{\varepsilon\searrow 0}\int_{h^+_{\varepsilon,\infty}}\di\bar{z}(1-\bar{z}^2)^b=-\frac{\I}{4}\frac{\Gamma(\tfrac{1}{2})\Gamma(a+1)}{\Gamma(a+\tfrac{3}{2})}\lim_{\varepsilon\searrow 0}\int_{h^+_{\varepsilon,\infty}}\di{z}(1-{z}^2)^b.
\end{equation}
We have thus reduced our original integral \eref{Iint} to an ordinary analytic line integral. This we calculate by the usual methods and find
\begin{equation}
\lim_{\varepsilon\searrow 0}\int_{h^+_{\varepsilon,\infty}}\di{z}(1-{z}^2)^b=-2\I\sin(\pi b)\frac{\Gamma(-\tfrac{1}{2}-b)\Gamma(b+1)}{\Gamma(\tfrac{1}{2})}.
\end{equation}
With
$
\Gamma(z)\Gamma(1-z)\sin\pi z=\pi
$
we further simplify and eventually get
\begin{equation}
\text{for $j>0$:}\quad I_{\beta,j}=-\frac{\I\pi}{\beta-\I}\frac{1}{4j}\frac{\Gamma(-\I\beta j+j)\Gamma(+\I\beta j+j+\tfrac{1}{2})}{\Gamma(+\I\beta j+j)\Gamma(-\I\beta j+j+\tfrac{1}{2})} = \frac{\pi}{\sqrt{1+\beta^2}}\frac{1}{4j} \E^{\I\Delta(\beta,j)}.
\label{fin}
\end{equation}
In the large spin limit we can use Stirling's asymptotic formula to approximate the ratio of gamma functions
by $({1+\I\beta})^{1/2} / ({1-\I\beta})^{1/2},$
and 
\begin{equation}
I_{\beta, j}\approx\frac{\pi}{\sqrt{1+\beta^2}}\frac{1}{4j}, \qquad \text{for $j\gg 1$}.
\end{equation}

The integral \Ref{Iint} is ill-defined for $j=0$. This is not directly relevant for the spin foam model, where the amplitude at $j=0$ is assigned independently requiring cylindrical consistency. Nonetheless, the integral can be regularized to vanish at this singular value.
Let us first introduce polar coordinates, and perform a partial fraction expansion to find:
\begin{align}
\nonumber I_{\beta,0}&=\frac{\I}{2}\int_{\mathbb{C}}\di z\wedge \di\bar{z}\frac{1}{(1-z^2)(1-\bar{z}^2)}=\int_{-\pi}^{\pi}\di\varphi\int_0^\infty\di r\frac{r}{(1-r^2\E^{2\I\varphi})(1-r^2\E^{-2\I\varphi})}\\
\nonumber &=\frac{1}{2}\int_{-\pi}^\pi\di\varphi\int_0^\infty\di
r\frac{1}{(\E^{-\I\varphi}-r)(\E^{\I\varphi}-r)}=-\frac{1}{2}\int_{-\pi}^\pi\di\varphi\int_0^\infty\di r\frac{1}{\E^{\I\varphi}-\E^{-\I\varphi}}\Big[(\E^{\I\varphi}-r)^{-1}-(\E^{-\I\varphi}-r)^{-1}\Big]\\
&=\frac{1}{2}\int_{-\pi}^\pi\frac{\di\varphi}{\E^{\I\varphi}-\E^{-\I\varphi}}\Big[\ln(\E^{\I\varphi}-r)-\ln(\E^{-\I\varphi}-r)\Big]_{r=0}^\infty
=\frac{1}{2}\int_{-\pi}^{\pi}\frac{\di\varphi}{\E^{\I\varphi}-\E^{-\I\varphi}}\Big[2\I\pi\mathrm{sign}(\varphi)-2\I\varphi\Big].
\end{align}
The last two integrals can readily be performed. Set $x=\cos \varphi$ to get:
\begin{equation}
\int_{-\pi}^\pi\di\varphi\frac{\mathrm{sign}(\varphi)}{\E^{\I\varphi}-\E^{-\I\varphi}}=-2\I\int_{0}^1\frac{\di x}{1-x^2}.\label{plusint}
\end{equation}
The second integral goes around the unit circle. We recognise the integrand is analytic along this path unless it reaches the point $z=-1$, and can thus smoothly deform the integration domain to find:
\begin{align}
\nonumber\int_{-\pi}^\pi\di\varphi\frac{\varphi}{\E^{\I\varphi}-\E^{-\I\varphi}}&=-\oint_{|z|=1}\di z\frac{\ln z}{z^2-1}=\\
\nonumber&=-\lim_{\varepsilon\searrow 0}\bigg[\int_0^1\di x\frac{\ln(-1+x-\I\varepsilon)}{(1-x)^2-1}-\int_0^1\di t\frac{\ln(-1+x+\I\varepsilon)}{(1-x)^2-1}+\I\int_{-\pi}^\pi\di\varphi\varepsilon\E^{\I\varphi}\frac{\ln(\varepsilon\E^{\I\varphi})}{\varepsilon^2\E^{2\I\varphi}-1}\bigg]=\\
&=2\I\pi\int_0^1\di x\frac{1}{(1-x)^2-1}=-2\I\pi\int_{0}^1\frac{\di x}{1-x^2}.\label{minusint}
\end{align}
We put the branch cut for the complex logarithm on the negative real axis, got for any $0<x<1$ that
\begin{equation}
\lim_{\varepsilon\searrow 0}\ln (-1+x\pm\I\varepsilon)=\pm\I\pi,
\end{equation}
and used that the integral around the small semicircle vanishes due to $\lim_{\varepsilon\rightarrow 0}\varepsilon\ln\varepsilon=0$. 
Hence, in both \eref{plusint} and \eref{minusint} there appears the very same integral. Each integral diverges logarithmically at its upper bound $x=1$:
\begin{equation}
2\int_0^1\frac{\di x}{x^2-1}=\Big[\ln|x-1|-\ln|x+1|\Big]_{x=0}^1=-\infty.
\end{equation}
But, if we were to take the limit $x\rightarrow 1$ in both \eref{plusint} and \eref{minusint} equally fast, both contributions would sum up to zero. We can thus put the regularised integral to vanish:
\begin{equation}
I_{\beta,0}\stackrel{\mathrm{reg}}=0.
\end{equation}

%--------------------------------------------------------------------------------------------------
%--------------------------------------------------------------------------------------------------


\begin{thebibliography}{10}

\bibitem{twigeo2}
L.~Freidel and S.~Speziale, {\it {From twistors to twisted geometries}},
  Phys.Rev. {\bf D82} (2010) 084041 [\href{http://arXiv.org/abs/1006.0199}{{\tt
  1006.0199}}].

\bibitem{WielandTwistors}
W.~M. Wieland, {\it {Twistorial phase space for complex Ashtekar variables}},
  Class.Quant.Grav. {\bf 29} (2012) 045007
  [\href{http://arXiv.org/abs/1107.5002}{{\tt 1107.5002}}].
%%CITATION = ARXIV:1107.5002;%%

\bibitem{IoHolo}
M.~Dupuis, L.~Freidel, E.~R. Livine and S.~Speziale, {\it {Holomorphic
  Lorentzian Simplicity Constraints}},
  \href{http://arXiv.org/abs/1107.5274}{{\tt 1107.5274}}.

\bibitem{IoTwistorNet}
E.~R. Livine, S.~Speziale and J.~Tambornino, {\it {Twistor Networks and
  Covariant Twisted Geometries}},  \href{http://arXiv.org/abs/1108.0369}{{\tt
  1108.0369}}.

\bibitem{EteraLifting}
M.~Dupuis and E.~R. Livine, {\it {Lifting SU(2) Spin Networks to Projected Spin
  Networks}},  Phys.Rev. {\bf D82} (2010) 064044
  [\href{http://arXiv.org/abs/1008.4093}{{\tt 1008.4093}}].

\bibitem{AlexandrovNewVertex}
S.~Alexandrov, {\it {The new vertices and canonical quantization}},  Phys.Rev.
  {\bf D82} (2010) 024024 [\href{http://arXiv.org/abs/1004.2260}{{\tt
  1004.2260}}].
%%CITATION = ARXIV:1004.2260;%%

\bibitem{IoCovariance}
C.~Rovelli and S.~Speziale, {\it {Lorentz covariance of loop quantum gravity}},
   Phys.Rev. {\bf D83} (2011) 104029
  [\href{http://arXiv.org/abs/1012.1739}{{\tt 1012.1739}}].

\bibitem{EPRL}
J.~Engle, E.~Livine, R.~Pereira and C.~Rovelli, {\it {LQG vertex with finite
  Immirzi parameter}},  Nucl.Phys. {\bf B799} (2008) 136--149
  [\href{http://arXiv.org/abs/0711.0146}{{\tt 0711.0146}}].
%%CITATION = ARXIV:0711.0146;%%

\bibitem{KKL}
W.~Kaminski, M.~Kisielowski and J.~Lewandowski, {\it {Spin-Foams for All Loop
  Quantum Gravity}},  Class.Quant.Grav. {\bf 27} (2010) 095006
  [\href{http://arXiv.org/abs/0909.0939}{{\tt 0909.0939}}].
%%CITATION = ARXIV:0909.0939;%%

\bibitem{CarloGenSF}
Y.~Ding, M.~Han and C.~Rovelli, {\it {Generalized Spinfoams}},  Phys.Rev. {\bf
  D83} (2011) 124020 [\href{http://arXiv.org/abs/1011.2149}{{\tt 1011.2149}}].
%%CITATION = ARXIV:1011.2149;%%

\bibitem{EPR}
J.~Engle, R.~Pereira and C.~Rovelli, {\it {The Loop-quantum-gravity
  vertex-amplitude}},  Phys.Rev.Lett. {\bf 99} (2007) 161301
  [\href{http://arXiv.org/abs/0705.2388}{{\tt 0705.2388}}].
%%CITATION = ARXIV:0705.2388;%%

\bibitem{EPRlong}
J.~Engle, R.~Pereira and C.~Rovelli, {\it {Flipped spinfoam vertex and loop
  gravity}},  Nucl.Phys. {\bf B798} (2008) 251--290
  [\href{http://arXiv.org/abs/0708.1236}{{\tt 0708.1236}}].
%%CITATION = ARXIV:0708.1236;%%

\bibitem{LS}
E.~R. Livine and S.~Speziale, {\it {A New spinfoam vertex for quantum
  gravity}},  Phys.Rev. {\bf D76} (2007) 084028
  [\href{http://arXiv.org/abs/0705.0674}{{\tt 0705.0674}}].
%%CITATION = ARXIV:0705.0674;%%

\bibitem{LS2}
E.~R. Livine and S.~Speziale, {\it {Consistently Solving the Simplicity
  Constraints for Spinfoam Quantum Gravity}},  Europhys.Lett. {\bf 81} (2008)
  50004 [\href{http://arXiv.org/abs/0708.1915}{{\tt 0708.1915}}].
%%CITATION = ARXIV:0708.1915;%%

\bibitem{FK}
L.~Freidel and K.~Krasnov, {\it {A New Spin Foam Model for 4d Gravity}},
  Class.Quant.Grav. {\bf 25} (2008) 125018
  [\href{http://arXiv.org/abs/0708.1595}{{\tt 0708.1595}}].
%%CITATION = ARXIV:0708.1595;%%

\bibitem{BahrOperatorSF}
B.~Bahr, F.~Hellmann, W.~Kaminski, M.~Kisielowski and J.~Lewandowski, {\it
  {Operator Spin Foam Models}},  Class.Quant.Grav. {\bf 28} (2011) 105003
  [\href{http://arXiv.org/abs/1010.4787}{{\tt 1010.4787}}].
%%CITATION = ARXIV:1010.4787;%%

\bibitem{BarrettLorAsymp}
J.~W. Barrett, R.~Dowdall, W.~J. Fairbairn, F.~Hellmann and R.~Pereira, {\it
  {Lorentzian spin foam amplitudes: Graphical calculus and asymptotics}},
  Class.Quant.Grav. {\bf 27} (2010) 165009
  [\href{http://arXiv.org/abs/0907.2440}{{\tt 0907.2440}}].
%%CITATION = ARXIV:0907.2440;%%

\bibitem{HanZhangLor}
M.~Han and M.~Zhang, {\it {Asymptotics of Spinfoam Amplitude on Simplicial
  Manifold: Lorentzian Theory}},  \href{http://arXiv.org/abs/1109.0499}{{\tt
  1109.0499}}.
%%CITATION = ARXIV:1109.0499;%%

\bibitem{PerezLR}
A.~Perez, {\it {The Spin Foam Approach to Quantum Gravity}},  To appear in
  Living Reviews in Relativity (2012)
  [\href{http://arXiv.org/abs/1205.2019}{{\tt 1205.2019}}].
%%CITATION = ARXIV:1205.2019;%%

\bibitem{BojoPerez}
M.~Bojowald and A.~Perez, {\it {Spin foam quantization and anomalies}},
  Gen.Rel.Grav. {\bf 42} (2010) 877--907
  [\href{http://arXiv.org/abs/gr-qc/0303026}{{\tt gr-qc/0303026}}].
%%CITATION = GR-QC/0303026;%%

\bibitem{Carlodimj}
E.~Bianchi, D.~Regoli and C.~Rovelli, {\it {Face amplitude of spinfoam quantum
  gravity}},  Class.Quant.Grav. {\bf 27} (2010) 185009
  [\href{http://arXiv.org/abs/1005.0764}{{\tt 1005.0764}}].
%%CITATION = ARXIV:1005.0764;%%

\bibitem{ConradyFreidel2}
F.~Conrady and L.~Freidel, {\it {On the semiclassical limit of 4d spin foam
  models}},  Phys.Rev. {\bf D78} (2008) 104023
  [\href{http://arXiv.org/abs/0809.2280}{{\tt 0809.2280}}].
%%CITATION = ARXIV:0809.2280;%%

\bibitem{ConradyFreidel3}
F.~Conrady and L.~Freidel, {\it {Quantum geometry from phase space reduction}},
   J.Math.Phys. {\bf 50} (2009) 123510
  [\href{http://arXiv.org/abs/0902.0351}{{\tt 0902.0351}}].
%%CITATION = ARXIV:0902.0351;%%

\bibitem{BarrettEPRasymp}
J.~W. Barrett, R.~Dowdall, W.~J. Fairbairn, H.~Gomes and F.~Hellmann, {\it
  {Asymptotic analysis of the EPRL four-simplex amplitude}},  J.Math.Phys. {\bf
  50} (2009) 112504 [\href{http://arXiv.org/abs/0902.1170}{{\tt 0902.1170}}].
%%CITATION = ARXIV:0902.1170;%%

\bibitem{HanZhangEucl}
M.~Han and M.~Zhang, {\it {Asymptotics of Spinfoam Amplitude on Simplicial
  Manifold: Euclidean Theory}},  \href{http://arXiv.org/abs/1109.0500}{{\tt
  1109.0500}}.
%%CITATION = ARXIV:1109.0500;%%

\bibitem{Mikovic:2011zx}
  A.~Mikovic and M.~Vojinovic,
  ``Effective action and semiclassical limit of spin foam models,''
  Class.\ Quant.\ Grav.\  {\bf 28} (2011) 225004
  [arXiv:1104.1384 [gr-qc]].
  %%CITATION = ARXIV:1104.1384;%%
  
\bibitem{PenroseRindler1}
R.~Penrose and W.~Rindler, {\em {Spinors And Space-Time. 1. Two Spinor Calculus
  And Relativistic Fields}}.
\newblock CUP, 1985.
%%CITATION = INSPIRE-216889;%%

\bibitem{PenroseRindler2}
R.~Penrose and W.~Rindler, {\em {Spinors And Space-Time. Vol. 2: Spinor And
  Twistor Methods In Space-Time Geometry}}.
\newblock CUP, 1986.
%%CITATION = INSPIRE-238378;%%

\bibitem{Penrose72}
R.~Penrose and M.~A. MacCallum, {\it {Twistor theory: An Approach to the
  quantization of fields and space-time}},  Phys.Rept. {\bf 6} (1972) 241--316.
%%CITATION = PRPLC,6,241;%%

\bibitem{twigeo}
L.~Freidel and S.~Speziale, {\it {Twisted geometries: A geometric
  parametrisation of SU(2) phase space}},  Phys.Rev. {\bf D82} (2010) 084040
  [\href{http://arXiv.org/abs/1001.2748}{{\tt 1001.2748}}].

\bibitem{CorichiHolstBoundary10}
A.~Corichi and E.~Wilson-Ewing, {\it {Surface terms, asymptotics and
  thermodynamics of the Holst action}},  Class. Quantum Grav. {\bf 27} (2010),
  no.~20 [\href{http://arXiv.org/abs/1005.3298}{{\tt 1005.3298}}].

\bibitem{AshtekarBoundaryTerms}
A.~Ashtekar, J.~Engle and D.~Sloan, {\it {Asymptotics and Hamiltonians in a
  First order formalism}},  Class.Quant.Grav. {\bf 25} (2008) 095020
  [\href{http://arXiv.org/abs/0802.2527}{{\tt 0802.2527}}].
%%CITATION = ARXIV:0802.2527;%%

\bibitem{BianchiWielandSurf}
E.~Bianchi and W.~Wieland, {\it {Horizon energy as the boost boundary term in
  general relativity and loop gravity}},
  \href{http://arXiv.org/abs/1205.5325}{{\tt 1205.5325}}.
%%CITATION = ARXIV:1205.5325;%%

\bibitem{SmolinGRasSFstate}
L.~Smolin, {\it {General relativity as the equation of state of spin foam}},
  \href{http://arXiv.org/abs/1205.5529}{{\tt 1205.5529}}.
%%CITATION = ARXIV:1205.5529;%%

\bibitem{Wieland1}
W.~Wieland, {\it {Complex Ashtekar variables and reality conditions for Holst's
  action}},  Annales Henri Poincare {\bf 13} (2012) 425--448
  [\href{http://arXiv.org/abs/1012.1738}{{\tt 1012.1738}}].
%%CITATION = ARXIV:1012.1738;%%

\bibitem{Capo2}
R.~Capovilla, T.~Jacobson, J.~Dell and L.~Mason, {\it {Selfdual two forms and
  gravity}},  Class.Quant.Grav. {\bf 8} (1991) 41--57.
%%CITATION = CQGRD,8,41;%%

\bibitem{MikeLR}
M.~P. Reisenberger, {\it {Classical Euclidean general relativity from
  'left-handed area = right-handed area'}},  Class. Quantum Grav. {\bf 16}
  (1998) 1357 [\href{http://arXiv.org/abs/gr-qc/9804061}{{\tt gr-qc/9804061}}].
%%CITATION = GR-QC/9804061;%%

\bibitem{Iobimetric}
S.~Speziale, {\it {Bi-metric theory of gravity from the non-chiral Plebanski
  action}},  Phys.Rev. {\bf D82} (2010) 064003
  [\href{http://arXiv.org/abs/1003.4701}{{\tt 1003.4701}}].
%%CITATION = ARXIV:1003.4701;%%

\bibitem{AshtekarBook}
A.~Ashtekar, {\em {Lectures on Non-Pertubative Canonical Gravity}}.
\newblock World Scientific, 1991.

\bibitem{Barros}
N.~Barros~e Sa, {\it {Hamiltonian analysis of general relativity with the
  Immirzi parameter}},  Int.J.Mod.Phys. {\bf D10} (2001) 261--272
  [\href{http://arXiv.org/abs/gr-qc/0006013}{{\tt gr-qc/0006013}}].
%%CITATION = GR-QC/0006013;%%

\bibitem{BuffenoirPleb}
E.~Buffenoir, M.~Henneaux, K.~Noui and P.~Roche, {\it {Hamiltonian analysis of
  Plebanski theory}},  Class.Quant.Grav. {\bf 21} (2004) 5203--5220
  [\href{http://arXiv.org/abs/gr-qc/0404041}{{\tt gr-qc/0404041}}].
%%CITATION = GR-QC/0404041;%%

\bibitem{AlexandrovSO4cov}
S.~Alexandrov, {\it {SO(4,C) covariant Ashtekar-Barbero gravity and the Immirzi
  parameter}},  Class.Quant.Grav. {\bf 17} (2000) 4255--4268
  [\href{http://arXiv.org/abs/gr-qc/0005085}{{\tt gr-qc/0005085}}].

\bibitem{AlexandrovReality}
S.~Alexandrov, {\it {Reality conditions for Ashtekar gravity from
  Lorentz-covariant formulation}},  Class.Quant.Grav. {\bf 23} (2006)
  1837--1850 [\href{http://arXiv.org/abs/gr-qc/0510050}{{\tt gr-qc/0510050}}].
%%CITATION = GR-QC/0510050;%%

\bibitem{AlexandrovChoice}
S.~Alexandrov, {\it {On choice of connection in loop quantum gravity}},
  Phys.Rev. {\bf D65} (2002) 024011
  [\href{http://arXiv.org/abs/gr-qc/0107071}{{\tt gr-qc/0107071}}].
%%CITATION = GR-QC/0107071;%%

\bibitem{Cianfrani}
F.~Cianfrani and G.~Montani, {\it {Towards Loop Quantum Gravity without the
  time gauge}},  Phys.Rev.Lett. {\bf 102} (2009) 091301
  [\href{http://arXiv.org/abs/0811.1916}{{\tt 0811.1916}}].
%%CITATION = ARXIV:0811.1916;%%

\bibitem{GeillerNewLook}
M.~Geiller, M.~Lachieze-Rey and K.~Noui, {\it {A new look at Lorentz-Covariant
  Loop Quantum Gravity}},  Phys.Rev. {\bf D84} (2011) 044002
  [\href{http://arXiv.org/abs/1105.4194}{{\tt 1105.4194}}].

\bibitem{Ashtekar:1987gu}
A.~Ashtekar, {\it {New Hamiltonian Formulation of General Relativity}},
  Phys.Rev. {\bf D36} (1987) 1587--1602.
%%CITATION = PHRVA,D36,1587;%%

\bibitem{Barbero}
J.~Barbero~G., {\it {Real Ashtekar variables for Lorentzian signature space
  times}},  Phys.Rev. {\bf D51} (1995) 5507--5510
  [\href{http://arXiv.org/abs/gr-qc/9410014}{{\tt gr-qc/9410014}}].
%%CITATION = GR-QC/9410014;%%

\bibitem{Immirzi96real}
G.~Immirzi, {\it {Real and complex connections for canonical gravity}},
  Class.Quant.Grav. {\bf 14} (1997) L177--L181
  [\href{http://arXiv.org/abs/gr-qc/9612030}{{\tt gr-qc/9612030}}].
%%CITATION = GR-QC/9612030;%%

\bibitem{ThiemannBook}
T.~Thiemann, {\em {Modern canonical quantum general relativity}}.
\newblock Cambridge University Press, 2001.
%%CITATION = GR-QC/0110034;%%

\bibitem{AshtekarReport}
A.~Ashtekar and J.~Lewandowski, {\it {Background independent quantum gravity: A
  Status report}},  Class.Quant.Grav. {\bf 21} (2004) R53
  [\href{http://arXiv.org/abs/gr-qc/0404018}{{\tt gr-qc/0404018}}].
%%CITATION = GR-QC/0404018;%%

\bibitem{AshtekarQG3}
A.~Ashtekar, A.~Corichi and J.~A. Zapata, {\it {Quantum theory of geometry III:
  Noncommutativity of Riemannian structures}},  Class.Quant.Grav. {\bf 15}
  (1998) 2955--2972 [\href{http://arXiv.org/abs/gr-qc/9806041}{{\tt
  gr-qc/9806041}}].

\bibitem{FreidelGeiller}
L.~Freidel, M.~Geiller and J.~Ziprick, {\it {Continuous formulation of the Loop
  Quantum Gravity phase space}},  \href{http://arXiv.org/abs/1110.4833}{{\tt
  1110.4833}}.

\bibitem{BarrettCrane}
J.~W. Barrett and L.~Crane, {\it {Relativistic spin networks and quantum
  gravity}},  J.Math.Phys. {\bf 39} (1998) 3296--3302
  [\href{http://arXiv.org/abs/gr-qc/9709028}{{\tt gr-qc/9709028}}].
%%CITATION = GR-QC/9709028;%%

\bibitem{BarrettCraneLor}
J.~W. Barrett and L.~Crane, {\it {A Lorentzian signature model for quantum
  general relativity}},  Class.Quant.Grav. {\bf 17} (2000) 3101--3118
  [\href{http://arXiv.org/abs/gr-qc/9904025}{{\tt gr-qc/9904025}}].
%%CITATION = GR-QC/9904025;%%

\bibitem{BaezBarrett}
J.~C. Baez and J.~W. Barrett, {\it {The Quantum tetrahedron in three-dimensions
  and four-dimensions}},  Adv.Theor.Math.Phys. {\bf 3} (1999) 815--850
  [\href{http://arXiv.org/abs/gr-qc/9903060}{{\tt gr-qc/9903060}}].
%%CITATION = GR-QC/9903060;%%

\bibitem{IoSigmaDiscrete}
M.~Dupuis, J.~P. Ryan and S.~Speziale, {\it {Discrete gravity models and Loop
  Quantum Gravity: a short review}},
  \href{http://arXiv.org/abs/1204.5394}{{\tt 1204.5394}}.
%%CITATION = ARXIV:1204.5394;%%

\bibitem{IoZako}
M.~Dupuis, S.~Speziale and J.~Tambornino, {\it {Spinors and Twistors in Loop
  Gravity and Spin Foams}},  \href{http://arXiv.org/abs/1201.2120}{{\tt
  1201.2120}}.
%%CITATION = ARXIV:1201.2120;%%

\bibitem{AlexandrovSigma}
S.~Alexandrov, M.~Geiller and K.~Noui, {\it {Spin Foams and Canonical
  Quantization}},  \href{http://arXiv.org/abs/1112.1961}{{\tt 1112.1961}}.
%%CITATION = ARXIV:1112.1961;%%

\bibitem{AlexandrovHilbert02}
S.~Alexandrov, {\it {Hilbert space structure of covariant loop quantum
  gravity}},  Phys.Rev. {\bf D66} (2002) 024028
  [\href{http://arXiv.org/abs/gr-qc/0201087}{{\tt gr-qc/0201087}}].
%%CITATION = GR-QC/0201087;%%

\bibitem{Anishetty}
R.~Anishetty and A.~Vytheeswaran, {\it {Gauge invariance in second class
  constrained systems}},  J.Phys.A {\bf A26} (1993) 5613--5620.
%%CITATION = JPAGB,A26,5613;%%

\bibitem{Mitra}
P.~Mitra and R.~Rajaraman, {\it {Gauge invariant reformulation of theories with
  second class constraints}},  Class.Quant.Grav. {\bf 7} (1990) 2131.
%%CITATION = CQGRD,7,2131;%%

\bibitem{Bodendorfer1}
N.~Bodendorfer, T.~Thiemann and A.~Thurn, {\it {New Variables for Classical and
  Quantum Gravity in all Dimensions I. Hamiltonian Analysis}},
  \href{http://arXiv.org/abs/1105.3703}{{\tt 1105.3703}}.
%%CITATION = ARXIV:1105.3703;%%

\bibitem{BodendorferSimpl}
N.~Bodendorfer, T.~Thiemann and A.~Thurn, {\it {On the Implementation of the
  Canonical Quantum Simplicity Constraint}},
  \href{http://arXiv.org/abs/1105.3708}{{\tt 1105.3708}}.
%%CITATION = ARXIV:1105.3708;%%

\bibitem{DittrichRyan}
B.~Dittrich and J.~P. Ryan, {\it {Phase space descriptions for simplicial 4d
  geometries}},  Class.Quant.Grav. {\bf 28} (2011) 065006
  [\href{http://arXiv.org/abs/0807.2806}{{\tt 0807.2806}}].

\bibitem{DittrichRyan2}
B.~Dittrich and J.~P. Ryan, {\it {Simplicity in simplicial phase space}},
  Phys.Rev. {\bf D82} (2010) 064026 [\href{http://arXiv.org/abs/1006.4295}{{\tt
  1006.4295}}].

\bibitem{DittrichSpeziale}
B.~Dittrich and S.~Speziale, {\it {Area-angle variables for general
  relativity}},  New J.Phys. {\bf 10} (2008) 083006
  [\href{http://arXiv.org/abs/0802.0864}{{\tt 0802.0864}}].
%%CITATION = ARXIV:0802.0864;%%

\bibitem{IoPoly}
E.~Bianchi, P.~Dona and S.~Speziale, {\it {Polyhedra in loop quantum gravity}},
   Phys.Rev. {\bf D83} (2011) 044035
  [\href{http://arXiv.org/abs/1009.3402}{{\tt 1009.3402}}].

\bibitem{IoCarloGraph}
C.~Rovelli and S.~Speziale, {\it {On the geometry of loop quantum gravity on a
  graph}},  Phys.Rev. {\bf D82} (2010) 044018
  [\href{http://arXiv.org/abs/1005.2927}{{\tt 1005.2927}}].

\bibitem{Ruhl}
W.~Ruhl, {\em {The Lorentz Group and Harmonic Analysis}}.
\newblock W. A. Benjamin, 1970.

\bibitem{GelfandLorentz}
I.~M. Gelfand, R.~A. Minlos and Z.~Y. Shapiro, {\em {Representations of the
  rotation and Lorentz groups and their applications}}.
\newblock Pergamon Press, Oxford, 1963.

\bibitem{DucLorentz}
D.~vong Duc and N.~van Hieu, {\it {On the theory of unitary representations of
  the $SL(2,\mathbb{C})$ group}},  {Acta Physica Academiae Scientiarum
  Hungaricae} {\bf 222} (1967), no.~1-4 201--219.

\bibitem{CarloEdParity}
C.~Rovelli and E.~Wilson-Ewing, {\it {Discrete Symmetries in Covariant LQG}},
  \href{http://arXiv.org/abs/1205.0733}{{\tt 1205.0733}}.
%%CITATION = ARXIV:1205.0733;%%

\bibitem{Thiemannfermihiggs}
T.~Thiemann, {\it Kinematical hilbert spaces for fermionic and higgs quantum
  field theories},  Class. Quantum Grav. {\bf 15} (1997) 1487--1512
  [\href{http://arXiv.org/abs/gr-qc/9705021}{{\tt gr-qc/9705021}}].

\bibitem{EteraProj}
E.~R. Livine, {\it {Projected spin networks for Lorentz connection: Linking
  spin foams and loop gravity}},  Class.Quant.Grav. {\bf 19} (2002) 5525--5542
  [\href{http://arXiv.org/abs/gr-qc/0207084}{{\tt gr-qc/0207084}}].
%%CITATION = GR-QC/0207084;%%

\bibitem{EteraTamboSpinor}
E.~R. Livine and J.~Tambornino, {\it {Spinor Representation for Loop Quantum
  Gravity}},  \href{http://arXiv.org/abs/1105.3385}{{\tt 1105.3385}}.

\bibitem{AlexandrovSimplClosure}
S.~Alexandrov, {\it {Simplicity and closure constraints in spin foam models of
  gravity}},  Phys.Rev. {\bf D78} (2008) 044033
  [\href{http://arXiv.org/abs/0802.3389}{{\tt 0802.3389}}].
%%CITATION = ARXIV:0802.3389;%%

\bibitem{MikeLeft}
M.~P. Reisenberger, {\it {A Left-handed simplicial action for Euclidean general
  relativity}},  Class.Quant.Grav. {\bf 14} (1997) 1753--1770
  [\href{http://arXiv.org/abs/gr-qc/9609002}{{\tt gr-qc/9609002}}].
%%CITATION = GR-QC/9609002;%%

\bibitem{EnglePereiraFiniteness}
J.~Engle and R.~Pereira, {\it {Regularization and finiteness of the Lorentzian
  LQG vertices}},  Phys.Rev. {\bf D79} (2009) 084034
  [\href{http://arXiv.org/abs/0805.4696}{{\tt 0805.4696}}].
%%CITATION = ARXIV:0805.4696;%%

\bibitem{CarloZako}
C.~Rovelli, {\it {Zakopane lectures on loop gravity}},
  \href{http://arXiv.org/abs/1102.3660}{{\tt 1102.3660}}.
%%CITATION = ARXIV:1102.3660;%%

\bibitem{Regge}
T.~Regge, {\it General relativity without coordinates},  Nuovo Cim. {\bf 19}
  (1961) 558--571.
%%CITATION = NUCIA,19,558;%%

\bibitem{Bonzom:2008ru}
V.~Bonzom and E.~R. Livine, {\it {A Lagrangian approach to the Barrett-Crane
  spin foam model}},  Phys.Rev. {\bf D79} (2009) 064034
  [\href{http://arXiv.org/abs/0812.3456}{{\tt 0812.3456}}].
%%CITATION = ARXIV:0812.3456;%%

\bibitem{Bonzom:2009hw}
V.~Bonzom, {\it {Spin foam models for quantum gravity from lattice path
  integrals}},  Phys.Rev. {\bf D80} (2009) 064028
  [\href{http://arXiv.org/abs/0905.1501}{{\tt 0905.1501}}].
%%CITATION = ARXIV:0905.1501;%%

\bibitem{Bonzom:2009wm}
V.~Bonzom, {\it {From lattice BF gauge theory to area-angle Regge calculus}},
  Class.Quant.Grav. {\bf 26} (2009) 155020
  [\href{http://arXiv.org/abs/0903.0267}{{\tt 0903.0267}}].
%%CITATION = ARXIV:0903.0267;%%

\bibitem{BaratinOriti}
A.~Baratin and D.~Oriti, {\it {Group field theory and simplicial gravity path
  integrals: A model for Holst-Plebanski gravity}},  Phys.Rev. {\bf D85} (2012)
  044003 [\href{http://arXiv.org/abs/1111.5842}{{\tt 1111.5842}}].
%%CITATION = ARXIV:1111.5842;%%

\bibitem{EteraHoloEucl}
M.~Dupuis and E.~R. Livine, {\it {Holomorphic Simplicity Constraints for 4d
  Spinfoam Models}},  Class.Quant.Grav. {\bf 28} (2011) 215022
  [\href{http://arXiv.org/abs/1104.3683}{{\tt 1104.3683}}].
%%CITATION = ARXIV:1104.3683;%%

\bibitem{abramstegun}
M.~Abramowitz and I.~A. Stegun, {\em Handbook of Mathematical Functions with
  Formulas}.
\newblock Dover Publications, New York, 1965.

\end{thebibliography}
\end{document}